\documentclass[twocolumn,english,rmp,aps,longbibliography]{revtex4-2}
\usepackage[T1]{fontenc}
\usepackage[utf8]{inputenc}
\usepackage{physics}
\usepackage{amsmath}
\usepackage{amssymb}
\usepackage{graphicx}
\usepackage{esint}
\usepackage{changes}
\usepackage{comment}
\usepackage{makecell}
\usepackage{mathtools}
\usepackage{hyperref}
\hypersetup{
    colorlinks=true,
    linkcolor=blue,
    filecolor=magenta,      
    urlcolor=cyan,
}

\makeatletter

\@ifundefined{textcolor}{}
{
 \definecolor{BLACK}{gray}{0}
 \definecolor{WHITE}{gray}{1}
 \definecolor{RED}{rgb}{1,0,0}
 \definecolor{GREEN}{rgb}{0,1,0}
 \definecolor{BLUE}{rgb}{0,0,1}
 \definecolor{CYAN}{cmyk}{1,0,0,0}
 \definecolor{MAGENTA}{cmyk}{0,1,0,0}
 \definecolor{YELLOW}{cmyk}{0,0,1,0}
 }

\usepackage{amsfonts}
\providecommand{\U}[1]{\protect\rule{.1in}{.1in}}

\usepackage{lmodern}
\usepackage[T1]{fontenc}

\makeatother

\usepackage{babel}
\begin{document}

\selectlanguage{english}

\title{Frequency combs and 
coherent dissipative structures in nonlinear optical microresonators}

\author{Tobias Herr}
\affiliation{Deutsches Elektronen-Synchrotron DESY, Notkestr. 85, 22607 Hamburg, Germany}
\affiliation{Institute of Experimental Physics, Universität Hamburg, Luruper Chaussee 149, 22607 Hamburg, Germany}
\email{tobias.herr@desy.de}

\author{Alexey Tikan}
\affiliation{Laboratoire Temps-Fréquence, Université de Neuchâtel, Avenue de Bellevaux 51, Neuchâtel, Switzerland}

\email{alexey.tikan@unine.ch}

\author{Tobias J. Kippenberg}
\affiliation{Swiss Federal Institute of Technology (EPFL), Institute of Condensed
Matter Physics CH-1015 Lausanne, Switzerland }
\email{tobias.kippenberg@epfl.ch}

\selectlanguage{english}

\begin{abstract}
Laser-driven high-$Q$ Kerr-nonlinear optical microresonators enable parametric oscillation with low-power continuous-wave lasers and host a variety of coherent dissipative structures, including dissipative Kerr solitons and switching waves. These time-periodic structures constitute coherent optical frequency combs, and photonic-chip integration has miniaturized them to the chip scale. Such photonic-integrated, microresonator-based frequency combs -- often termed ``microcombs'' or ``Kerr combs'' -- have been demonstrated in various system-level and scientific applications. They complement femtosecond-laser-based frequency combs when high repetition rates, broad bandwidths, or high power per comb line are needed. This review introduces the field of microcombs and outlines the fundamental physical principles governing the generation of coherent frequency combs in microresonators.\\
\\
\textcolor{cyan}{Comments welcome: microcomb-review@desy.de}
\end{abstract}
\maketitle

\tableofcontents{}

\begin{table*}
\caption{\textbf{List of symbols.} Key symbols and formulas used in this review.}
\centering
\begin{ruledtabular}
\begin{tabular}{ll}
Symbol & Meaning \\
\hline

$A (t,\phi)=\sqrt{\kappa/2g_\mathrm{K}} \Psi(\tau,\theta)$ & Normalized complex field envelopes (photon number normalization) \\
$a_\mu(t)=\sqrt{\kappa/2g_\mathrm{K}} \psi_\mu(\tau)$ & Normalized complex mode field amplitude (photon number normalization) \\
$D_{\mathrm{int}}(\mu) = \omega_\mu - (\omega_0 + \mu D_1) = \sum_{n=2}^\infty D_n \frac{\mu^n}{n!}$ & Integrated dispersion \\

$d_2 = 2 D_2 / \kappa$ & Normalized dispersion coefficient \\

$\mathcal{F} = 2\pi \mathrm{FSR}/\kappa$ & Finesse of the resonator \\
$\mathrm{FSR}(\mu) = \frac{1}{2\pi}  \partial \omega_\mu / \partial \mu $ & Free spectral range \\
$f^2 = |s_{\mathrm{in}}|^2 (8 \eta g_\mathrm{K}/\kappa^{2}  ) $ & Normalized pump power \\

$   g_\mathrm{K} = \hbar\omega_0^{2}c n_2/(n_\mathrm{eff}^2 V_\mathrm{eff} )$ & Kerr single-photon coupling strength \\

$n_2= 3  \chi^{(3)}/ (4 n_\mathrm{eff}^2(\omega_0) \varepsilon_{0} c)$ & Kerr nonlinear index \\

$Q = \omega_0 / \kappa$ & Quality factor \\

$|s_{\mathrm{in}}|^2 = P_{\mathrm{in}} / (\hbar \omega_\mathrm{p})$ & Pump photon flux \\

$\gamma = (\omega_\mathrm{p} / c) \cdot (n_2 / A_\mathrm{eff})$ & Effective nonlinear coefficient for Kerr nonlinearity \\

$\zeta_0 = 2 (\omega_0 - \omega_\mathrm{p}) / \kappa$ & Normalized detuning \\

$\kappa = \kappa_{\mathrm{ex}} + \kappa_\mathrm{0}$ & Total loss rate, sum of external and intrinsic losses \\
$\mu = m - m_0$ & Mode number relative to the pump \\

$\tau=2t/\kappa$ & Normalized (slow) time\\

$\omega_\mu = \omega_0 + D_1\mu + \tfrac{1}{2} D_2 \mu^2 + ...$ & Circular eigenfrequency of mode $m$ \\

\end{tabular}
\end{ruledtabular}
\end{table*}

\begin{minipage}{\columnwidth}
\noindent
\textbf{List of abbreviations} 

\vspace{10pt}
\begin{tabular}{@{}ll@{}}
\textbf{AMX} & Avoided mode crossing \\
\textbf{CME} & Coupled mode equations \\
\textbf{CW} & Continuous wave \\
\textbf{DKS} & Dissipative Kerr soliton \\
\textbf{DW} & Dispersive wave \\
\textbf{FI} & Faraday instability \\
\textbf{FSR} & Free spectral range \\
\textbf{FWM} & Four-wave mixing \\
\textbf{GVD} & Group velocity dispersion \\
\textbf{HOD} & Higher-order dispersion \\
\textbf{LLE} & Lugiato--Lefever Equation \\
\textbf{MI} & Modulation instability \\
\textbf{NDR} & Nonlinear dispersion relation \\
\textbf{NLSE} & Nonlinear Schrödinger equation \\
\textbf{OPO} & Optical parametric oscillation \\
\textbf{PSD} & Power spectral density \\
\textbf{PQS} & Pure quartic solitons \\
\textbf{PhCR} & Photonic crystal resonator \\
\textbf{RBW} & Resolution bandwidth \\
\textbf{RIN} & Relative intensity noise \\
\textbf{SBS} & Stimulated Brillouin scattering \\
\textbf{SIL} & Self-injection locking \\
\textbf{SNR} & Signal-to-noise ratio \\
\textbf{SPM} & Self-phase modulation \\
\textbf{SRS} & Stimulated Raman scattering \\
\textbf{SW} & Switching wave \\
\textbf{TOD} & Third-order dispersion \\
\textbf{TPA} & Two-photon absorption \\
\textbf{TRN} & Thermo-refractive noise \\
\textbf{WGM} & Whispering gallery mode \\
\textbf{XPM} & Cross-phase modulation \\
\textbf{ZDS} & Zero-dispersion soliton \\
\end{tabular}

\end{minipage}

\section{Introduction}
\label{sec:Introduction}

\subsection{Dissipative patterns and microcombs}
A remarkable phenomenon in driven-dissipative nonlinear systems is the spontaneous formation of stationary patterns from uniform states—a concept introduced by Ilya Prigogine as \emph{dissipative structures}~\cite{nicolis1977SelforganizationNonequilibriumSystemsa}. These self-organizing patterns emerge in open, nonequilibrium systems, where the onset of a Turing instability \cite{turing1990ChemicalBasisMorphogenesis} triggers the self-organization of the system, and have been observed across chemistry, biology, hydrodynamics, and even social systems.

Over the past decades, significant advances in high-$Q$ optical microresonators~\cite{armani2003ultra} have enabled access to traditionally weak nonlinear effects for the first time with lower power (sub mW level) and continuous wave (CW) lasers. 
These systems support new classes of optical dissipative structures --``microcombs'' (also known as ``Kerr (frequency) combs'') -- which arise from parametric oscillations in continuously driven nonlinear resonators~\cite{delhaye2007OpticalFrequencyComb, kippenberg2004KerrNonlinearityOpticalParametrica, matsko2005optical}.
Unlike spatial pattern formation, optical microresonators exhibit temporal dissipative structures, such as dissipative Kerr solitons (DKS) in the anomalous dispersion regime~\cite{herr2014TemporalSolitonsOptical} and switching waves (SW) in the normal dispersion regime~\cite{xue2015ModelockedDarkPulse}. These waveforms correspond in the frequency domain to coherent optical frequency combs: broadband spectra of discrete, coherent, and equidistant frequency components (comb lines)
\begin{equation}
	f_\mu = f_\mathrm{p} + \mu \cdot f_\mathrm{rep},
 \label{eq:ch2:comb_equation}
\end{equation}
where $f_{\mathrm{p}}$ is the CW pump laser frequency (which is part of the microcomb), $\mu$ is an integer line index (counted relative to the central mode), and $f_\mathrm{rep}$ the spacing between the frequency components (repetition rate). 

In contrast to traditional optical frequency combs from femtosecond mode-locked lasers~\cite{udem2002OpticalFrequencyMetrology,cundiff2003ColloquiumFemtosecondOpticala,fortier201920YearsDevelopments,picque2019FrequencyCombSpectroscopy,diddams2020OpticalFrequencyCombs} which rely on atomic gain media and saturable absorbers to achieve pulse formation, microcombs offer several distinct characteristics that complement conventional frequency comb technology.

Owing to the short roundtrip time in microresonators, the repetition rate of microcombs is very high and usually in the range from 10 to 1000~GHz. Compared to conventional frequency comb sources, microcombs exhibit lower per-pulse energy, but relatively high power per comb line. Additionally, the wide spacing between microcomb lines allows them to be easily resolved and separated using gratings or wavelength-division multiplexers -- an essential requirement for many of their applications. 

As opposed to traditional mode-locked lasers, the microcomb generation process relies on nonlinear parametric gain. The phase-sensitive nature of these gain processes enables frequency comb generation with low noise and the broadband nature of parametric gain enables direct generation of octave-spanning spectra. 
Finally, microcombs based on photonic integrated circuits offer a pathway to miniaturization and scaling of frequency comb technology.

\subsection{Historical development of microcombs}

\begin{figure}
\includegraphics[width=1\columnwidth]{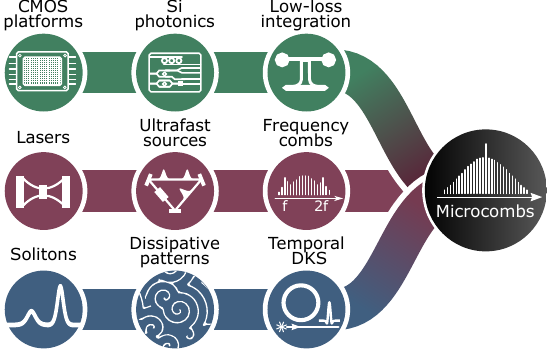}
\caption{\textbf{The research field of microcombs.} The study of solitons and frequency combs in driven nonlinear resonators unifies integrated photonics, frequency metrology, and pattern formation.
} \label{fig:ch1:circles}
\end{figure}

Microcombs emerged from the confluence of previously distinct fields: nonlinear optical microresonators, frequency metrology, and nonlinear spatial pattern formation as well as soliton Physics as depicted in Fig.~\ref{fig:ch1:circles}. A key enabling factor was the fabrication of high quality factor ($Q$) resonators, dramatically reducing the power threshold for nonlinear optical effects, such as parametric oscillations via four-wave mixing (FWM). An early starting point was the demonstration of high-$Q$ whispering-gallery-mode (WGM) resonators, first reported by Braginsky et al. in 1989~\cite{braginsky1989QualityfactorNonlinearPropertiesb}. Subsequent experiments observed low-pump power nonlinear phenomena, including thermal bistability~\cite{ilchenko1992thermal}, intrinsic Kerr nonlinearity at cryogenic temperatures~\cite{treussart1998EvidenceIntrinsicKerr}, and stimulated Raman scattering~\cite{spillane2002UltralowthresholdRamanLasera}. Kerr-induced optical parametric oscillations (OPO), also called ``hyper-parametric oscillations'', although previously known in nonlinear optics~\cite{bloembergen2010NonlinearOptics}, were first observed experimentally in high-$Q$ toroidal and crystalline resonators in 2004 ~\cite{kippenberg2004KerrNonlinearityOpticalParametrica,savchenkov2004LowThresholdOptical}, by employing CW lasers at only sub-mW of pump laser power. These studies highlight the dramatic reduction of threshold power for nonlinear oscillations due to the $1/Q^{2}$ scaling. 

\begin{figure}[h]
	\includegraphics[width=1.0\columnwidth]{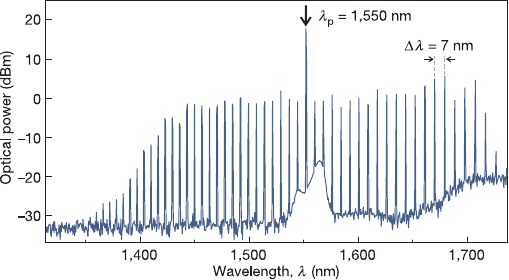}
	\caption{\textbf{Observation of an optical microcomb.} A CW-pumped high-Q toroid microcavity produces a frequency comb via four-wave mixing \cite{delhaye2007OpticalFrequencyComb}.} \label{fig:ch1:first_comb}
\end{figure}

In 2007, it was experimentally demonstrated that Kerr-nonlinear optical microresonators can generate broadband, optical frequency combs~\cite{delhaye2007OpticalFrequencyComb} through cascaded FWM parametric oscillation (see Fig.~\ref{fig:ch1:first_comb}). By comparison with a conventional mode-locked fiber laser comb, the equidistance of the comb lines could be proven at the level of 1 part in $10^{17}$.

Over a short duration of time, optical broadband spectra were demonstrated in numerous platforms, including crystalline resonators\cite{savchenkov2008TunableOpticalFrequency}, 
CMOS-compatible platforms such as silicon nitride ($\mathrm{Si_{3}N_{4}}$)~\cite{foster2011SiliconbasedMonolithicOptical,levy2010CMOScompatibleMultiplewavelengthOscillator,razzari2010CMOScompatibleIntegratedOptical, johnson2012ChipbasedFrequencyCombs}, Hydex glass~\cite{moss2013NewCMOScompatiblePlatforms}, as well as aluminum nitride~\cite{jung2016AluminumNitrideNonlinear} or diamond~\cite{hausmann2014DiamondNonlinearPhotonics}; in some cases octave spanning spectra were observed~\cite{delhaye2011OctaveSpanningTunable, okawachi2011OctavespanningFrequencyComb}.

While initial observations suggested cascaded FWM leads to equidistant combs, 
later experiments found that equidistant combs were not always generated and often spectra of non-equidistant lines of incommensurate `sub-combs' or incoherent spectral components with high-noise were observed~\cite{savchenkov2008tunable, ferdous2011SpectralLinebylinePulse, delhaye2011OctaveSpanningTunable, herr2012UniversalFormationDynamics, li2012LowPumpPowerLowPhaseNoiseMicrowave}. The role of strong anomalous dispersion (at least around the pump laser) 
was understood in 2012~\cite{herr2012UniversalFormationDynamics}. However, such strong dispersion was only possible in a few platforms with very large free spectral range (FSR), or systems where unpredictable avoided mode crossings (AMX) with higher order modes would induce strong anomalous dispersion.

\begin{figure}
\includegraphics[width=1\columnwidth]{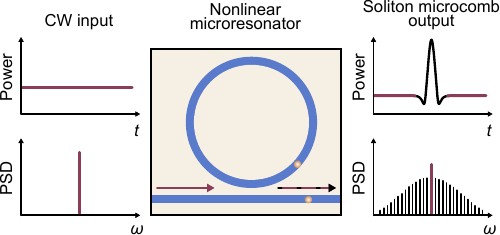}
\caption{\textbf{The basic principle of the dissipative soliton microcomb generation.} The CW input is swept from the blue to the red side of the microcavity resonance, which leads to the formation of temporal nonlinear dissipative structures corresponding to a coherent frequency comb in the spectral domain.} \label{fig:ch1:microcomb_generaion}
\end{figure}

The theoretical understanding significantly advanced when the process of microcomb generation was connected to the Lugiato–Lefever Equation (LLE)~\cite{lugiato1987SpatialDissipativeStructures,matsko2009WHISPERINGGALLERYMODE,leo2010TemporalCavitySolitons,matsko2011ModelockedKerrFrequency}, inspired by earlier works~\cite{haelterman1992DissipativeModulationInstability,haelterman1992AdditivemodulationinstabilityRingLaser, wabnitz1993SuppressionInteractionsPhaselocked} that studied temporal DKSs theoretically~\cite{haelterman1992DissipativeModulationInstability,wabnitz1993SuppressionInteractionsPhaselocked} in the context of optical cavities or more generally as a solution to the driven-dissipative nonlinear Schr\"odinger equation (NLSE)~\cite{kaup1978SolitonsParticlesOscillators,nozaki1986LowdimensionalChaosDriven,barashenkov1996ExistenceStabilityChart}. The first observation of temporal DKS was made in optical fibers~\cite{leo2010TemporalCavitySolitons}, where they could be triggered by `writing pulses'. These works implied the existence of temporal DKS i.e., pulsed waveforms stably circulating in the microresonator. An overview of the DKS microcomb generation is depicted in Fig.~\ref{fig:ch1:microcomb_generaion}.

However, they were not observed in microresonators, where, due to their high finesse, external writing pulses could not easily be applied, and accessing \emph{red-detuned} pumping -- the condition under which DKS exist~\cite{chembo2013SpatiotemporalLugiatoLefeverFormalism,coen2013ModelingOctavespanningKerra,godey2014StabilityAnalysisSpatiotemporal} -- is usually associated with thermal instability of the microresonator. 

A breakthrough came with the observation of discrete 'steps' in the transmission spectrum of crystalline resonators, which were shown to correspond to formation of stable single and multiple DKSs that spontaneously formed without external writing pulses; importantly, the system was shown to be thermally stable, provided the pump laser was tuned into the resonance sufficiently fast to achieve a thermal steady-state~\cite{herr2014TemporalSolitonsOptical}. These step features are a hallmark of DKS formation in microresonators. Remarkably, a similar behavior is present in nonlinear microwave resonators~\cite{gasch1984MultistabilitySolitonModes} highlighting their universality. 

Following the initial observation of DKS in crystalline $\mathrm{MgF_2}$ resonators, their formation was subsequently demonstrated in other microresonator platforms, including high-$Q$ silica wedge resonators~\cite{yi2015SolitonFrequencyComb,yang2016BroadbandDispersionengineeredMicroresonator}, confirming the earlier findings.
A crucial technological advancement was the demonstration that DKS can also be accessed in $\mathrm{Si_{3}N_{4}}$ foundry-compatible photonic integrated circuits~\cite{brasch2016PhotonicChipBased}, supporting the integration of complex photonic structures on a single chip. Silicon nitride, a CMOS-compatible material with a large 5~eV bandgap, avoids two-photon absorption -- a limitation of earlier integrated platforms, particularly those based on silicon. Improved fabrication methods~\cite{pfeiffer2016PhotonicDamasceneProcess, ji2017UltralowlossOnchipResonators} drastically lowered the propagation losses in thick $\mathrm{Si_{3}N_{4}}$ waveguides with anomalous group velocity dispersion -- a prerequisite for DKS -- and led to a dramatic reduction in pump power requirements, from watt-level in chip-integrated system to milliwatt levels. This enabled more efficient DKS generation and facilitated their integration with chip-scale pump lasers~\cite{stern2018BatteryoperatedIntegratedFrequency,raja_electrically_2019}.

\begin{figure}
\includegraphics[width=1\linewidth]{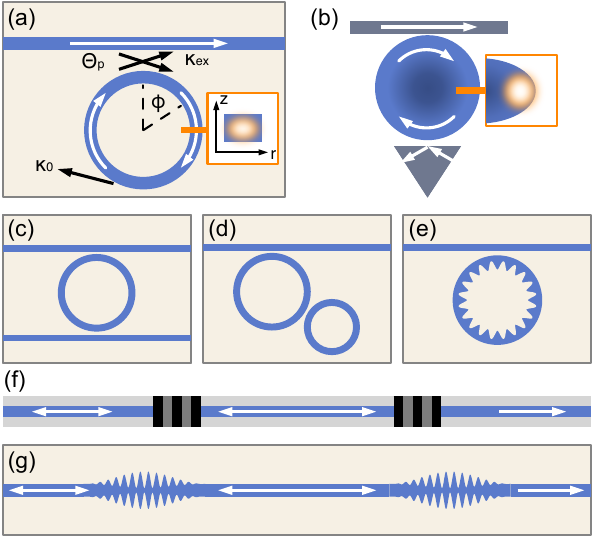}
\caption{\textbf{Microresonator geometries}. 
(a) Waveguide microresonators with \textit{bus} waveguide for evanescent coupling. Inset: optical intensity profile (white/orange indicates higher/lower intensity). (b) Whispering-gallery mode resonators with coupling waveguide and coupling prism. (c) In addition to the main (\textit{add}) input-output coupling waveguide, a second (\textit{drop}) coupling waveguide can be implemented. (d) Resonators can be evanescently coupled. (e) Photonic crystal ring resonator. (f,g) Fabry-P\'erot resonators based on dielectric mirrors or photonic crystal reflectors. }
\label{fig:ch1:platforms_schematics}
\end{figure}

\begin{figure*}
\includegraphics[width=1\linewidth]{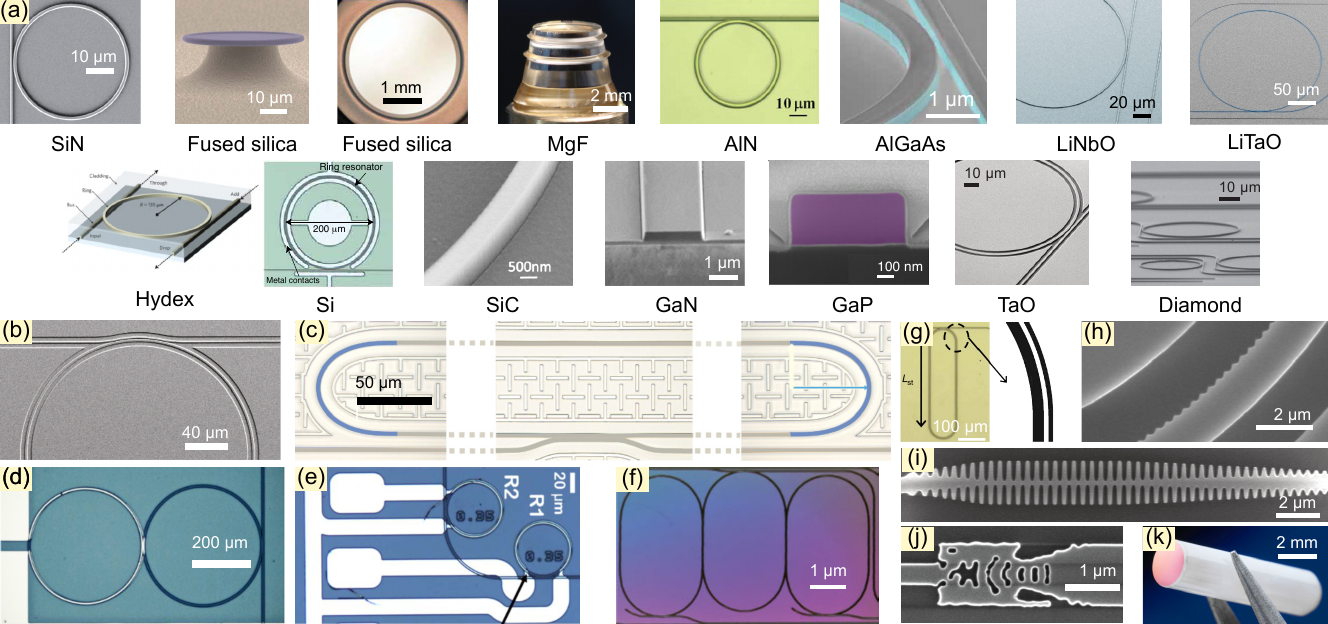}
\caption{\textbf{Examples of microresonator platforms}. (a) Resonators made of various materials: SiN~\cite{levy2010cmos}, Fused silica~\cite{delhaye2007OpticalFrequencyComb, yi2015SolitonFrequencyComb}, MgF~\cite{herr2014ModeSpectrumTemporal}, AlN~\cite{jung2013OpticalFrequencyComb}, AlGaAs~\cite{pu2016EfficientFrequencyComb}, LiNbO~\cite{zhang2017MonolithicUltrahighQLithium}, LiTaO~\cite{wang2024LithiumTantalatePhotonic}, Hydex~\cite{razzari2010CMOScompatibleIntegratedOptical}, Si~\cite{griffith2015SiliconchipMidinfraredFrequency}, SiC~\cite{zheng20194HSiCMicroringResonators}, GaN~\cite{zheng2022IntegratedGalliumNitridea}, GaP~\cite{wilson2020integrated}, TaO~\cite{jung2021TantalaKerrNonlinear}, Diamond~\cite{hausmann2014DiamondNonlinearPhotonics}.  
(b) Pulley coupler on a ring resonator~\cite{he2019SelfstartingBichromaticLiNbO3}. (c) Racetrack resonator with Euler-bends~\cite{ji2022CompactSpatialmodeinteractionfreeUltralowloss}. (d) Two coupled ring resonators~\cite{helgason2021DissipativeSolitonsPhotonic}. (e) Two ring resonators coupled to the same bus waveguide~\cite{dutt2018OnchipDualcombSource}. (f) Three coupled ring resonators~\cite{yuan2023SolitonPulsePairs}. (g) Coupled concentric resonators~\cite{kim2017DispersionEngineeringFrequency}. (h) Photonic crystal ring resonator~\cite{yu2021SpontaneousPulseFormation}. (i) Photonic crystal waveguide reflector of an integrated Fabry-P\'erot resonator~\cite{yu2019PhotonicCrystalReflectorNanoresonatorsKerrFrequency}. (j) Inverse design waveguide reflector of an integrated Fabry-P\'erot resonator~\cite{ahn2022PhotonicInverseDesign}. (k) Fiber-based Fabry-P\'erot resonator with dielectric mirrors~\cite{brasch2019NonlinearFilteringOptical}. }
\label{fig:ch1:platforms}
\end{figure*}

By now the field of microcombs has expanded, revealing numerous complex nonlinear phenomena as pathways to comb generation, including switching waves (or ``platicons'')~\cite{lobanov2015FrequencyCombsPlaticons,xue2015ModelockedDarkPulse}, soliton crystals~\cite{cole2017SolitonCrystalsKerr,karpov2019DynamicsSolitonCrystals}, and novel dynamics in coupled resonators~\cite{tikan2021EmergentNonlinearPhenomena,mittal2021TopologicalFrequencyCombsb,tusnin2020NonlinearStatesDynamics}, and a variety of microresonator geometries (Fig.~\ref{fig:ch1:platforms_schematics}) and platforms (Fig.~\ref{fig:ch1:platforms}).
Applications have grown extensively, encompassing optical atomic clocks~\cite{newman2019ArchitecturePhotonicIntegration,papp2014MicroresonatorFrequencyComb}, microwave photonics~\cite{liu2020PhotonicMicrowaveGeneration,xie2017PhotonicMicrowaveSignals} and frequency division~\cite{kudelin2024PhotonicChipbasedLownoise, sun2024IntegratedOpticalFrequency, zhao2024AllopticalFrequencyDivision}, terabit optical communications~\cite{pfeifle2014CoherentTerabitCommunications}, dual-comb spectroscopy~\cite{suh2016MicroresonatorSolitonDualcomb}, astronomical spectrograph calibration~\cite{suh2019SearchingExoplanetsUsing, obrzud2019MicrophotonicAstrocomb},  the interaction of microcombs with electron beams~\cite{yang2023free}, and photonic neuromorphic computing~\cite{feldmann2021ParallelConvolutionalProcessing,xu202111TOPSPhotonic}.
All these developments in themselves are challenging to review comprehensively and in-depth, as they encompass various fields of research, e.g., nonlinear dynamics, nonlinear optics, and material science.
Several reviews that cover microresonator frequency combs have been written to date. They include general reviews~\cite{nie2022PhotonicFrequencyMicrocombs,hermans2022OnchipOpticalComb,chang2022IntegratedOpticalFrequency,wang2020AdvancesSolitonMicrocomb,gaeta2019PhotonicchipbasedFrequencyCombs,fortier201920YearsDevelopments,pasquazi2018MicrocombsNovelGeneration,lugiato2018LugiatoLefeverEquation,kippenberg2018DissipativeKerrSolitons,lin2017NonlinearPhotonicsHighQ,savchenkov2016FrequencyCombsMonolithic,hansson2016DynamicsMicroresonatorFrequency,chembo2016KerrOpticalFrequency,kippenberg2011MicroresonatorBasedOpticalFrequency,zhu2021IntegratedPhotonicsThinfilm}, as well as specialized reviews on applications~\cite{okawachi2023ChipscaleFrequencyCombs,hu2021ChipbasedOpticalFrequency,sun2023ApplicationsOpticalMicrocombs,picque2019FrequencyCombSpectroscopy,wu2018RFPhotonicsOptical,xue2016MicrowavePhotonicsConnected, chang2022IntegratedOpticalFrequency}, design and fabrication~\cite{kim2023DesignFabricationAlGaAsonInsulator,kovach2020EmergingMaterialSystems,fujii2020DispersionEngineeringMeasurement,jung2016AluminumNitrideNonlinear,moss2013NewCMOScompatiblePlatforms}, various aspects of nonlinear dynamics~\cite{li2016ModelingFrequencyComb,xue2016NormaldispersionMicroresonatorKerr,kondratiev2023RecentAdvancesLaser,jiang2020OptothermalDynamicsWhisperinggallery,smirnov2023SelfStartingSolitonComb}, and quantum optics~\cite{kues2019QuantumOpticalMicrocombs,strekalov2016NonlinearQuantumOptics,caspani2016MultifrequencySourcesQuantum}. 

\subsection{Goal of this review}
This review aims to provide an introduction to the field of microresonator-based frequency combs. It introduces the fundamental linear and nonlinear dynamics of these driven-dissipative systems and presents key principles underlying the generation of microcombs. As such, the review is intended to serve both as an entry point for newcomers and as a reference for experienced researchers.

\section{Linear optical microresonators}
\label{sec:basics_lin}
This section discusses microresonators in the linear optical regime, as a foundation for the later extension to nonlinear optical effects and generation of frequency combs.

\subsection{Microresonator platforms}
\label{sec:basics_lin:res_platf}
Optical microresonators for frequency comb generation are made from transparent dielectric materials and confine light on a closed trajectory inside the resonator material through a refractive index that is higher than that of the surrounding.
In a \textit{whispering-gallery-mode} (WGM) resonator, light is guided by total internal reflection along the circumference of the resonator. In \textit{ring- or racetrack} resonators, light is guided in a waveguide that closes on itself. For WGM resonators, the surrounding material is usually air, whereas in waveguide-based resonators the waveguide core is surrounded by substrate and cladding materials.
In addition to those \textit{traveling wave} resonators, a \textit{standing wave resonator} topology can be formed by terminating a waveguide with reflecting optical elements. Examples of different resonator geometries and topologies are illustrated in Fig.~\ref{fig:ch1:platforms_schematics}.
Input and output coupling to the resonators can be achieved through evanescent field coupling between a prism, tapered fiber, or waveguides. A standing wave-resonator may in addition be coupled through a weak transmission in their reflecting elements.

\begin{table}
\caption{\textbf{Material properties of various materials for the fabrication of microresonators.} List of the key materials used for the fabrication of microresonators and associated material properties: refractive index $n_0$, nonlinear refractive index $n_2$, maximum second-order nonlinear coefficient d$_{ij}$, and energy bandgap. References used in the table: Si~\cite{dinu2003ThirdorderNonlinearitiesSilicon}, $\mathrm{Al}_{0.17} \mathrm{Ga}_{0.83} \mathrm{As}$~\cite{pu2016EfficientFrequencyComb}, $\mathrm{GaP}$~\cite{wilson2020IntegratedGalliumPhosphide}, $\mathrm{4H\text{-}SiC}$~\cite{wang2021HighQMicroresonators4Hsiliconcarbideoninsulator}, Diamond \cite{hausmann2014DiamondNonlinearPhotonics}, $\mathrm{LiNbO}_3$~\cite{zhu2021IntegratedPhotonicsThinfilm}, $\mathrm{LiTaO}_3$~\cite{wang2023lithium},  AlN~\cite{jung2013OpticalFrequencyComb}, Ta$_2$O$_5$~\cite{jung2021TantalaKerrNonlinear}, $\mathrm{Si}_3\mathrm{N}_4$~\cite{liu2021HighyieldWaferscaleFabrication}, Hydex~\cite{razzari2010CMOScompatibleIntegratedOptical},  $\mathrm{SiO}_2$~\cite{delhaye2007OpticalFrequencyComb}. }
\begin{ruledtabular}
\label{tab:ch1:materials}
\begin{tabular}{|c|c|c|c|c|}
\hline Material & \begin{tabular}[c]{@{}c@{}}$n_0$ \\1550~nm\end{tabular}& \begin{tabular}[c]{@{}c@{}}$n_2$ \\ $\left[10^{-18} \mathrm{~m}^2/\mathrm{W}\right]$\end{tabular}  & \begin{tabular}[c]{@{}c@{}} |d$_{ij}$| \\ $\left[\mathrm{pm}/\mathrm{V}\right]$ \end{tabular}   &   \begin{tabular}[c]{@{}c@{}} Bandgap \\ $\left[ \mathrm{eV} \right]$\end{tabular}  \\

\hline  $\mathrm{SiO}_2$ & 1.4 & 0.022 & - & 8.9  \\

\hline Hydex  & 1.7 & 0.12 & - & 8.9  \\

\hline $\mathrm{Si}_3\mathrm{N}_4$ & 2.0 & 0.25 & - & 5.4 \\

\hline Ta$_2$O$_5$ & 2.0 & 0.62 & - & 3.9 \\

\hline AlN & \makecell {2.12(o) \\ 2.16(e)}  & 0.23 & 4.3 & 6.2 \\

\hline $\mathrm{LiTaO}_3$ & \makecell {2.119(o) \\ 2.123(e)} & 0.146 & 13.8 & 3.93  \\

\hline $\mathrm{LiNbO}_3$ & \makecell {2.21(o) \\ 2.14(e)} & 0.18 & 27 & 3.78  \\

\hline Diamond & 2.4 & 0.082 & - & 5.47\\

\hline $\mathrm{4H\text{-}SiC}$ & 2.6 & 1 & 30 & 3.3 \\

\hline $\mathrm{GaP}$ & 3.1 & $11$ & 41 & 2.26 \\

\hline $\mathrm{Al}_{0.17} \mathrm{Ga}_{0.83} \mathrm{As}$  & 3.3 & 26 & 120 & 1.64 \\

\hline Si & 3.5 & 4 & - & 1.12 \\

\hline
\end{tabular}
\end{ruledtabular}
\end{table}

Dielectric microresonators have been made from many different materials. A key prerequisite is that the material is transparent and that in addition nonlinear absorption processes, such as two-photon absorption (TPA) are low in the frequency range of interest. TPA is one of the key limitations of the silicon-on-insulator platform, which exhibits a strong TPA in the telecommunication C-band. Therefore, other materials with larger energy bandgaps, such as silicon nitride, gain increasing interest. Another important aspect is the nonlinear properties of the materials. Strong mode confinement through large index contrast with the surrounding material and highly nonlinear materials enables achieving effective nonlinearities orders of magnitude greater than those possible in silica optical fibers. Frequency comb generation usually relies on the material's third-order nonlinearity characterized by the nonlinear index $n_2=\frac{3\chi^{(3)}}{4n_0^2\varepsilon_0 c}$, microresonators have also been fabricated from materials with second-order nonlinearity $\chi^{(2)}$. 
Another consideration is thermal material properties, which enter the picture when the inevitable absorption of laser light changes the resonator's temperature. Notably, thermal expansion and the temperature-dependent refractive index contribute to a change in cavity frequency, which may result in instability (e.g. in CaF$_2$) if they occur on different time scales and with opposite effects.   A more detailed discussion on the thermal effects is presented in Sec.~\ref{sec:exp_DKS:Res_thermal}.

\subsection{Resonance condition, free spectral range, and dispersion}
\label{sec:basics_lin:res_FSR_disp}
Laser light of frequency $\omega$ is resonant when the resonator's roundtrip length $L$ is a positive integer $m\in \mathbb{N}$ multiple of the wavelength $\lambda$ in the resonator, or equivalently, when the phase required in one resonator roundtrip is an $m$-multiple of $2\pi$, 
\begin{equation}
\label{eq:ch3:resonance_cond}
 2\pi m = \beta(\omega) L(\omega) \, ,
\end{equation}
where $\beta(\omega)=\frac{\omega}{c}n_\mathrm{eff}(\omega)$ is the wave's propagation constant, $n_\mathrm{eff}$ is the effective refractive index (see Section~\ref{sec:basics_lin_disp_eng} for more details). Both $n_\mathrm{eff}$ and $L$ can be frequency dependent\footnote{Note that the definition of $L$ can be ambiguous in curved geometries where a laterally extended mode extends over paths with different physical lengths. This is unproblematic as long as $L$ and $n_\mathrm{eff}$ are computed consistently with each other. In waveguide-based resonators $L$ is usually the length of the path traced by the waveguide's center; in WGM resonators is usually defined via the resonator radius $R$, i.e. $L=2\pi R$.}. The resonance frequencies are then solutions to the following equation
\begin{equation}
\label{eq:ch3:resonance_freq}
\omega = m\frac{2\pi c}{n_\mathrm{eff}(\omega) L(\omega)}\, .
\end{equation}

\begin{figure}
\includegraphics[width=0.99\columnwidth]{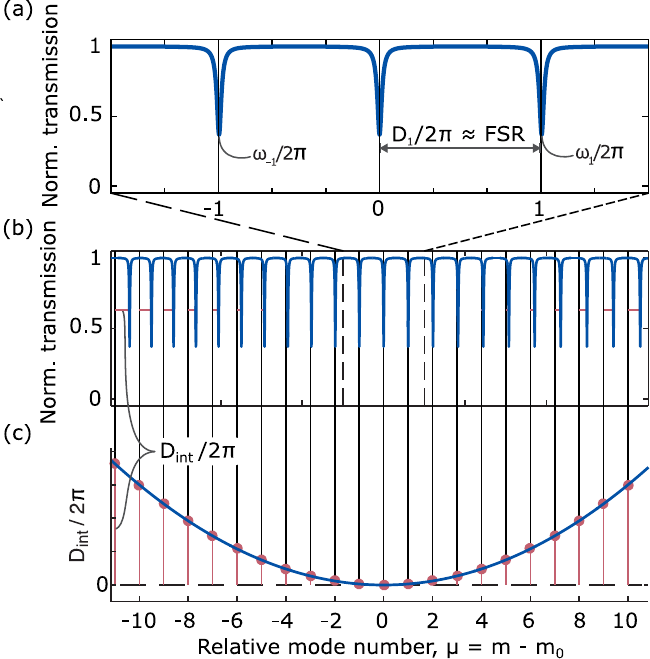}
\caption{\textbf{Integrated dispersion in microresonators.} 
(a) One adjacent resonance to the reference resonance with relative mode number $\mu = 0$ at $\omega_0$ defines the free spectral range (FSR) of the resonator at this wavelength, which is equivalent
to D$_1/2\pi$. (b) Other resonances of the same mode family with higher and lower mode numbers deviate from the equidistant grid with FSR spacing (red vertical lines) by the integrated
dispersion $D_{\mathrm{int}}$ (red horizontal lines). (c) Plotting the integrated dispersion $D_{\mathrm{int}}$ over the relative mode number allows fitting it with a polynomial (blue solid line) in order to derive
the values of the dispersion parameters $D_2$, $D_3$ ... $D_{\mathrm{n}}$.}
\label{fig:ch3:dispersion}
\end{figure}

Due to frequency dependence (chromatic dispersion) of $n_\mathrm{eff}$ and $L$, the resonance frequencies are usually not equidistantly spaced. In many cases, the resonance frequencies may be described through a Taylor expansion around a central resonance with mode number $m_0$ that is probed with a pump laser operating at frequency $\omega_\mathrm{p}$. We index the resonance by a \textit{relative (longitudinal) mode number} measured relative to a central mode number $m_0$ 
\begin{equation}
    \mu = m-m_0 \in \mathbb{Z}
    \label{eq:ch3:mu}
\end{equation}
so that the resonance frequencies are given by
\begin{align}
\label{eq:ch3:resonance_expansion}
\omega_{\mu}= & \omega_{0}+D_{1} \mu+\frac{D_{2}}{2 !} \mu^{2}+\ldots, \, \\  &\text {where } \, D_{n}=\left.\frac{\partial^{n} \omega_{\mu}}{\partial \mu^{n}}\right|_{\mu=0},\nonumber
\end{align}
we assume that $D_1 \gg D_2, D_3,\cdots$. The frequency separation between two adjacent resonances, e.g. $\mu$ and $\mu+1$ is called \textit{free-spectral range}, FSR. 
Interpreting the mode number $\mu$ as a continuous variable and the resonance frequencies and the FSR as a continuous variable of $\mu$, we see that the FSR is frequency (or mode-number) dependent
\begin{equation}
    \mathrm{FSR}(\mu) = \frac{1}{2\pi}\frac{\partial \omega_\mu}{\partial \mu} = D_1/(2\pi) + \mu D_2/(2\pi) + ...\, 
\end{equation}
where we can identify $D_1$ with the FSR at the central mode with $\mu=0$.
The higher order dispersion coefficients $D_2, D_3, ...$ describe the deviation of the resonance frequencies $\omega_\mu$ from an equidistant $D_1$-spaced frequency grid (Fig.~\ref{fig:ch3:dispersion}). This deviation is called \textit{integrated dispersion} 
\begin{equation}
D_{\mathrm{int}}\left(\mu\right)=\omega_{\mu}-\left(\omega_{0}+\mu D_{1}\right)=\sum_{n=2}^{\infty} D_{n} \frac{\mu^{n}}{n !} \, .
   \label{eq:ch3:dint}
\end{equation}

The resonator dispersion coefficients can be related to the expansion of the propagation constant $\beta(\omega)=\beta_0 + \beta_1 (\omega-\omega_0) + \tfrac{1}{2}\beta_2 (\omega-\omega_0)^2 + ...$ in waveguide optics by considering the derivatives $\partial^n/\partial \mu^n$ of Eq.~\ref{eq:ch3:resonance_cond}. Assuming that the resonator length $L$ is independent of frequency, we find
\begin{align}
\label{eq:ch3:Dn}
    D_1 & = \frac{2\pi}{\beta_1 L}  = \frac{2\pi v_\mathrm{g}}{L} =\frac{2\pi}{T_\mathrm{R}},\\
    D_2 & = -\frac{D_1^2 \beta_2}{\beta_1} =  -\frac{4\pi^2\beta_2}{\beta_1^3 L^2},
\end{align}
where $\beta_1=\partial \beta/\partial\omega|_{\omega=\omega_0} = v_g^{-1}$ is the inverse group velocity, $\beta_2=(\partial^2 \beta/\partial \omega^2)|_{\omega=\omega_0}$ the group velocity dispersion (GVD), and $T_\mathrm{R} = L/v_\mathrm{g} = 2\pi/D_1$ the resonator roundtrip time.

\subsection{Electric field, intensity, and normalization}
To describe the optical dynamics in the microresonator, we express the real electric field $\boldsymbol{\mathcal{E}}$ as a superposition of 
discrete modes with complex field amplitudes $\mathcal{E}_\mu$, frequency $\omega_\mu$, and propagation constant $k_\mu$, with the longitudinal mode number $\mu$ as introduced in Eq.~\ref{eq:ch3:mu}. In a geometry with axial symmetry as in a ring resonator or a WGM resonator, we can write using cylindrical coordinates $(r,\phi,z)$
\begin{equation}\label{sec4:eq:field_fwm}
       \boldsymbol{\mathcal{E}}(r,\phi, z, t)=\frac{1}{2}\sum_{\mu} \mathcal{E}_{\mu}(t) \boldsymbol{u}_\mu(r,z)\, e^{i\left((\mu+m_0)\phi-\omega_{\mu} t\right)}+\text { c.c.},
\end{equation}
where $\mathcal{E}_{\mu}$ is the complex field amplitude, $\boldsymbol{u}_\mu$ is the unity magnitude normalized electric field vector, defined in the 2-dimensional plane orthogonal to the propagation direction. Often $\boldsymbol{u}_\mu$ is referred to as the transverse mode profile as indicated in Fig.~\ref{fig:ch1:platforms_schematics}a,b.  
This description is also useful for resonators without axial symmetric geometry, e.g., in racetrack resonators, where then $\phi(\boldsymbol{r})\in[0,2\pi)$ is used to describe the position along one resonator roundtrip path and usually 
$\boldsymbol{u}_\mu(x,y)$ is given in Cartesian coordinates in a plane orthogonal to the propagation direction.

The field intensity in the resonator can be expressed as 
\begin{equation}
    I_\mu(r,\phi, z, t)=\frac{1}{2} n_\mathrm{eff}(\omega_\mu) \varepsilon_{0} c\left|\mathcal{E}_\mu(t)\right|^{2} \| \boldsymbol{u}_\mu(r,z)\|^2 \, .
    \label{eq:ch3:intensity}
\end{equation}
To describe the dynamics in the microresonator, it is convenient to define a normalized complex field amplitude
\begin{equation}
    a_\mu(t) = \sqrt{\frac{\varepsilon_0 n_\mathrm{eff}^2(\omega_\mu) V_\mathrm{\mu}}{2\hbar \omega_\mu}} \mathcal{E}_\mu(t) \, e^{-i\left(\omega_{\mu}- \omega_\mathrm{p} - \mu D_1\right) t},
    \label{eq:ch3:zpe}
\end{equation}
so that $|a_\mu|^2$ is equal to the number of photons in the mode with relative mode number $\mu$. Moreover, the amplitudes are described in rotating frames of frequency $\omega_\mathrm{p} + \mu D_1$. This represents an equidistant
$D_{1}$ spaced grid, centered on a pump laser frequency $\omega_\mathrm{p}$. 

For strongly confining resonators, where the field is almost fully contained in the homogeneous resonator material,  $V_\mathrm{\mu} = \int \| \boldsymbol{u}_\mu (\boldsymbol{r}) \|^2 \mathrm{d}V$ is the mode volume; if the field extends substantially into a lower refractive index cladding material, or if the resonator material is not homogeneous, then the different material properties need to be taken into account~\cite{koos2007NonlinearSilicononinsulatorWaveguides}.

For a given relative longitudinal mode number $\mu$, there can be more than one possible transverse field profile $\boldsymbol{u}_\mu(\boldsymbol{r})$, i.e. different transverse modes. The mode with the highest $n_\mathrm{eff}$ (for a given polarization) is called the fundamental mode, the others a higher-order transverse mode. Although comb generation usually occurs in the longitudinal modes of one transverse mode-family, linear coupling between transverse mode-families can influence the resonator's mode-structure, as discussed in Sec.~\ref{sec:basics_lin_AMX}.

\subsection{Input-output coupling for an optical resonator}\label{sec:basics_lin_in-ourt}
To describe how light is coupled into and out of the resonator via coupling optics, we derive the input-output relations for an optical resonator, considering resonator intrinsic loss mechanisms such as absorption, scattering, and radiation loss~\cite{gorodetsky2000RayleighScatteringHighQ}. 
There are several ways for deriving the input-output coupling relation for a resonator, for instance, based on energy conservation and symmetry considerations ~\cite{haus1984WavesFieldsOptoelectronics,yariv2000UniversalRelationsCoupling}. Here, we follow an open quantum system approach~\cite{gardiner1985InputOutputDamped, gardiner2004QuantumNoiseHandbook, scully1997QuantumOptics} and transition from classical fields (according to Eq.~\ref{eq:ch3:zpe}) to the annihilation and creation operators $a_\mu \rightarrow \hat{a}_\mu$ and  $a_\mu^* \rightarrow \hat{a}_\mu^\dagger$.
    
The system Hamiltonian of the resonator mode $a_0$ coupled to the continuum of modes of a coupling waveguide $b_\omega$ is 
\begin{equation}   
\begin{aligned}
    &\hat{H} = \hat{H}_{\mathrm{res}, \,0}  + \hat{H}_\mathrm{bath}+ \hat{H}_\mathrm{int}
\end{aligned}    
\end{equation}
with the free Hamiltonians of the resonator and bath, as well, as their interaction Hamiltonian:
\begin{equation}   
\begin{aligned}
    &\hat{H}_{\mathrm{res}, \,0}  =  \hbar (\omega_0-\omega_\mathrm{p}) \hat{a}_0^\dagger \hat{a}_0 \\ &\hat{H}_\mathrm{bath} = \hbar \int \mathrm{d}\omega \,(\omega-\omega_\mathrm{p})\, \hat{b}_\omega^\dagger \hat{b}_\omega\\
     &\hat{H}_\mathrm{int} =  \hbar \int \mathrm{d}\omega \, \left( g_\mathrm{ex}(\omega) \hat{b}_\omega^\dagger \hat{a}_0 + g_\mathrm{ex}(\omega)^* \hat{a}_0^{\dagger}\hat{b}_\omega \right).
\end{aligned}    
\end{equation}
Here, $\hbar$ is the Planck constant and $[\hat{b}_\omega, \hat{b}_{\omega'}^\dagger] =\delta(\omega-\omega')$. Consistent with Eq.~\ref{eq:ch3:zpe}, we use the rotating frame description of $\omega_\mathrm{p}$ and neglected rapidly rotating terms in the interaction Hamiltonian (rotating wave approximation).

Deriving the Heisenberg equation of motion $\dot{\hat{a}}_0 = -\frac{i}{\hbar}[\hat{a}_0, \hat{H}]$, and making the first Markov Approximation that assumes frequency independence of $g_\mathrm{ex}$, we obtain:
\begin{equation}
\begin{aligned}
    \dot{\hat{a}}_0(t)= & -\left(\frac{\kappa_\mathrm{ex}}{2} + i(\omega_0-\omega_\mathrm{p}) \right) \hat{a}_0(t) + \sqrt{\kappa_\mathrm{ex}} \, \hat{b}_\mathrm{in}(t),
\end{aligned}
\end{equation}
where we have defined the \textit{external} coupling rate  $\kappa_\mathrm{ex} =2\pi|g_\mathrm{ex}|^2$ and $\hat{b}_\mathrm{in}=\sqrt{1/(2\pi)} \int \mathrm{d}  \omega  \,  \hat{b}_\omega(t_0) \, e^{-i\left(\omega-\omega_\mathrm{p}\right)t}$. A continuous-wave laser drive delivered through the waveguide with frequency $\omega_\mathrm{p}$ is then represented by a coherent state $\hat{b}_\mathrm{in}(t) = s_\mathrm{in} + \hat{\xi}_\mathrm{ex}(t)$, where the operator $\hat \xi_\mathrm{ex}$ describes quantum noise. Following a similar approach, we can describe resonator loss from absorption or scattering as a coupling to another continuum bath (all possible modes except for the coupling waveguide that is already described by $\kappa_\mathrm{ex}$), leading to an additional $intrinsic$ decay rate $\kappa_\mathrm{0}=2\pi|g_0|^2$. With the definition of the 
\textit{total} cavity decay rate 
\begin{equation}
    \kappa = \kappa_\mathrm{0} + \kappa_\mathrm{ex}
\end{equation}
we obtain the Langevin equations for all resonator modes 
\begin{equation}
    \begin{aligned}
    	\dot{\hat{a}}_\mu(t)  = &  -\left( \frac{\kappa}{2} + i(\omega_\mu-\omega_\mathrm{p}-\mu D_1) \right)\hat{a}_\mu(t) +  \sqrt{\kappa_\mathrm{ex}}    \delta_{0\mu}s_\mathrm{in}\\
        & +  \sqrt{\kappa_\mathrm{0}}\, \hat{\xi}_\mathrm{0,\mu}(t) + \sqrt{\kappa_\mathrm{ex}}\, \hat{\xi}_{\mathrm{ex},\mu}(t) ,
    \end{aligned}
	\label{eq:sec3:one_mode_CME}
\end{equation}
where, for simplicity, we have assumed that the coupling rates for all modes are the same, $\kappa_{\mathrm{ex},\mu} = \kappa_\mathrm{ex}$ and $\kappa_{0,\mu} = \kappa_0$. 
The driving field amplitude, here driving mode $\mu=0$, is related to the pump power by $|s_\mathrm{in}|^2=P_{\mathrm{in}}/(\hbar\omega_\mathrm{p})$.
The impact and description of quantum noise $\hat{\xi}_{0,\mu}(t)$ and $\hat{\xi}_{\mathrm{ex},\mu}(t)$, as well as other sources of noise is discussed in Section~\ref{sec:exp_DKS:noise}.

The total cavity decay rate $\kappa$ represents the rate with which energy $W$ stored in the resonator is dissipated following an exponential decay  $\mathrm{d} W/\mathrm{d} t = -\kappa W$, so that $\tau_\mathrm{ph}=1/\kappa$ can be identified with the lifetime of photons in the cavity. The internal decay rate is related to the attenuation constant $\alpha$ via $\kappa_\mathrm{0}= \alpha L / T_{\mathrm{R }}$. The external decay rate is related to the relative power coupling coefficient at the coupler $\Theta_\mathrm{p}$ (i.e., the fraction of the power that couples into and out of the resonator) via $\kappa_\mathrm{ex} = \Theta_\mathrm{p} / T_\mathrm{R}$, provided that the loss rate is small ($\kappa_\mathrm{ex}T_R<<1$). 

The quality factor ($Q$-factor) of a low-loss (high-$Q$) resonator indicates the number of field oscillations occurring during the photon lifetime
\begin{equation}
Q=\omega_{0} \tau_{\mathrm{ph}} = \frac{\omega_0}{\kappa}.
\end{equation}
To permit comparison of resonators regardless of coupling optics, one often uses the intrinsic $Q$-factor $Q_\mathrm{0} = \omega_{0} / \kappa_\mathrm{0}$ to compare different resonators.
The finesse of a resonator is defined as the ratio of its FSR to resonance width:
\begin{equation}
    \mathcal{F} = \frac{2\pi \mathrm{FSR}}{\kappa},
\end{equation}
and $\mathcal{F}/(2\pi)$ is the average number of roundtrips a photon completes in the resonator during the photon lifetime $\tau_\mathrm{ph}$.

\textit{Stationary solutions.} Searching for the stationary solutions of Eq.~\ref{eq:sec3:one_mode_CME} for the driven mode with $\mu=0$ in the classical limit ($\hat{a}_0 \rightarrow \langle \hat{a}_0 \rangle = a_0$, $\hat{s}_\mathrm{in} \rightarrow \langle \hat{s}_\mathrm{in} \rangle = s_\mathrm{in}$ and $\delta \hat{s}_\mathrm{0} \rightarrow \langle \delta \hat{s}_\mathrm{0} \rangle = 0$), we obtain:
\begin{equation}
    \label{eq:resonance_shape}
	|a_0|^2=\frac{\kappa_\mathrm{ex}}{(\omega_0-\omega_\mathrm{p})^2 + (\frac{\kappa}{2})^2} |s_\mathrm{in}|^2.
\end{equation}
The intra-cavity power for critical coupling $\kappa_{\mathrm{ex}}=\kappa_0$, and zero cavity detuning $\omega_{\mathrm{p}} = \omega_0$ is given by
\begin{equation}
\label{eq:ch3:intrcav_power}
    P_\mathrm{cav}=\hbar\omega |a_0|^2 \mathrm{FSR}  = \frac{\mathcal{F}}{\pi}P_\mathrm{in}.
\end{equation}
Thus $\mathcal{F}/\pi$ is the maximal power enhancement factor in the resonator.

Using time reversal symmetry~\cite{gardiner1985InputOutputDamped,haus1984WavesFieldsOptoelectronics},  and thus linking the forward and backward Langevin equation, it can be shown that the input and output fields obey the relation\footnote{To include quantum noise, the output relation is  $\hat s_\mathrm{out}(t) = \hat s_\mathrm{in}(t) - \sqrt{\kappa_\mathrm{ex}} \hat{a}_0(t)$, where $\hat s_\mathrm{in}(t)= s_\mathrm{in} + \hat{\xi}_{\mathrm{ex,0}}(t)$ with the commutators and correlators as described in Sec.~\ref{sec:ch6:noise:quantum}.}: 
\begin{equation}
s_\mathrm{out} = s_\mathrm{in} - \sqrt{\kappa_\mathrm{ex}} a_0. 
\label{eq:ch3:out_of_cavity}
\end{equation}

Thus, the transmission $T$ for a ring-resonator coupled to a bus waveguide is given by\footnote{A general treatment of the case of multiple coupled resonators with add and drop ports can be found in~\cite{van2016OpticalMicroringResonators}.}:

\begin{equation}\label{eq:ch3:line_shape}
\mathrm{T}(\omega_\mathrm{p})  =\left|\frac{s_{\mathrm{out}}}{s_{\mathrm{in}}}\right|^{2}
=1-\frac{4 \eta(1-\eta)}{1+\left(\frac{\omega_0-\omega_\mathrm{p}}{\kappa / 2}\right)^{2}},
\end{equation}
where we introduced the coupling ratio $\eta=\kappa_{\mathrm{ex}} / \kappa$, which can be tuned by increasing or decreasing the mode-overlap integral between resonator and coupling optics. The second term in Eq.~\ref{eq:ch3:line_shape} defines the Lorentzian shape of the resonance with full-width at half-maximum (FWHM) linewidth of $\kappa$. In a Fabry-P\'erot resonator, Eq.~\ref{eq:ch3:line_shape} describes the reflected signal. Note that in Eq.~\ref{eq:ch3:line_shape}, we have neglected contributions from neighboring resonances at frequencies $\omega_{\pm1}$, which is justified for high-finesse resonators ($\kappa\ll 2\pi\mathrm{FSR}$) and not too large laser detuning ($|\omega_0-\omega_\mathrm{p}|\ll 2\pi\mathrm{FSR}$). 

\begin{figure}
\includegraphics[width=0.95\columnwidth]{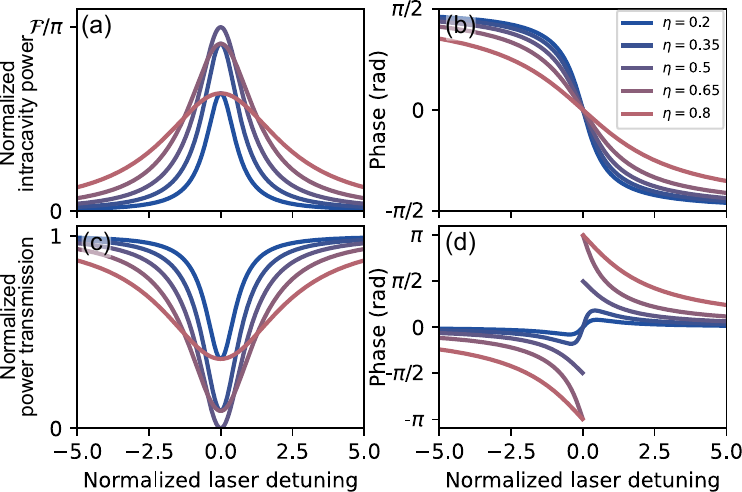}
\caption{\textbf{Amplitude and phase response of a cavity at different coupling regimes.} (a) 
Dependence of the normalized intracavity power on the pump laser detuning according to Eq.~\ref{eq:ch3:intrcav_power}, for different coupling conditions: from undercoupled (blue) to overcoupled (red). $\eta =0.5$ corresponds to the critical coupling. The intrinsic linewidth $\kappa_0$ is constant. (b) Evolution of the field phase. (c,d) Similar dependence for the normalized transmitted power T($\omega$), colors are preserved. The horizontal axis is normalized as $(\omega_0-\omega_\mathrm{p})/\kappa_0$.
}\label{fig:ch3:crit_coup}
\end{figure}

Depending on the ratio $\eta$, three distinct regimes of coupling can be distinguished~\cite{cai2000ObservationCriticalCoupling}: \textit{under-coupling} ($\eta<1/2$), \textit{critical coupling} ( $\eta=1 /2)$\footnote{An alternative way to achieve the critical coupling is to excite the resonator with time-shaped signals that result in the variable virtual coupling efficiency~\cite{hinney2024EfficientExcitationControl}}, and \textit{over-coupling} $(\eta>1 /2)$. Fig.~\ref{fig:ch3:crit_coup} shows the power and phase of the intracavity and transmitted wave, assuming a ring resonator with a coupling waveguide. Stronger coupling (larger $\eta$) leads to a wider linewidth and the maximal intracavity power may be reached for critical coupling. Depending on the detuning, there is a characteristic phase response.

\subsection{Phase matching, ideality and frequency dependence of input-output coupling}
\label{sec:basics_lin:ideality}
\textit{Phase matching.}
Input-output coupling occurs efficiently if the \textit{phase-matching} condition between light in the waveguide and in the resonator is fulfilled, i.e. the phase mismatch across the coupling region is small $\Delta k L_{c}<\pi/2$, where $\Delta k$ is the difference of the propagation constants and $L_\mathrm{c}$ the length of the coupler. For this reason, in high repetition rate microresonators (>500 GHz), waveguide dimensions and coupling regions have to be carefully designed to optimize the coupling ideality. 

\begin{figure}[h]
\includegraphics[width=0.9\columnwidth]{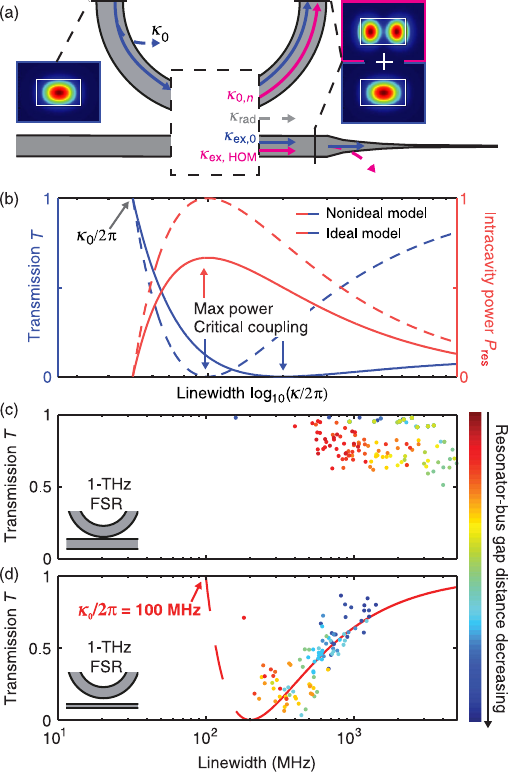}
\caption{\textbf{Resonator coupling ideality.} (a) Schematic representation of the coupling rates in an integrated microresonator with multimode waveguides. $\kappa_{\mathrm{ex}, 0}$ represents the coupling rate to the fundamental bus waveguide mode, $\kappa_{\mathrm{ex}, \mathrm{HOM}}$ - the coupling rate to the higher-order bus waveguide modes. (b) Plot of the transmission $T$ (blue) and the intracavity power $P_{\text {res }}$ (red) as a function of the total linewidth $\kappa / 2 \pi$ for the ideal $(I=1$, dashed lines) and nonideal $(I=0.67$, solid lines) cases. (c,d) Example of the coupling ideality dependence on the bus waveguide configuration for rings with 1 THz FSR showing a multimode waveguide with low ideality (c) and a single mode bus waveguide with high ideality(d). From~\cite{pfeiffer2017CouplingIdealityIntegrated} \label{fig:ch3:ResonatorIdeality}}
\end{figure}

\textit{Coupling ideality.}
In the presence of multiple transverse modes, the coupling optics may couple light not only to the desired mode (usually the fundamental mode) but also to parasitic (usually higher-order) transverse modes. This is especially true for the relaxed phase-matching condition of short $L_\mathrm{c}$, as those obtained when using \textit{point couplers}, straight waveguides in conjunction with small radius ring resonators.  
This process is schematically shown in Fig.~\ref{fig:ch3:ResonatorIdeality}(a). To quantitatively describe the performance of the coupling optics, we introduce the so-called coupling ideality~\cite{pfeiffer2017CouplingIdealityIntegrated,cai2000ObservationCriticalCoupling,spillane2003IdealityFiberTaperCoupledMicroresonator}:
\begin{equation}
I_{\mathrm{c}}=\frac{\kappa_{\mathrm{ex}}}{\kappa_{\mathrm{ex}}+\kappa_\mathrm{p}},
\end{equation}
where $ \kappa_\mathrm{ex}$ is the external coupling rate to the fundamental mode and $\kappa_\mathrm{p}=\sum \kappa^{(i)}_{\mathrm{ex}} $ is the external total coupling rate to parasitic higher-order modes with index $i$. For a well-designed coupler, the coupling ideality is close to 1. For the resonator transmission on resonance with finite ideality we obtain the following expression:
\begin{equation}
 T=\left(\frac{\kappa_\mathrm{p}+\kappa_\mathrm{0}-\kappa_{\mathrm{ex}}}{\kappa_\mathrm{p}+\kappa_\mathrm{0}+\kappa_{\mathrm{ex}}}\right)^{2},
\end{equation}
which is plotted in Fig.~\ref{fig:ch3:ResonatorIdeality}(b) for both ideal and finite cases. High coupling ideality, i.e. selective coupling to the desired mode, may be obtained, for instance when using a pulley-coupler configuration -- a bus waveguide following the circumference of the resonator for a given angle~\cite{moille2019BroadbandResonatorwaveguideCouplinga}, creating a long $L_\mathrm{c}$.

\textit{Frequency dependence of coupling.}
The coupling rate $\kappa_\mathrm{ex}$ is generally frequency dependent. This is a result of the wavelength dependence of the exponential spatial decay of the evanescent field, and hence the field overlap between evanescently coupled modes. In addition, for long coupling length $L_\mathrm{c}$ the wavelength dependence of the phase-matching conditions contributes to a wavelength-dependent coupling. Hence, long $L_\mathrm{c}$, as those in a pulley-coupler configuration can be leveraged to tailor the wavelength-dependent coupling across a wide bandwidth. \cite{moille2019BroadbandResonatorwaveguideCouplinga}, whereas short $L_\mathrm{c}$ couplers offer broadband coupling.

\subsection{Linear mode coupling, mode hybridization and avoided mode crossings}
\label{sec:basics_lin_AMX}

\begin{figure}
\includegraphics[width=1.\columnwidth]{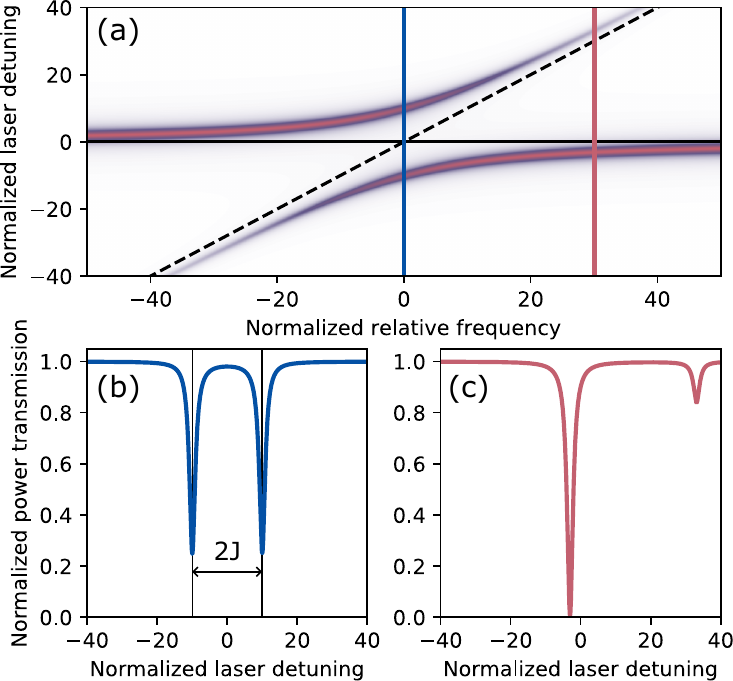}
\caption{\textbf{Avoided mode crossing.} (a) A normalized power transmission for different relative frequencies calculated with driven-dissipative coupled-mode equation ($J = 10 \kappa/2$). Loss rates of both modes are set to be equal. The black lines show uncoupled bright and dark (dashed line) mode positions. (b,c) Cross-sections of plot (a) as indicated by vertical lines of the corresponding colors.}\label{fig:ch3:amx}
\end{figure}

Approximately frequency degenerate resonator modes can exchange energy through linear coupling. The resulting mode-hybridization modifies the resonance frequencies and linewidths in the resonator, and leads to an AMX. Modes of different polarization, different transverse mode families, counter-propagating modes in a ring-resonator, or modes of a coupled second resonator may be involved. Often, the coupling arises from random scattering imperfections in the microresonator, such as surface roughness, where a reciprocal space Fourier-component of the roughness provides phase matching via a refractive index modulation with the angular period $\Lambda_\phi = \frac{2\pi}{\Delta m}$ ($\Delta m$ is the mode number difference of the coupling modes); coupling may also result from non-adiabatic geometry modifications, such as a changing waveguide cross-section or a change in waveguide bend radius~\cite{gorodetsky2000RayleighScatteringHighQ,
ji2022CompactSpatialmodeinteractionfreeUltralowloss}, and may even be engineered deliberately as discussed in Section~\ref{sec:coupled_mode-families_and_resonators}.

Following the procedure of Section~\ref{sec:basics_lin_in-ourt} for the two modes $\mu$ and $\nu$ and describing their coupling via the interaction Hamiltonian
\begin{equation}
\hat{H}_\mathrm{cpl} = \hbar (J\hat{a}_\mu^\dagger \hat{a}_\nu + J^*\hat{a}_\mu \, \hat{a}_\nu^\dagger),
\end{equation} 
where $J\in\mathbb{C}$ denotes the coupling rate between the modes, we obtain (in matrix form):
\begin{align}
  \begin{pmatrix} \dot{a}_\mu \\ \dot{a}_\nu \end{pmatrix}
=
\left(
\begin{smallmatrix}
 -i(\omega_\mu-\omega_\mathrm{p}) -\tfrac{\kappa_\mu}{2} & iJ \\
 iJ^* & -i(\omega_\nu-\omega_\mathrm{p}) -\tfrac{\kappa_\nu}{2}
\end{smallmatrix}
\right)
\begin{pmatrix} a_\mu \\ a_\nu \end{pmatrix}
+ \mathbf{S}\, ,  
\label{eq:ch3:amx_eq_simple}
\end{align}

where $\omega_{\mu,\nu}$ and $\kappa_{\mu,\nu}$ denote the respective mode frequencies and linewidth of the two modes, and $\mathbf{S} = [\sqrt{\kappa_{\mathrm{ex},\mu}}s_{\mathrm{in},\mu};\sqrt{\kappa_{\mathrm{ex},\nu}}s_{\mathrm{in},\nu}]$ is a pump field vector. The coupling matrix is diagonalized in the basis of hybrid (or super-) modes with eigenvalues

\begin{equation}
\begin{split}
     \sigma_{\pm} & =-i\omega_\mathrm{avg} - \kappa_\mathrm{avg}/2 \\
    & \pm i\sqrt{|J|^2 + \left(\omega_\mathrm{diff}/2 - i\kappa_\mathrm{diff}/4\right)^2} \, ,
    \label{eq:lincoupling}  
\end{split}
\end{equation}
where the (negative) imaginary and (negative) real parts represent the frequencies and linewidth of the hybrid modes, respectively; $\omega_\mathrm{avg} = (\omega_\mu + \omega_\nu)/2-\omega_\mathrm{p}$ and $\kappa_\mathrm{avg} = (\kappa_\mu + \kappa_\nu)/2$ are average values, $\omega_\mathrm{diff} = \omega_\mu - \omega_\nu$ and $\kappa_\mathrm{diff} = \kappa_\mu - \kappa_\nu$ the differences between the uncoupled mode frequencies and linewidths. A typical microresonator mode interaction is depicted in Fig.~\ref{fig:ch3:amx}. If the two modes exhibit different spatial profiles, the interaction leads to a hybridization of those profiles as well~\cite{carmon2008StaticEnvelopePatterns}. This phenomenon is versatile and can be found all across Physics. For example, level repulsion has been predicted for molecular energy surfaces~\cite{wightman1993CollectedWorksEugene}, for coupled circuits~\cite{frank1994ClassicalAnalogyQuantum}, and different classical systems~\cite{novotny2010StrongCouplingEnergy}. If the hybrid basis is used to describe the system, the pump vector must also be transformed to the hybrid-mode basis~\cite{komagata2021DissipativeKerrSolitonsa, ulanov2024SyntheticReflectionSelfinjectionlocked}. 

For $\kappa_\mathrm{diff}\neq0$ the square root in Eq.~\ref{eq:lincoupling} can be complex valued, implying not only a modification of the resonance frequency but also of the hybrid modes' linewidths. Considering the case of $\omega_\mathrm{diff}=0$, we note that the hybridized resonance splitting occurs only after passing a threshold in $|J|$, corresponding to a so-called exceptional point~\cite{ozdemir2019ParityTimeSymmetry,miri2019ExceptionalPointsOptics}. In the case when the crossing cavity modes are dissipatively coupled
through a common decay channel, a quality factor enhancement via the bound state in the continuum is observed~\cite{lei2023HyperparametricOscillationBound}.

Coupling to other mode families can be a major and often unwanted effect.
To avoid unwanted coupling to higher order modes, multiple strategies exist including reduction of scattering defects and surface roughness, small waveguide cross-section or mode filtering sections \cite{kordts2016HigherOrderMode}, large FSR to reduce the density of modes, as well as pulley-couplers~\cite{pfeiffer2017CouplingIdealityIntegrated, moille2019BroadbandResonatorwaveguideCouplinga} and Euler-bends (in racetrack resonators)~\cite{cherchi2013DramaticSizeReduction,vogelbacher2019AnalysisSiliconNitride, ji2022CompactSpatialmodeinteractionfreeUltralowloss}.

In many cases, one can approximate $\kappa_\mu\approx\kappa_\nu\approx\kappa$ so that the frequencies of the hybrid modes are given by
\begin{equation}
\begin{split}
     \omega_{\pm} & =\omega_\mathrm{avg} \pm \sqrt{|J|^2 + \omega_\mathrm{diff}^2/4} \, ,
    \label{eq:ch3:amx}  
\end{split}
\end{equation}
For instance, this is the case when considering the coupling between resonators of comparable linewidth~\cite{xue2015NormaldispersionMicrocombsEnabled,tikan2021EmergentNonlinearPhenomena}, where by tuning $\omega_\mathrm{diff}$ one can obtain adjustable hybrid mode frequencies. In ring resonators, where forward and backward propagating modes may be coupled, we have $\kappa_\mathrm{diff}=0$, and in addition $\omega_\mathrm{diff}=0$, so that the frequency splitting between the hybridized modes is exactly $2|J|$. 
Section~\ref{sec:coupled_mode-families_and_resonators} discusses opportunities accessible through deliberate linear coupling.

\begin{figure*}
	\includegraphics[width=1\linewidth]{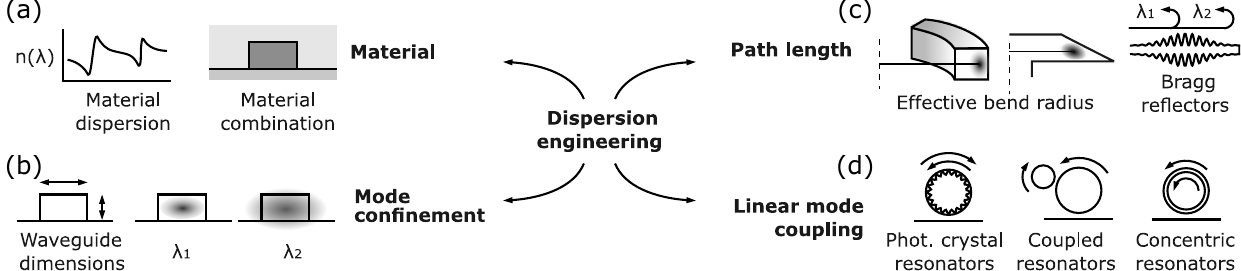} 
	\caption{\textbf{Dispersion engineering.} The dispersion in a microresonator is influenced by (a) the intrinsic (bulk) material dispersion of the resonator materials, (b) the geometric dimensions of the resonator impacting mode confinement and wavelength-dependent overlap with the higher-refractive index guiding material, (c) the wavelength-dependent geometric path length (or effective bend radius in circular resonators) along which the light travels, and (d) linear coupling between different modes, e.g. counter-propagating modes in a photonic crystal resonator, or nearly degenerate modes in coupled or concentric resonators.} \label{fig:disp_engineering}
\end{figure*}

\subsection{Microresonator dispersion engineering}
\label{sec:basics_lin_disp_eng}

\begin{figure}[h]
\includegraphics[width=1.0\columnwidth]{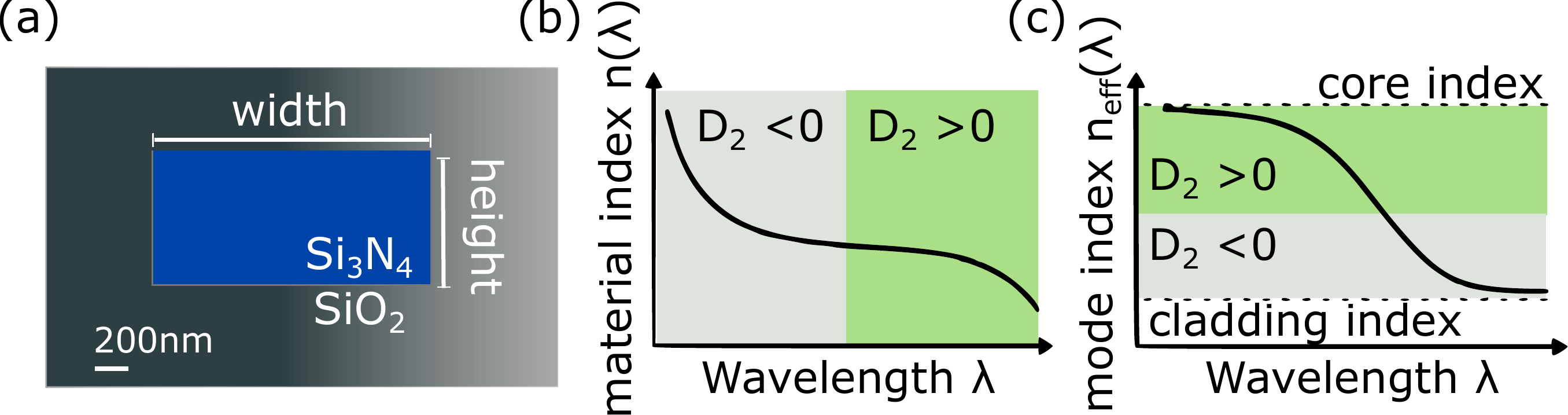}
\caption{\textbf{Dispersion in waveguide-based resonator.} (a) A schematic of typical Si$_3$N$_4$ resonator waveguide profile. (b) A material refractive index showing the transition from normal (D$_2$<0) to anomalous (D$_2$>0) dispersion. (c) By tailoring the waveguide cross-section to highly confine the optical mode, the anomalous waveguide dispersion can overcompensate the normal material dispersion. From~\cite{pfeiffer2016PhotonicDamasceneProcess}. \label{fig:ch3:DispersionEngineering} 
}
\end{figure}

Due to the chromatic dispersion of the effective refractive index $n_\mathrm{eff}(\omega)$ the resonance frequencies $\omega_\mu$ are not generally equidistant (cf. Eq.~\ref{eq:ch3:resonance_expansion}). This has a decisive impact on the nonlinear dynamics and frequency comb formation in a microresonator. As Fig.~\ref{fig:disp_engineering} illustrates, one can distinguish different contributions to the overall dispersion of $n_\mathrm{eff}$, which may be leveraged for \textit{dispersion engineering} to influence and control the nonlinear optical processes in the resonator:

\textit{Material dispersion}.
Each resonator material that is part of the resonator is characterized by a unique refractive index $n(\lambda)$ and its curvature $\partial^2 n(\lambda)/\partial \lambda^2$ defines the sign of $D_2$ as shown in Fig.~\ref{fig:ch3:DispersionEngineering}(b).
While $n(\lambda)$ exhibits strong dispersion in the spectral vicinity of an absorption band, it is only weakly dispersive in the low-loss regime and can be well modeled by the \textit{Sellmeier equation} whose coefficients can be found for example in~\cite{weber2018HandbookOpticalMaterials}. Most optical materials absorb in the ultraviolet domain due to electronic resonances and hence, as follows from the Kramers-Kronig relation, possess normal GVD in visible and near IR regions. While usually the resonator materials are predominantly chosen for their low-loss or nonlinear qualities, modifying the material composition can be used to impact the dispersion of a resonator \cite{riemensberger2012DispersionEngineeringThick,moille2021ImpactPrecursorGas}.

\textit{Mode confinement induced dispersion}.
All resonators are made of at least two materials that result in a guided optical mode e.g., a waveguide with a higher refractive index core $n_\mathrm{core}$ surrounded by a lower index cladding $n_\mathrm{clad}$ (which may be air or vacuum), as in Fig.~\ref{fig:ch3:DispersionEngineering}(a).  Generally, $n_\mathrm{clad} < n_\mathrm{eff} < n_\mathrm{core}$, where more strongly confined modes exhibit higher $n_\mathrm{eff}$. As mode confinement reduces with longer wavelength $n_\mathrm{eff}(\lambda)$ will show the wavelength dependence as shown in Fig.~\ref{fig:ch3:DispersionEngineering}(c), creating anomalous $D_2>0$ (normal $D_2<0$) dispersive contributions for tightly (weakly) confined modes \cite{pfeiffer2016PhotonicDamasceneProcess}.

\textit{Path length difference induced dispersion}.
In resonators where light is propagating on a curved trajectory, the `center of gravity' of the mode-profile will be shifted towards larger radii of curvature for shorter wavelength \cite{delhaye2011OctaveSpanningTunable,zhang2011AnalysisEngineeringChromatic, yang2016BroadbandDispersionengineeredMicroresonator} resulting in a chromatic dispersion of the cavity length $L(\omega)$.
Alternatively, we can formally consider the radius (or length $L$) of the resonator as a wavelength-independent geometrical resonator property, implying that shorter wavelengths experience a larger $n_\mathrm{eff}$. The effect of bending waveguides generally results in a normally dispersive contribution ($D_2<0$). While resonator curvature and bending are essential in whispering-gallery mode resonators, they can usually be neglected in waveguide-based resonators, unless the bend radii are very small (typically < 25 $\mu$m).
Another example of resonators with path length induced dispersion are Fabry-P\'erot microresonators where the wavelength-dependent reflection depth in a Bragg-reflector can induce a path length difference \cite{yu2019PhotonicCrystalReflectorNanoresonatorsKerrFrequency, ahn2022PhotonicInverseDesign, wildi2023DissipativeKerrSolitons}.  Normal, anomalous, and more complex dispersion characteristics may be implemented in this way, similar to what has been shown with chirped Bragg-mirrors~\cite{szipocs1994ChirpedMultilayerCoatings,kartner1997DesignFabricationDoublechirped}.

\textit{Dispersion from coupled resonances}.
As discussed in Section~\ref{sec:coupled_mode-families_and_resonators:AMX} linear coupling between (approximately degenerate) resonances can modify the resonance frequencies (of the emerging hybrid modes). This mechanism can hence be used to modify the resonator dispersion. Practical implementation includes photonic crystal resonators (coupling between clockwise and counter-clockwise propagating modes) ~\cite{lu2014SelectiveEngineeringCavity,yu2021SpontaneousPulseFormation,yu2022ContinuumBrightDarkpulse,lucas2023TailoringMicrocombsInversedesigned}, coupling between distinct resonators \cite{xue2015NormaldispersionMicrocombsEnabled,kim2019TurnkeyHighefficiencyKerr,helgason2021DissipativeSolitonsPhotonic,ji2023EngineeredZerodispersionMicrocombsa}, or concentric resonators ~\cite{soltani2016EnablingArbitraryWavelength,kim2017DispersionEngineeringFrequency}, as illustrated in Fig.~\ref{fig:disp_engineering}. 
While material dispersion, mode confinement and path length difference usually impact the entire resonance spectrum, coupling between resonances can be utilized to induce a targeted shift also on one individual resonance. This enables, for instance, the generation of strong spectrally-local anomalous dispersion to induce comb formation in the normal dispersion regime~\cite{ji2023EngineeredZerodispersionMicrocombsa,xue2015NormaldispersionMicrocombsEnabled} (also see: Sections~\ref{sec:switching_waves},\ref{sec:basics_lin_AMX}, \ref{sec:coupled_mode-families_and_resonators:AMX}).

In some cases, the total dispersion can be estimated analytically~\cite{demchenko2013AnalyticalEstimatesEigenfrequencies} but usually is calculated numerically. For waveguide-based resonators with negligible impact of bending, one can simulate $n_\mathrm{eff}(\omega)$ for a given $\omega$ using an electromagnetic mode solver on the 2D waveguide cross-section (assuming infinite length of the waveguide). For resonators where bending plays an important role, e.g. waveguide ring-resonators with small bend-radius or WGM resonators, one can solve for the eigenfrequencies for a given longitudinal mode number $m$, where the resonator cross-section is again defined in 2D and axial symmetry is assumed. The simulation of Fabry-P\'erot resonators combines a simulation of waveguide dispersion with a simulation of the reflection characteristics (bandgap, back-coupling rate) of the Bragg reflectors.

\begin{figure}[h]
\includegraphics[width=0.99\columnwidth]{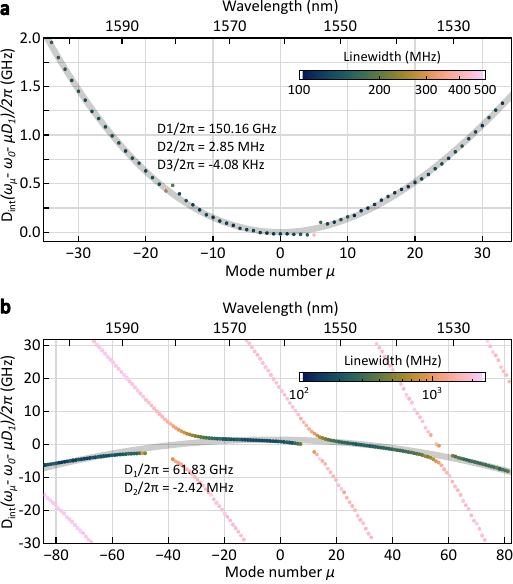}
\caption{\textbf{ Dispersion profile of chip-integrated microresonators.} Every point represents a resonance. Colors indicate the resonance linewidth. (a) Anomalous dispersion microresonator. (b) Normal dispersion microresonator with strong AMXs caused by lower-$Q$, higher-order transverse modes.
\label{fig:ch3:real_dispersion}}
\end{figure}

Fig.~\ref{fig:ch3:real_dispersion}(a,b) show the cases of anomalous and normal dispersion, respectively. The normal dispersion resonator exhibits numerous AMXs. The resonator dispersion can be measured by frequency comb assisted spectroscopy ~\cite{delhaye2009FrequencyCombAssisted,twayana2021FrequencycombcalibratedSweptwavelengthInterferometry} and various other techniques~\cite{fujii2020DispersionEngineeringMeasurement}.

\section{Kerr-Nonlinear optical microresonators and microcombs}
\label{sec:basics_nln}
This section introduces the formalism for the description of microresonators in the nonlinear optical regime. For a general account of nonlinear optical processes, we refer to comprehensive textbooks~\cite{bloembergen2010NonlinearOptics,boyd2020NonlinearOptics}.

\subsection{Description of Kerr-nonlinear optical effects}

We extend our description of a microresonator by including the nonlinear interactions. Here, we assume that the second-order nonlinearity is absent and hence consider only the third-order Kerr nonlinearity. We consider only one polarization and treat the electric field as a scalar field $\boldsymbol{\mathcal{E}}(\boldsymbol{r},t)=\mathcal{E}(\boldsymbol{r},t)$ and $\boldsymbol{u}_\mu(\boldsymbol{r})=u_\mu(\boldsymbol{r})$. We also assume that the resonator material is almost lossless so that we can treat the $\chi^{(3)}$-tensor as a real-valued scalar. These approximations are well-justified in most microresonators. Under these assumptions, 
the nonlinear polarization takes the form
\begin{equation}\label{sec4:eq:polarization}
\begin{aligned}
& \mathcal{P}_{\mathrm{NL}}(\boldsymbol{r}=(r,z,\phi), t) = \\
& \varepsilon_{0} \frac{3}{8}\sum_{k, l, m}  \chi^{(3)}(\omega_m-\omega_k-\omega_l; -\omega_k, -\omega_l, \omega_m)\\
& \times \mathcal{E}_{k}^* \mathcal{E}_{l}^* \mathcal{E}_{m} u_k(r,z)^* u_l(r,z)^* u_m(r,z)\, \\
& \times e^{i\left((m-k-l-m_0)\phi-(\omega_{m} -\omega_{k} - \omega_{l}) t\right)} + \text{ c.c.},
\end{aligned}
\end{equation}
where we have neglected those terms that correspond to third harmonic, triple sum, and their inverse processes. The 
interaction energy is 
\begin{equation}
\begin{aligned}
    H_\mathrm{Kerr} 
    = & -\frac{1}{4}\int  \mathcal{P}_{\mathrm{NL}}(\boldsymbol{r}, t) \mathcal{E}(\boldsymbol{r}, t)    \, \mathrm{d}V  
\end{aligned} 
\end{equation}
and performing the volume integral with the definition of the fields as in Eq.~\ref{eq:ch3:zpe},  and making the transition to creation and annihilation operators similar to Section~\ref{sec:basics_lin_in-ourt}, we obtain the Kerr-Hamiltonian 
\begin{equation}
    \label{eq:ch4:kerr_hamilt}
    \hat{H}_\mathrm{Kerr} 
    =  - \frac{\hbar g_\mathrm{K}}{2} \, \sum_{k, l, m} \hat{a}_{k}^\dagger \hat{a}_{l}^\dagger \hat{a}_{m} \hat{a}_{k+l-m} \, \Lambda_{k,l,m}
\end{equation}
The coupling constant is
\begin{equation}
        g_\mathrm{K} =\frac{\hbar\omega_0^{2}c n_2(\omega_0)}{n_\mathrm{eff}^2(\omega_0) V_\mathrm{eff}(\omega_0)}
        \label{eq:ch4:g0}
\end{equation}
where \begin{equation}
    V_\mathrm{eff}(\omega_\mu) 
     =\frac{\left(\int u_\mu^2(\boldsymbol{r}) \, \mathrm{d} V\right)^2}{\int u_\mu^4(\boldsymbol{r}) \,\mathrm{d} V} = \frac{V_\mu^2}{\int u_\mu^4(\boldsymbol{r}) \,\mathrm{d} V}\, .
\end{equation}
is the {\em effective (nonlinear)} mode volume and $V_\mu$ the mode-volume defined in the context of Eq.~\ref{eq:ch3:zpe}.
The nonlinear refractive index $n_2$, here for fields with frequency $\omega_0$, is related to the third order nonlinear susceptibility via
\begin{equation}
n_{2}=\frac{3}{4 n_\mathrm{eff}^2(\omega_0) \varepsilon_{0} c} \chi^{(3)}(\omega_0; \omega_0, \omega_0, -\omega_0).
\end{equation}
For a waveguide resonator with length $L$ we have $V_\mathrm{eff}=A_\mathrm{eff}L$, where $A_\mathrm{eff}$ is the effective nonlinear mode area of the waveguide, which then also permits the definition of the effective nonlinear coefficient $\gamma=\tfrac{\omega_\mathrm{p}}{c}\tfrac{n_2}{A_\mathrm{eff}}$ often used in fiber optics~\cite{agrawal2013NonlinearFiberOptics}. 

The factor $\Lambda_{k,l,m}$ accounts for the frequency dependence in the nonlinear susceptibility, the mode field normalization, the refractive index and the effective mode volume\footnote{
\begin{equation}
    \begin{aligned}
        \Lambda_{k,l,m} 
        = & \frac{\sqrt{\omega_k \omega_l \omega_m \omega_{k+l-m}}}{\omega_0^2}\frac{n_\mathrm{eff}^4(\omega_0)}{n_\mathrm{eff}(\omega_k)n_\mathrm{eff}(\omega_l)n_\mathrm{eff}(\omega_m)n_\mathrm{eff}(\omega_{k+l-m})}   \\
        &  \frac{\chi^{(3)}(\omega_m-\omega_k-\omega_l; -\omega_k, -\omega_l, \omega_m)}{\chi^{(3)}(-\omega_0, \omega_0, -\omega_0, -\omega_0)} \, \\
        & \frac{V_\mathrm{eff}(\omega_0)\,\int  \,
    u_k^*(\boldsymbol{r})
    u_l^*(\boldsymbol{r})
    u_m(\boldsymbol{r})
    u_{k+l-m}(\boldsymbol{r}) \, \mathrm{d}V }{\sqrt{V_k V_l V_m V_{k+l-m} } } 
    \end{aligned}
\end{equation}}. For combs localized around the pump, we can assume to good approximation $\Lambda_{k,l,m}\approx1$. Despite the assumptions made, this describes well (at least qualitatively) the nonlinear optical interactions, even for very broadband and octave-spanning spectra.

\begin{figure}[h!]
\includegraphics[width=0.8\columnwidth]{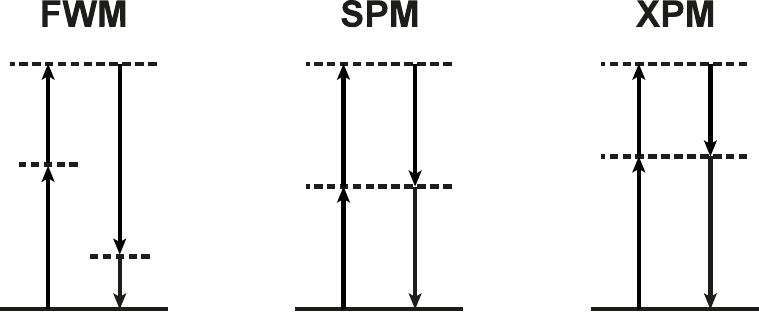}
\caption{\textbf{Different FWM schemes}. From left to right: general FWM - four-wave mixing, SPM - self-phase modulation, XPM - cross-phase modulation. } \label{fig:ch4:FWM}
\end{figure}

Each term in the Kerr Hamiltonian (Eq.~\ref{eq:ch4:kerr_hamilt}) describes 
an interaction between four optical modes, i.e. a FWM process, in which two photons of frequency $\omega_k$ and $\omega_l$ are created (annihilated) and two photons of frequency $\omega_m$ and $\omega_{k+l-m}$ are annihilated (created) (Fig.~\ref{fig:ch4:FWM}).  
We can identify two important special cases, where we have set for simplicity $\Lambda_{m, k, l}=1$:

\textit{Self-phase modulation} (SPM)  
involves photons from only one mode, so the corresponding part of Kerr Hamiltonian is:
\begin{equation}
        \label{eq:ch4:H_SPM}
       \hat{H}_\mathrm{SPM}=  -\frac{\hbar g_\mathrm{K}}{2} \hat{a}_\mu^{\dagger} \hat{a}_\mu^{\dagger} \hat{a}_\mu \hat{a}_\mu 
\end{equation}
As $-\tfrac{i}{\hbar}[\hat{a}_\mu,\hat{H}_\mathrm{SPM}] = ig_\mathrm{K} \hat{a}_\mu^\dagger \hat{a}_\mu \hat{a}_\mu$, this leads to an additional term $ig_\mathrm{K}|a_\mu|^2 a_\mu$ on the RHS of Eq.~\ref{eq:sec3:one_mode_CME}, implying a Kerr shift in the effective resonance frequency proportional to the intensity (i.e. an additional intensity-dependent phase increment per resonator roundtrip). The coupling constant $g_\mathrm{K}$ corresponds to the amount of Kerr shift per photon~\cite{matsko2005optical}. 

\textit{Cross-phase modulation} (XPM) involves interaction of photons from two different microresonator modes ($\mu\neq\nu$). 
Considering possible permutations of the fields, this part of the Kerr Hamiltonian is given by
\begin{equation}
       \hat{H}_\mathrm{XPM}=  -2 \hbar g_\mathrm{K}\hat{a}_\mu^{\dagger} \hat{a}_\nu^{\dagger} \hat{a}_\mu \hat{a}_\nu
\end{equation}
On the RHS of Eq.~\ref{eq:sec3:one_mode_CME} XPM leads to the additional term $2ig_\mathrm{K}|a_\nu|^2 a_\mu$, as $-\tfrac{i}{\hbar}[\hat{a}_\mu,\hat{H}_\mathrm{XPM}] = 2ig_\mathrm{K} \hat{a}_\nu^\dagger \hat{a}_\nu \hat{a}_\mu$. Thus, XPM leads to a Kerr resonance shift for mode $\mu$ caused by the intensity of mode $\nu$, and this shift is twice as large as the corresponding SPM shift of mode $\nu$. The differential shifts in effective resonance frequency between a strong pump mode (experiencing SPM) and a weak mode (experiencing XPM) play an important role in phase-matching and first sideband generation that is covered in Sec~\ref{sec:basics_kcomb:MI}, imposing requirements on dispersion engineering for the parametric gain control~\cite{pidgayko2023VoltagetunableOpticalParametric} and highly efficient comb generation~\cite{black2022OpticalparametricOscillationPhotoniccrystal}.

\subsection{Bistability and hysteresis}
\label{sec:basics_nln:bistab}
Considering only one cavity mode and neglecting quantum noise, we extend the classical evolution equation described by Eq.~\ref{eq:sec3:one_mode_CME} to the nonlinear case by including SPM. Adding $\hat{H}_\mathrm{SPM}$ of Eq.~\ref{eq:ch4:H_SPM} to the system's Hamiltonian we obtain:
\begin{equation}\label{eq:ch4:one_mode_eq}
    \dot{a}_0 = -\left(\frac{\kappa}{2}+i(\omega_0-\omega_\mathrm{p})\right)a_0 + i g_\mathrm{K} \left|a_0\right|^2a_0 + \sqrt{\kappa_{\mathrm{ex}}}s_{\mathrm{in}}.
\end{equation}
Through regrouping of the terms, one can see that the nonlinear term has the same effect as introducing a shifted effective resonance frequency: 
\begin{equation}
    \label{eq:eff_resonance}
    \omega_\mathrm{eff,0}=\omega_0 - g_\mathrm{K}|a_0|^2
\end{equation}

\begin{figure}[h]
\includegraphics[width=\columnwidth]{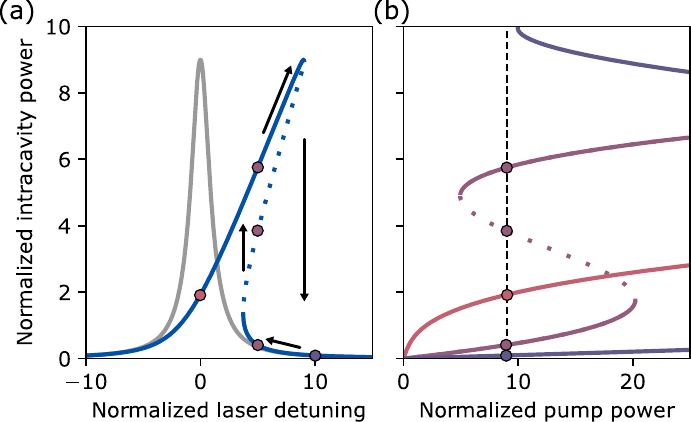}
\caption{ \textbf{Kerr tilt and bistability.} Intracavity power as a function of (a) normalized laser detuning and (b) normalized pump power. The dashed lines show the unstable branch of the solution. Normalized pump power $f^2 = 9$. Black arrows show the directions while tuning from the blue and red sides of the resonance.
\label{fig:ch4:Bistability}}
\end{figure}

Setting $\dot{a}_0=0$ and introducing normalized variables $\zeta_0=2 (\omega_0 - \omega_\mathrm{p}) / \kappa$,  $\rho = \left|\psi_0\right|^2  =  2 g_\mathrm{K} / \kappa \left| a_0\right|^2 $, $f =\sqrt{\frac{8 \eta g_\mathrm{K}}{\kappa^2}}s_{\mathrm{in}} $, we obtain the following algebraic equation of the third order describing the stationary state of the system:
\begin{equation}\label{eq:ch4:cubic}
\rho^3-2 \zeta_0 \rho^2+\left(\zeta_0^2+1\right) \rho= f^2.
\end{equation}
It can have either one or three roots depending on the parameters $\zeta_0$ and $|f|^2$. 
Taking a derivative of the left-hand side of Eq.~\ref{eq:ch4:cubic} with respect to $\rho$, we obtain $\rho_{\pm}= \left( 2\zeta_0 \pm \sqrt{\zeta_0^2-3}\right)/3$
which indicates that there is a critical value of the normalized detuning $\zeta_\mathrm{crit} = \sqrt{3}$. Below this value the system has only one solution. while above it there are three solutions. A stability analysis shows that two out of the three solutions are stable as indicated in Fig.~\ref{fig:ch4:Bistability}.
The region, where the system has two stable solutions is called the \textit{bistability region}. We can define the range of normalized pump power and detuning values corresponding to the bistability region substituting $\rho_{\pm}$ in Eq.~\ref{eq:ch4:cubic}:
\begin{equation}
\label{eq:ch5:bistability_area}
f_{\pm}^{2}=\frac{2 \zeta_0 \mp \sqrt{\zeta_0^2-3}}{3}\left[1+\left(\frac{\sqrt{\zeta_0^2-3} \pm \zeta_0}{3}\right)^2\right].
\end{equation}
The intra-cavity power-dependent Kerr nonlinear resonance shift modifies the apparent resonance shape (despite being shifted, the underlying resonance shape for a specific intracavity power remains Lorentzian). If a pump laser is frequency-scanned towards increasing normalized detuning $\zeta_0$, the system follows the upper branch of the tilted resonance, before falling down to the lower branch stable solution. The resonance shape observed in a laser scan resembles a triangle (Fig.~\ref{fig:ch4:Bistability}a). Conversely, for decreasing normalized detuning $\zeta_0$, the system follows the lower branch of the bistable region and then jumps to the monostable one (Fig.~\ref{fig:ch4:Bistability}a, black arrows).  A similar dynamics exists in the context of thermal effect (Sec.~\ref{sec:exp_DKS:Res_thermal}). 

Usually, coherent Kerr combs are operated in the bistable regime and require pump power of the same order of magnitude needed to achieve the bistability as we describe in Sec.~\ref{sec:basics_kcomb:models_of_OFC}. In this case, the lower branch solution of Eq.~\ref{eq:ch4:cubic} defines the power of the background in the cavity.
Thus, it is useful to find an approximate expression for $\zeta_{0}>\sqrt{3}$ (bistability criterion) and $\zeta_{0}>\zeta_{0}^\mathrm{min}$ ~\cite{barashenkov1996ExistenceStabilityChart, herr2014TemporalSolitonsOptical}: 
\begin{eqnarray}
\Psi =  \frac{if}{|\Psi|^{2}-\zeta_{0}+i}\simeq\frac{f}{\zeta_{0}^{2}}-i\frac{f}{\zeta_{0}}.\label{eq:ch4:lower_br_approx}
\end{eqnarray}

\subsection{Models of microresonator dynamics}
\label{sec:basics_kcomb:models_of_OFC}

The nonlinear dynamics in microresonators can be described equivalently in the frequency domain by coupled mode equations (CMEs)
or in the time domain by the LLE, a driven damped NLSE. Depending on the context one or the other description may be more practical. Further, we present a more general model represented by an infinite-dimensional map known as the Ikeda map, based on the 1-D NLSE with prescribed conditions at the resonator coupling interface. In this section, we neglect quantum noise, which is discussed in Sec.~\ref{sec:exp_DKS:noise}.

\subsubsection{Coupled-mode equations}
The intracavity field may be described by its frequency components in the so-called modal expansion approach~\cite{chembo2010ModalExpansionApproach,chembo2010SpectrumDynamicsOptical,herr2012UniversalFormationDynamics}. It follows from Eq.~\ref{eq:ch4:kerr_hamilt} and constitutes a generalization of Eq.~\ref{eq:ch4:one_mode_eq} to the case of multimode nonlinear interactions. A general form of the CMEs can be expressed as 
\begin{eqnarray}
\dot{a}_{\mu}= & - & [i\left(\omega_\mu - \omega_\mathrm{p} -\mu D_1\right)+\frac{\kappa_{\mu}}{2}]a_{\mu}+\delta_{\mu,0}\sqrt{\kappa_\mathrm{ex,\mu}} s_\mathrm{in}\nonumber \\
 & + & ig_\mathrm{K}\sum_{k,l} \Lambda_{k,l,\mu}\, a_{k}a_{l}a_{k+l-\mu}^{*}
 \label{eq:ch5:CME_rf}
\end{eqnarray}
where $\kappa_{\mu}$ denotes the cavity decay rate of mode $\mu$. One mode ($\mu=0$) is driven by a classical field with optical power $P_\mathrm{in}$ and frequency $\omega_{p}$.
 The pump term is written as $\sqrt{\kappa_\mathrm{ex,\mu}} s_\mathrm{in}$ where $|s_\mathrm{in}|=\sqrt{P_\mathrm{in}/ \hbar\omega_{p}}$.

Through the definition of the $\omega_\mu$ the CMEs~\ref{eq:ch5:CME_rf} allow for modeling of arbitrary mode dispersion in numerical simulations~\cite{herr2014ModeSpectrumTemporal}\footnote{Eq.~\ref{eq:ch5:CME_rf} describes a ring resonator with unidirectional propagation. For a Fabry-P\'erot resonator an additional cross-phase modulation effect resulting from counter-propagating light needs to be considered \cite{obrzud2017TemporalSolitonsMicroresonators,cole2018TheoryKerrFrequency}: 
$$
    (\omega_{\mu}-D_{1}\mu-\omega_{p}) \rightarrow (\omega_{\mu}-D_{1}\mu-\omega_{p} -2g_\mathrm{K}\sum_k|a_k|^2)
$$
Formal equivalence between a Fabry-P\'erot resonator and a ring with unidirectional propagation is obtained by a small intensity dependent offset in the laser detuning.}.
The photon output rates are 
\begin{equation}
    s_\mathrm{out,\mu} =  \delta_{\mu,0} s_\mathrm{in} - \sqrt{\kappa_\mathrm{ex,\mu}} a_\mu.
\end{equation}

Introducing the dimensionless quantities 
\begin{align}\label{eq:ch5:normalization_cme}
    \tau=\frac{\kappa}{2} t \quad \psi_\mu=\sqrt{\frac{2 g_\mathrm{K}}{\kappa}} a_\mu \quad \quad f=\sqrt{\frac{8 \eta g_\mathrm{K}}{\kappa^{2}}} s_{\text {in }} \nonumber\\ 
    \zeta_{\mu}=\frac{2}{\kappa} (\omega_\mu-\omega_{p} -\mu D_1) \quad \quad
\end{align}
we can write Eqs.~\ref{eq:ch5:CME_rf} in dimensionless form (also setting $\Lambda_{k,l,m}=1$ for simplicity)
\begin{equation}
    \label{eq:ch5:cme_normalized}
    \frac{\partial \psi_\mu}{\partial \tau} = -(1+i\zeta_\mu)\psi_\mu + i\sum_{k,l} \psi_k \psi_l \psi_{k+l-\mu}^* + \delta_{0\mu}f.
\end{equation}
Considering only pure second-order dispersion (without AMX or higher order dispersion $D_i=0$ for $i>2$) then 
\begin{equation}
\label{eq:ch5:gen_detuning}
\zeta_{\mu}=\zeta_0 + \frac{d_2 \mu^2}{2}, \quad \quad d_2 = \frac{2 D_2}{\kappa} \, .
\end{equation}

\subsubsection{Lugiato-Lefever equation}
Another way to describe the microresonator dynamics is to use a nonlinear partial-differential equation, the LLE~\cite{lugiato1987SpatialDissipativeStructures,castelli2017LLEPatternFormationa} which can be seen as a driven-dissipative version of the 1-D NLSE~\cite{haelterman1992DissipativeModulationInstability,barashenkov1996ExistenceStabilityChart}. 
The LLE and CMEs (Eqs.~\ref{eq:ch5:CME_rf}) are equivalent and directly related via the Fourier transform~\cite{hansson2014NumericalSimulationKerr,chembo2013SpatiotemporalLugiatoLefeverFormalism}. For the nonlinear terms, this can be seen by introducing an index $k = \mu''- \mu$ in Eq.~\ref{eq:ch5:CME_rf}, resulting in two convolutions~\cite{hansson2014NumericalSimulationKerr} transforming according to:
\begin{equation}
\sum_{k,\mu'} a_{\mu+k} a_{\mu'+k}^{*} a_{\mu'}=\mathcal{F}^{-1}\left[\left|A\right|^{2} A\right]\, .\end{equation}
The LLE describes the evolution of a slowly varying complex field envelope in a frame co-rotating with the group velocity around the circumference of the microresonator~\cite{matsko2011ModelockedKerrFrequency,chembo2013SpatiotemporalLugiatoLefeverFormalism}.
It is often formulated as 
\begin{eqnarray}
\frac{\partial A}{\partial t}  = -\left(\frac{\kappa}{2}+i(\omega_{0}-\omega_\mathrm{p})\right)A + i\frac{D_2}{2}\frac{\partial^{2}A}{\partial\phi^{2}} \nonumber \\ + ig_\mathrm{K}|A|^{2}A +\sqrt{\kappa_\mathrm{ex}}s_\mathrm{in},
\label{eq:ch5:LLE}
\end{eqnarray}
where  $A(\phi, t)=\sum_\mu a_\mu(t) e^{i \mu \phi}$

is the slowly varying amplitude normalized with respect to the electric field as given by Eq.~\ref{eq:ch3:zpe}, where

$\phi=\varphi-D_1 t$ is the co-rotating angular coordinate inside the resonator, $\varphi\in[0,2\pi)$ is the regular polar angle, and $ s_\mathrm{in} = \sqrt{P_{\mathrm{in}}/\hbar\omega_\mathrm{p}}$.

The choice of the co-rotating (angular) velocity $D_1$ does not make an assumption on the actual (angular) group velocity of the waveform. If the waveform's (angular) group velocity differs from $D_1$ then the obtained solution will slide through the co-moving time frame. The time dependent photon output rate is
\begin{equation}
\label{eq:ch5:sout}
    s_\mathrm{out}(t) = s_\mathrm{in} - \sqrt{\kappa_\mathrm{ex}} A(\phi_0,t)
\end{equation}
where $\phi_0$ is the coupler position.

With the following change of variables:
\begin{align}\label{eq:ch5:normalization}
    \Psi&=\sqrt{\frac{2 g_\mathrm{K}}{\kappa}} A = \sum_\mu \psi_\mu e^{i \mu \phi} \nonumber\\  \tau&=\frac{\kappa}{2} t \quad d_{2}= 2 D_2/\kappa \nonumber\\ 
    \zeta_{0}&=\frac{2 (\omega_0-\omega_\mathrm{p})}{\kappa}  \quad f=\sqrt{\frac{8 \eta g_\mathrm{K}}{\kappa^{2}}} s_{\text {in }}=\sqrt{\frac{P_{\text {in }}}{P_{\text {th }}}},
\end{align}
where $P_\mathrm{th}$ is a threshold power of the parametric process, which is discussed in Sec.~\ref{sec:basics_kcomb:MI}. We can obtain the LLE (Eq.~\ref{eq:ch5:LLE}) in its dimensionless form:
\begin{equation}\label{eq:ch5:LLE_norm}
\frac{\partial \Psi}{\partial \tau}=-\left(1+i \zeta_{0}\right) \Psi+i \frac{d_2}{2} \frac{\partial^{2} \Psi}{\partial \phi^{2}}+i|\Psi|^{2} \Psi+f
\end{equation}
or with the additional change of variable 
$\theta=\sqrt{\frac{1}{|d_{2}|}} \phi$, which represents the normalized space (or fast time in the fiber optics setting):
\begin{equation}\label{eq:ch5:LLE_norm2}
\frac{\partial \Psi}{\partial \tau}=-\left(1+i \zeta_{0}\right) \Psi + i\frac{ \text{sgn}(d_2)}{2} \frac{\partial^{2} \Psi}{\partial \theta^{2}}+i|\Psi|^{2} \Psi+f.
\end{equation}

In this form, it becomes clear that solutions of the LLE can be mapped to 2-D space: effective cavity detuning $\zeta_0$ vs normalized pump power $f^2$.

\begin{figure}[h]
\includegraphics[width=0.99\columnwidth]{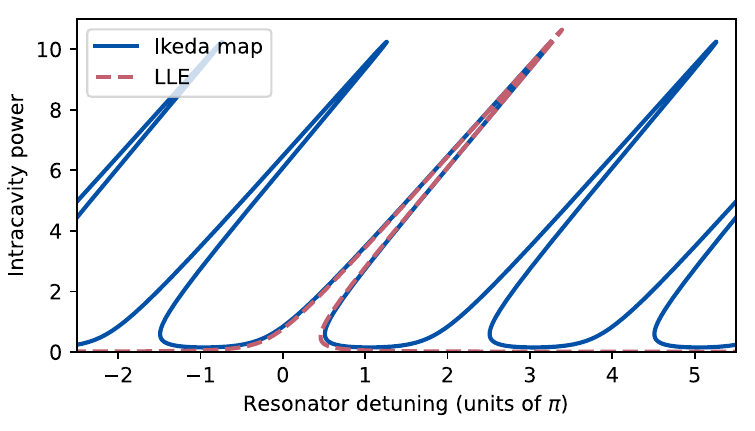}
\caption{\textbf{Nonlinear cavity resonances.} Comparison of the LLE and the Ikeda map intracavity power as a function of the detuning phase (pump power is adjusted to match the amplitudes), showing that only the latter model captures multistable states. Similar plot can be found in~\cite{anderson2017CoexistenceMultipleNonlinear}.} \label{fig:ch5:LLE_vs_Ikeda}
\end{figure}

\subsubsection{Ikeda map}
The Ikeda map is more general than the CMEs or the LLE of Eqs.~\ref{eq:ch5:CME_rf} and \ref{eq:ch5:LLE} and only requires the validity of the NLS equation along the circumference of the ring (i.e., it does not require that the changes per resonator roundtrip are small)~\cite{ikeda1979MultiplevaluedStationaryStatea}. It can be written in the following way\footnote{The Ikeda map formalism can be extended to Fabry-P\'erot resonator with XPM from counter-propagating fields~\cite{ziani2024TheoryModulationInstability}.}: 
\begin{equation}\label{Eq:ch5:Ikeda_map}
\begin{aligned}
E^{m+1}(t, 0)=& \sqrt{\Theta_\mathrm{P}} E_\mathrm{in}+\sqrt{1-\Theta_\mathrm{P}} e^{i \phi_{0}} E^{m}(t, L) \\
\frac{\partial E^{m}(t, z)}{\partial z}=&-\frac{\alpha_{\mathrm{loss}}}{2} E^{m}(t, z)-i \frac{\beta_{2}}{2} \frac{\partial^{2} E^{m}(t, z)}{\partial t^{2}} \\
&+i \gamma\left|E^{m}(t, z)\right|^{2} E^{m}(t, z),
\end{aligned}
\end{equation}
where $E_\mathrm{in}$ is the power-normalized pump field envelope, index $m$ indicates the roundtrip number, $E^{m}$ is the field envelope inside the microresonator at the m$^\mathrm{th}$ roundtrip, $\Theta_\mathrm{P}$ is the power coupling coefficient (i.e. the dropped power through the coupler), and $\alpha_\mathrm{loss} = \kappa_0 T_\mathrm{R}/L$ is the attenuation coefficient. The phase $\phi_0 = k_0 L$ is accumulated during the roundtrip propagation, where $k_0 = 2\pi n/\lambda$ denotes the conventional wavenumber. The output field is $E_{\mathrm{out}}^{m}(t)
= \sqrt{1-\Theta_\mathrm{P}}\, E_\mathrm{in}(t)
  - \sqrt{\Theta_\mathrm{P}}\, e^{i\phi_{0}}\, E^{m}(t,L)$.

In contrast to the LLE (Eq.~\ref{eq:ch5:LLE}), the Ikeda map accounts for the multi-resonant nature of the optical cavity and thus can describe the effect of \textit{multistability} - 
simultaneous stability of more than two CW solutions (when the apparent resonance tilt described in Sec.~\ref{sec:basics_nln:bistab} exceeds the FSR).  The steady state ($E^{m+1}(t,0)=E^{m}(t,0)$) homogenous solution ($\partial^{2} E^{m}(t, z)/\partial t^{2} = 0$) in this case satisfies the Airy equation~\cite{anderson2017CoexistenceMultipleNonlinear, hansson2015FrequencyCombGeneration}:
\begin{equation}
    P=\frac{\Theta_\mathrm{P} P_{\text {in}}}{(1-C)^2\left[1+F \sin ^2\left(\frac{\phi_{0}+\gamma L P}{2}\right)\right]},
    \label{eq:ch5:Ikeda_tilt}
\end{equation}
where the power $P = |E^{m}(t,0)|^2$, $P_\mathrm{in} = |E_\mathrm{in}|^2$, and $C = \sqrt{1-\Theta_\mathrm{P}}\exp[-\alpha_{\mathrm{loss}}L/2]$, $F = 4C/(1-C)^2$, coefficient $\gamma L P$ represents a nonlinear phase acquired due to the SPM. A direct comparison between the nonlinear resonances of the Ikeda map and the LLE is shown in Fig.~\ref{fig:ch5:LLE_vs_Ikeda}.
As multi-stability is usually only present in small FSR fiber-based resonators, we formulate the Ikeda map in the notation used primarily in fiber optics~\cite{agrawal2013NonlinearFiberOptics}.

The mean-field model (Eq.~\ref{eq:ch5:LLE}) can be derived from the Ikeda map
assuming small roundtrip losses and that both the nonlinear length $L_\mathrm{nl} = (\gamma |E|^2)^{-1}$ and the dispersion length $L_\mathrm{d} = t_0^2 / |\beta_2|$ (with $t_0$ the duration of the shortest
temporal feature in the waveform) are much larger than the resonator length $L$, and that the pump laser detuning from the nearest cavity resonance is small,
$|2\pi m_0 - \phi_0| \ll 1$, where $m_0$ denotes the pumped mode~\cite{haelterman1992DissipativeModulationInstability,agrawal2013NonlinearFiberOptics}.

In this way, the discrete roundtrip iterations turn into the partial (slow) time derivative: $\left[E_{m+1}(z=0, \tau)-E_m(z=0, \tau)\right] / T_{\mathrm{R}} \rightarrow~\partial_t E\left(t=m T_{\mathrm{R}}, \tau\right)$. 
Combining this approximation with the prescribed boundary conditions (Eq.~\ref{Eq:ch5:Ikeda_map}, first line) and applying the assumptions on the cavity length, we obtain the LLE. An extended version of the LLE that accounts for the effect of multistability has been demonstrated in Ref.~\cite{conforti2017MultiresonantLugiatoLefever,kartashov2017MultistabilityCoexistingSoliton}.

\begin{figure*}

\includegraphics[width=1\linewidth]{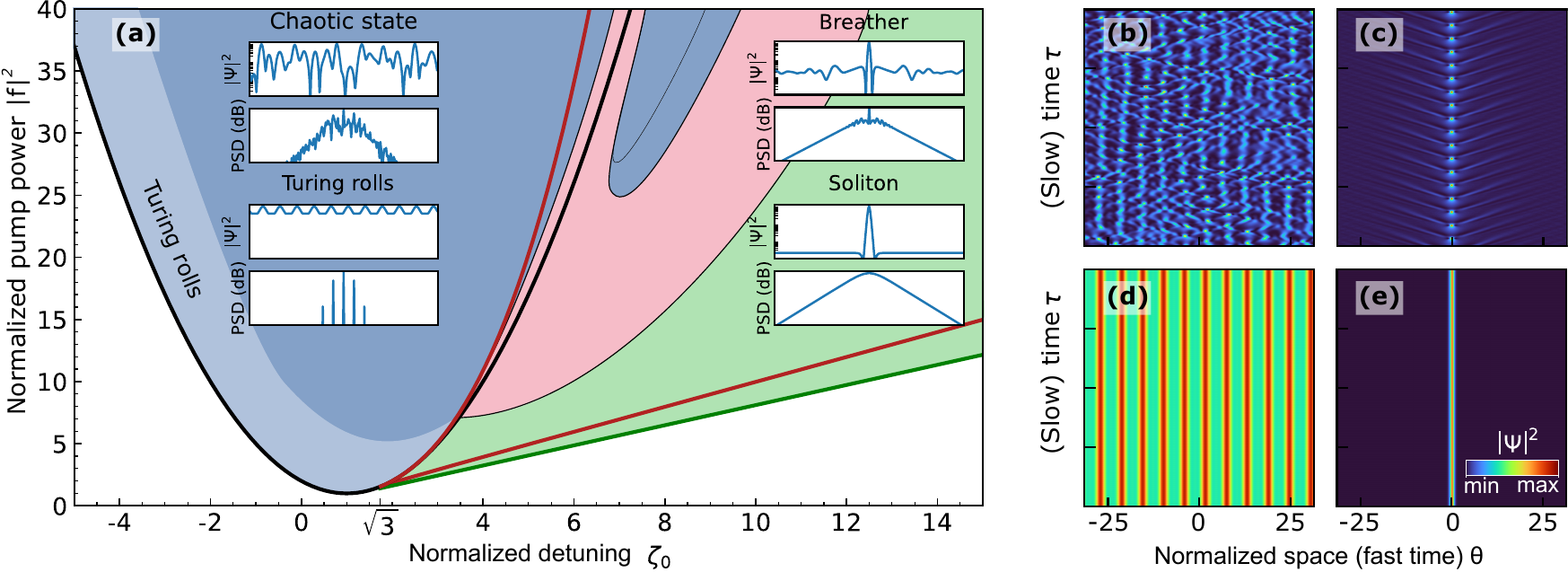}
\caption{\textbf{Phase diagram of the LLE.} (a) The phase diagram in $|f|^2-\zeta_0$ plane. Colors depict different states of the system. Green: stable DKS; Red: breather; Light blue: Turing patterns; Dark blue: chaotic regime; White: Only CW solution. The black curve (threshold of complexity) indicated parameters where MI occurs (and a stable CW waveform is not stable). Red curves delineate the bistable regime (b-e) Spatiotemporal diagrams of the key states (chaotic state, breather, Turing pattern, DKS) that can be observed in the system at $|f|^2 = 36$.
Similar diagrams can be found in~\cite{godey2014StabilityAnalysisSpatiotemporal,leo2013DynamicsOnedimensionalKerr,parra-rivas2014DynamicsLocalizedPatterned,karpov2019DynamicsSolitonCrystals}
\label{fig:ch5:phase_diagram} } 
\end{figure*}

\subsection{Phase diagram for the LLE with anomalous dispersion}
\label{sec:basics_kcomb:phase_diagram}

The regimes of a nonlinear microresonator with anomalous dispersion as described by the CMEs~\ref{eq:ch5:cme_normalized} or the LLE~\ref{eq:ch5:LLE_norm} can be mapped onto a 2D phase diagram as a function of normalized detuning $\zeta_0$ and normalized pump power $|f|^2$, as shown in Fig.~\ref{fig:ch5:phase_diagram}(a)~\cite{godey2014StabilityAnalysisSpatiotemporal,leo2013DynamicsOnedimensionalKerr,parra-rivas2014DynamicsLocalizedPatterned,karpov2019DynamicsSolitonCrystals}. Three main analytical curves separate the key areas in the phase diagram:

\textit{(i)} The first curve (black) describes the \textit{threshold of complexity} that is determined by the condition $\rho = \left|\psi_0\right|^2=1$; a derivation of the threshold is given by Eq.~\ref{eq:ch5:mi_threshold}. It separates the parameter space where the homogeneous (CW) solution is stable (below the black line) from the area where so-called modulation instability (MI, Sec.~\ref{sec:basics_kcomb:MI}) leads to a breakup of the CW solution into a temporally structured waveform. Depending on the parameters, MI will lead to the formation of Turing rolls (Sec.~\ref{sec:basics_kcomb:TuringRolls}), chaotic dynamics (Sec.~\ref{sec:basics_kcomb:chaotic_states}), dissipative Kerr solitons (Sec.~\ref{sec:basics_kcomb:DKS}), or breathers (Sec.~\ref{sec:basics_kcomb:Breathing}).

\textit{(ii)} The second defining curves (red) are the \textit{bistability curves} given by Eq.~\ref{eq:ch5:bistability_area}, enclosing the parameter space where a low intracavity power CW solution or a higher power solution can exist.
This includes the area where DKS - the ultrashort temporal structure underlying coherent and broadband microcombs - can coexist with the CW solution (Sec.~\ref{sec:basics_kcomb:DKS})\footnote{Note that DKSs can exist in a narrow range of detunings in the monostable regime~\cite{parra-rivas2014DynamicsLocalizedPatterned}, but this state is of limited use for the experimental microcomb generation.}. In addition, there is a variety of chaotic states (Sec.~\ref{sec:basics_kcomb:chaotic_states}) that include spatiotemporal chaos and transient chaos
as well as temporal chaos, all shown by dark blue in Fig.~\ref{fig:ch5:phase_diagram}a and  and discussed in Sec.~\ref{sec:basics_kcomb:chaotic_states}.

The red area indicates breathing states, i.e. non-stationary but periodically evolving temporal patterns shown in Fig.~\ref{fig:ch5:phase_diagram}(c) and further discussed in Sec.~\ref{sec:basics_kcomb:Breathing}.

\textit{(iii)} The third curve (green), marks the \textit{DKS existence range} give by Eq.\ref{eq:ch5:sol_ex_range} in Sec.~\ref{sec:basics_kcomb:DKS} (maximal detuning $\zeta_0$ for a given $|f|^2$ where DKS can exist). Thus, non-breathing DKS states can exist in the green area between this line and the breathing states. 
In the white area, only stable CW solutions can be observed. The DKS existence range and non-CW waveforms that fall below the black line cannot be directly accessed as the CW solution is stable. To access these states, a parameter trajectory that crosses above the black line is needed to perturb the CW solution via MI. 

\textit{Boundaries and bifurcations}. The boundaries between the different nonlinear regimes can be found by numerically investigating the eigenvalue structure $\{\lambda_i\}$ of the Jacobian matrix (see Sec.~\ref{sec:basics_kcomb:MI}) in the presence of a stationary solution of the LLE and applying boundary tracking algorithms~\cite{wangBoundaryTrackingAlgorithms2014a}. Points in the parameter space where the Re$(\lambda) > 0$ define the boundary of the linear stability of the solution and correspond to bifurcations. Different cases can be distinguished~\cite{qi2019DissipativeCnoidalWaves}: saddle-node (this type of bifurcation defines the boundaries of the bistability region given by Eq.~\ref{eq:ch5:bistability_area} and thus corresponds to the upper detuning limit of the DKS existence), transcritical (e.g. switching between different number of patterns), and Hopf temporal bifurcation corresponding to breathing or chaotic dynamics. More details about the analysis of Jacobian eigenvalues and the bifurcations structure in the nonlinear microresonators can be found in~\cite{parra-rivas2014DynamicsLocalizedPatterned,barashenkov1996ExistenceStabilityChart,gomila2007BifurcationStructureDissipative,godey2014StabilityAnalysisSpatiotemporal,miyaji2010BifurcationAnalysisLugiato,perinet2017EckhausInstabilityLugiatoLefever,parra-rivas2022DissipativeLocalizedStates}, also see~\cite{kapitula2013SpectralDynamicalStability,kuznetsov2023ElementsAppliedBifurcation,seydel2010PracticalBifurcationStability} for a general description. 
We note that the phase diagram in Fig.~\ref{fig:ch5:phase_diagram} has approximate contours found both with the analytical expressions, but also via the stationary or dynamical numerical methods (Sec.~\ref{sec:basics_kcomb:numerical_methods}) combined with eigenvalue tracking. The diagram can be substantially modified if the system is described by a generalized version of the LLE, see Sec.~\ref{sec:aspects_DKS}.

\begin{figure}[h]
\includegraphics[width=1.0\columnwidth]{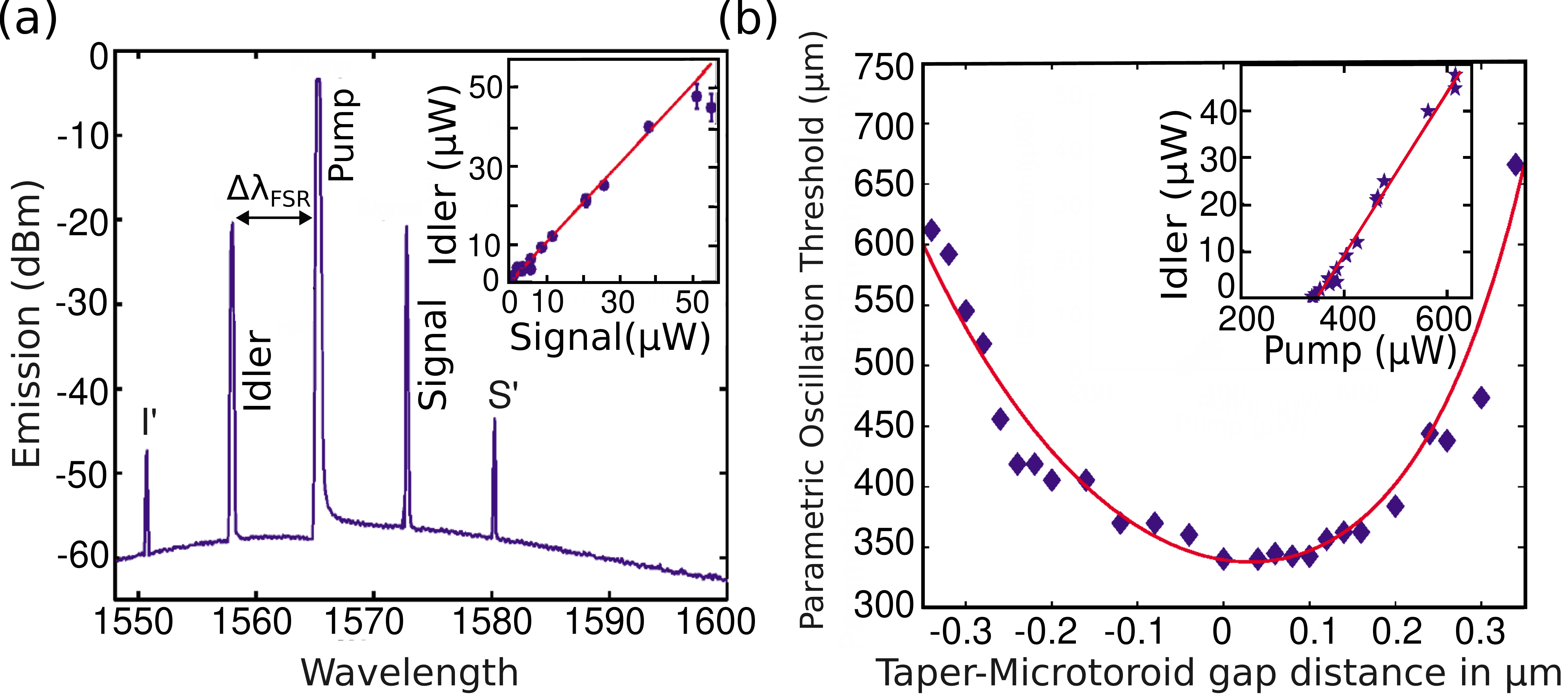}
\caption{\textbf{Observation of optical parametric oscillations in microresonators.} (a) Emission spectrum. (b) Oscillation threshold as a function of the microresonator-taper gap distance and hence the external coupling rate $\kappa_\mathrm{ex}$. From~\cite{kippenberg2004KerrNonlinearityOpticalParametrica}} \label{fig:ch4:FWM_obs}
\end{figure}

\subsection{Modulation instability and optical parametric oscillation threshold}
\label{sec:basics_kcomb:MI}
The stationary CW solution in the microresonator found in Sec.~\ref{sec:basics_nln:bistab} can be unstable towards perturbations, where certain frequency components of a small temporal modulation can experience positive gain. This effect is broadly called \textit{modulation instability}, MI~\cite{zakharov2009ModulationInstabilityBeginning}. MI has been independently observed in hydrodynamics~\cite{benjamin1967InstabilityPeriodicWavetrains,benjamin1967DisintegrationWaveTrains} and in optics~\cite{bespalov1966FilamentaryStructureLight}, almost in parallel with theoretical investigations~\cite{whitham1965GeneralApproachLinear,lighthill1965ContributionsTheoryWaves}. In contrast to conservative systems, such as the 1D NLSE, where (initially) exponential growth of perturbations can leads to the emergence of complex coherent structures~\cite{copie2020PhysicsOnedimensionalNonlinear}, MI in driven-dissipative systems can lead to both stationary (spatial and temporal patterns), limit cycles (breathing solutions) as well as chaotic behavior, as those shown in Fig.~\ref{fig:ch5:phase_diagram}(b). The connection between the linear instability of the solution and formation of patterns has been investigated by Alan Turing for the reaction-diffusion equations~\cite{Turing1952ChemicalBasis} and actively developed by I. Prigogine and others~\cite{cross1993PatternFormationOutside}. Later, as a spatial phenomenon, this has been predicted and observed in optical cavities~\cite{lugiato1987SpatialDissipativeStructures,ackemann2009ChapterFundamentalsApplications,abraham1990OverviewTransverseEffectsa,lugiato2018LugiatoLefeverEquation}.

In Kerr-nonlinear optical resonators, MI leads to the amplification of small fluctuation in the cavity field, or equivalently, amplification of spectral noise sidebands to monochromatic CW field. The sidebands can grow into coherent optical parametric oscillation (OPO), symmetric around the pump laser frequency and their amplitude and position depend on the pump power and laser detuning~\cite{haelterman1992DissipativeModulationInstability,herr2012UniversalFormationDynamics}. The emergence of sidebands marks the transition from the stable CW solution to temporally modulated waveform.
In microresonators, this was first observed in silica microtoroids~\cite{kippenberg2004KerrNonlinearityOpticalParametrica} (Fig.~\ref{fig:ch4:FWM_obs}) and in a CaF$_2$ whispering-gallery mode resonator~\cite{savchenkov2004LowThresholdOptical}. Such OPO has since been widely investigated~\cite{black2022OpticalparametricOscillationPhotoniccrystal,guidry2020OpticalParametricOscillation,okawachi2020DemonstrationChipbasedCoupled,pasquazi2013SelflockedOpticalParametric,reimer2015CrosspolarizedPhotonpairGeneration,zhou2022HybridModeFamilyKerrOptical} enabling low threshold and high efficiency of the process~\cite{ji2017UltralowlossOnchipResonators,lu2019MilliwattthresholdVisibleTelecom,perez2023HighperformanceKerrMicroresonator,stone2022ConversionEfficiencyKerrMicroresonator}, widely separated sidebands and their tunability~\cite{fujii2019OctavewidePhasematchedFourwave,lin2008ProposalHighlyTunable,sayson2017WidelyTunableOptical,sayson2019OctavespanningTunableParametric}, as well as, operation in the visible and infrared parts of the spectrum~\cite{lu2020OnchipOpticalParametric,lu2022KerrOpticalParametric,stone2022EfficientChipbasedOptical}.

The pump power threshold of MI can be found by considering three interacting cavity modes i.e., pump and two sidebands symmetric to the pump~\cite{hansson2013DynamicsModulationalInstability,godey2014StabilityAnalysisSpatiotemporal,chembo2010SpectrumDynamicsOptical,herr2012UniversalFormationDynamics}. The dynamic equation for the two sideband follows from Eq.~\ref{eq:ch5:cme_normalized} and in the undepleted pump approximation read:
\begin{equation}
    \label{eq:mi_dynamics}
    \frac{\partial}{\partial \tau}\begin{pmatrix} \psi_{+\mu} \\ \psi_{-\mu}^* \end{pmatrix}  = \mathbb{J}_\mathrm{MI}
     \begin{pmatrix} \psi_{+\mu} \\ \psi_{-\mu}^* \end{pmatrix}  
\end{equation}
\begin{equation}
    \label{eq:mi_jacobian}
    \mathbb{J}_\mathrm{MI} = \begin{pmatrix}  -(1 +i\zeta_\mu-2i|\psi_{0}|^2) & i\psi_{0}^2 \\ -i\psi_{0}^{*\,2} &  -(1 -i\zeta_\mu+2i|\psi_{0}|^2) \end{pmatrix},
\end{equation}
where $\zeta_\mu$ is given by Eq.~\ref{eq:ch5:normalization_cme}, matrix $\mathbb{J}_\mathrm{MI}$ is the system's Jacobian (Sec.~\ref{sec:num_tool_stable_stationary}), permitting stability analysis of the (temporally flat or homogeneous) CW solution. The real part of the eigenvalues $\lambda$ of $\mathbb{J}_\psi$ 
\begin{equation}
  \label{eq:ch5:mi_gain} 
    \lambda_\pm(\mu)=-1\pm\sqrt{ |\psi_0|^4 -(\zeta_\mu - 2|\psi_0|^2)^2} 
\end{equation}
describe the growth rate of the fields $\Psi_{\pm\mu}$ (for different phase relation), implying instability of the CW solution when $\operatorname{Re}\{\lambda_+\}>0$. The minimal $|\psi_0|^2$ for which $\lambda_\pm(\mu)$ can be non-negative is $|\psi_0|^2=1$ with $\zeta_\mu = 1$. With Eqs.~\ref{eq:ch4:cubic} and \ref{eq:ch5:normalization_cme} we obtain the MI threshold pump power and the mode number of the sideband that first reaches threshold for a given detuning
\begin{equation}
f_{\mathrm{th}}^2 = 1+\left(1-\zeta_0\right)^2; \;  \mu_{\mathrm{th}} =\sqrt{\frac{2\left(1+\sqrt{f_\mathrm{th}^2-1}\right)} { d_2}}
\label{eq:ch5:mi_threshold}
\end{equation}
The mode number $\mu_\mathrm{th}$ reflects the phase matching condition, when SPM and XPM induced by the pump laser are compensated by the dispersion $d_2$. The threshold power is minimal $f_{\mathrm{th}}^2 = 1$ for $\zeta_0=1$, which corresponds to a Kerr-frequency shift (SPM) of half-cavity linewidth $\kappa/2$. In non-normalized units the minimal (parametric/MI) threshold pump power $P_{\mathrm{th}}$ is~\cite{kippenberg2004KerrNonlinearityOpticalParametrica,savchenkov2004LowThresholdOptical,herr2012UniversalFormationDynamics}
\begin{equation}
    P_{\mathrm{th}} =  \frac{\pi}{4} \frac{n_0 V_\mathrm{eff}}{n_2 \eta \lambda Q^2}.
    \label{eq:ch5:threshold_power}
\end{equation}
As a higher-Q factor implies lower loss and higher intracavity power (i.e. higher gain), the threshold scales proportionally to 
$Q^{-2}$ thus significant reduction in threshold power is possible in high-Q cavities~\cite{ji2021ExploitingUltralowLoss}. The undepleted pump approximation used here correctly describes the onset of MI, however, the threshold for stable OPO is higher by a factor of approximately 1.54~\cite{matsko2005optical}. 

\begin{figure}[h]
\includegraphics[width=0.99\columnwidth]{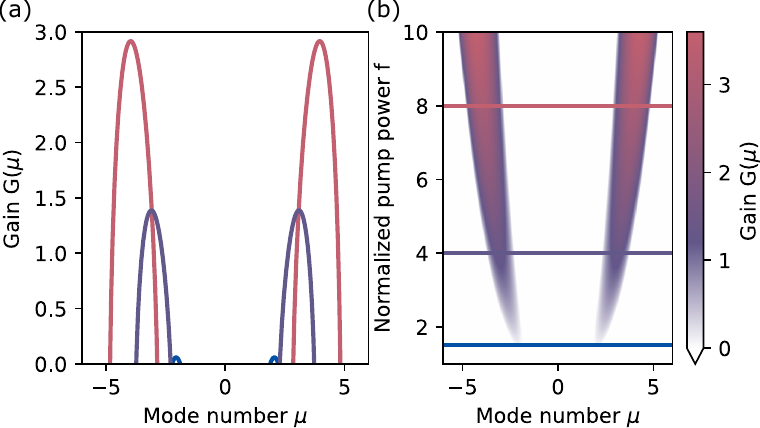}
\caption{\textbf{Modulation instability parametric gain lobes for anomalous dispersion cavity.} In all the plots $\eta = 0$ (a) Parametric gain as a function of mode number for f = 1.5, 4, 8 depicted by green, gray, and red, respectively. (b) The shape of the parametric gain lobes as a function of normalized pump power and mode number. Horizontal lines indicate cross-sections plotted in (a), colors are preserved.  }
\label{fig:ch5:ParametricGainLobes}
\end{figure}

For $\mu_\mathrm{th}$ in non-normalized units, we obtain
\begin{equation}
\label{eq:first_MI_sideband}
  \mu_{\mathrm{th}} = \sqrt{\frac{\kappa}{D_2}\left(1+\sqrt{\frac{P_{\mathrm{in}}}{P_{\mathrm{th}}}-1}\right)},
\end{equation}

The normalized MI gain $G(\mu)=\operatorname{Re}\{\lambda_+ + 1\}$ for anomalous dispersion case is depicted in Fig.~\ref{fig:ch5:ParametricGainLobes}, where $\mu$ is treated as a continuous variable. Due to the discrete nature of $\mu$, also the excitation diagram of the primary combs has a discrete structure~\cite{qi2019DissipativeCnoidalWaves,skryabin2021ThresholdComplexityArnold}, having the form of Arnold tongues~\cite{arnold1983RemarksPerturbationTheory} known from coupled nonlinear oscillators~\cite{wen2016SelforganizationKerrcavitysolitonFormationa,taheri2017SelfsynchronizationPhenomenaLugiatoLefever}.

\begin{figure}[h]
\includegraphics[width=1.0\columnwidth]{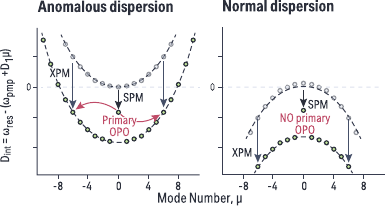}
\caption{\textbf{Phase-matching in microresonators.} FWM phase-matching in anomalous (left) and normal (right) dispersion microresonators. In the anomalous case, frequency matching can be fulfilled, allowing for a primary OPO cascading into a comb. In contrast, the normal dispersion precludes such a nonlinear state, preventing the soft excitation of frequency combs in this regime. From~\cite{moille2025BroadbandVisibleWavelength}.} \label{fig:ch4:FWM_PM_OPO}
\end{figure}

The amplitudes of self-phase and cross-phase modulation differ by a factor of two, leading to distinct frequency shifts between the pumped mode and the other modes, as illustrated in Fig.~\ref{fig:ch4:FWM_PM_OPO}. Consequently, anomalous dispersion is required to trigger the instability dynamics and thus generate bright DKS frequency combs, a conclusion that is also supported by the analytical derivations.
We note, that a discrete input-output coupler (sudden field addition/subtraction), as described by the Ikeda map, can create instability sidebands in both normal and anomalous dispersion regime~\cite{hansson2015FrequencyCombGeneration}.

\subsection{States of the nonlinear resonator.}
\label{sec:basics_kcomb:states}
\begin{figure*}
\includegraphics[width=\textwidth]{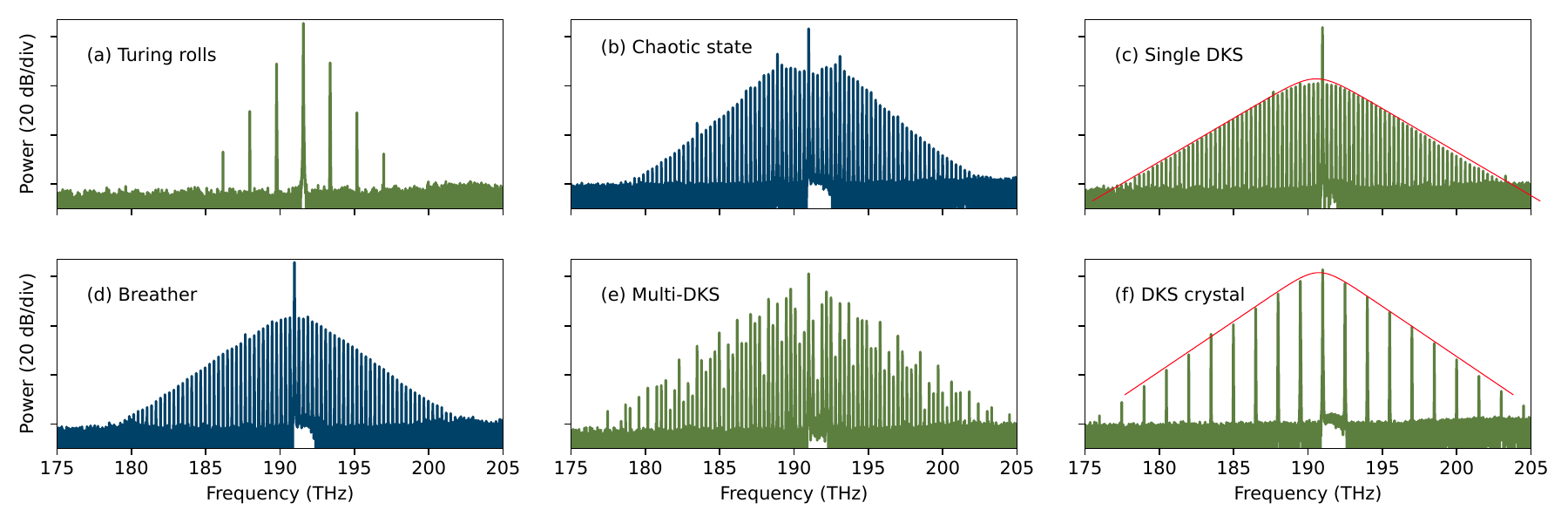}
\caption{\textbf{Experimentally observed spectra of different states.} (a) Turing rolls usually show widely spaced lines. (b) Chaotic state observed for high pump power, appearing as a dense comb-like spectrum. The characteristic `ears' are a hallmark of this state. (c) Single soliton (DKS) state with a smooth $\sech^2$-envelope (red line). (d) Breather state observed form a single DKS state when reducing the detuning. (e) Multi-DKS state, i.e. multiple DKS are present in the resonator. The not generally equal temporal separation of the DKS in the resonator creates a structured interference pattern. (f) Soliton crystal, i.e. a multi-DKS state where inside the resonators the DKSs are arranged with equal temporal separation, effectively resulting in a higher pulse repetition rate and wider line spacing. 
All spectra were measured in the same SiN 300~GHz ring-microresonator. Stationary states are shown in green, dynamic/chaotic states in blue.}  \label{fig:exp_comb_states}
\end{figure*}

Here we discuss the different dynamic and stationary 
states shown in Fig.~\ref{fig:ch5:phase_diagram} including DKS states~\cite{leo2010TemporalCavitySolitons,herr2014TemporalSolitonsOptical} which enable broadband coherent frequency combs and their applications~\cite{kippenberg2018DissipativeKerrSolitons,pasquazi2018MicrocombsNovelGeneration}. A selection of experimentally observed states, which we will discuss below in detail, is presented in Fig.~\ref{fig:exp_comb_states}.

\subsubsection{Turing rolls }
\label{sec:basics_kcomb:TuringRolls}
\begin{figure}[h]
\includegraphics[width=0.99\columnwidth]{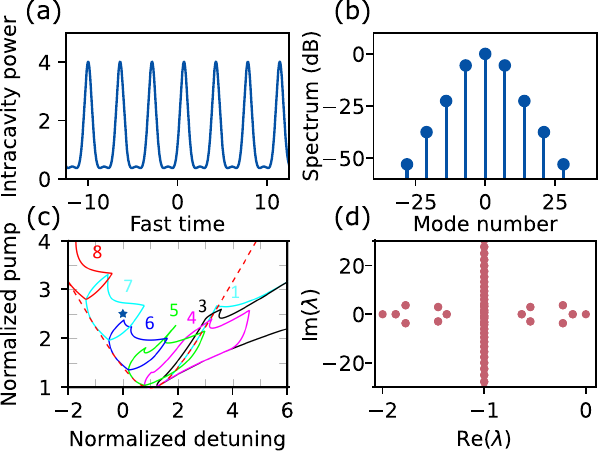}
\caption{\textbf{Turing rolls.}  (a,b) Typical temporal profile of a Turing roll having 7 maxima and corresponding power spectral density, respectively. (c) Stable regions of the cnoidal waves. Numbers indicate the number of maxima. The blue star shows the parameters of the Turing roll plotted in (a,b). The red-dashed curves show the limit below which continuous waves are stable. (d) Eigenvalues of the Jacobian matrix \{$\lambda_i$\} corresponding to a stable Turing roll state. Adapted from~\cite{qi2019DissipativeCnoidalWaves}. }\label{fig:ch5:cn_waves} 
\end{figure}

Turing rolls \footnote{Also known as cnoidal waves named after \textit{cn} Jacobi elliptic function~\cite{korteweg1895ChangeFormLong} - exact solutions of the 1D NLSE~\cite{agafontsev2016IntegrableTurbulenceGenerated}.} are periodic solutions that appear in a variety of nonlinear partial differential equations (PDEs). They emerge as a result of instability of the CW solution discussed in Sec.~\ref{sec:basics_kcomb:MI}. In experiments they can be observed when tuning across the resonance with increasing $\zeta_0$; a typical spectrum is shown in Fig.~\ref{fig:exp_comb_states}. Fig.~\ref{fig:ch5:cn_waves}(a,b) show the simulated temporal power and spectrum. The width of the spectrum increases with detuning and the number of the coherent structures in the Turing roll (i.e. the frequency separation between the sidebands in units of mode numbers) corresponds to the position of the primary sidebands given by the maximum parametric gain value Eq.~\ref{eq:ch5:mi_threshold}; the number of coherent structures may change if the pump laser parameters are varied.

At low pump powers, the Turing roll extend across the monostable and bistable regions of the phase diagram; in the bistable part of the phase diagram they are usually called soliton crystals and a spectrum is shown in Fig.~\ref{fig:exp_comb_states}. The key difference between these states is that the stability of the background allows the existence of an arbitrary number of not necessarily temporally equidistantly localized soliton structures~\cite{parra-rivas2014DynamicsLocalizedPatterned}; see Sec.~\ref{sec:basics_kcomb:DKS}) for more details on soliton crystals.

It has been found that the stability regions of Turing rolls with a different number of temporal maxima can overlap~\cite{qi2019DissipativeCnoidalWaves}, as shown in Fig.~\ref{fig:ch5:cn_waves}(c). 
The number of maxima and hence the corresponding frequency comb spacing can be changed once the stability boundary is crossed. The stability boundary can be tracked by examining the eigenvalue structure (spectrum) of the Jacobian matrix as mentioned in Sec.~\ref{sec:basics_kcomb:phase_diagram}, where eigenvalues with a positive real part imply instability. An example is depicted in Fig.~\ref{fig:ch5:cn_waves}(d). 
The eigenvalue structure of the unperturbed LLE exhibits symmetries along the real axis as well as the line parallel to the imaginary axis at Re$(\lambda) = -1$. The eigenvalue at $\lambda = (0,0)$ corresponds to the translational invariance of the solution, known as Goldstone mode~\cite{gomila2002FluctuationsCorrelationsHexagonal}.

\begin{figure*}
\centering
\includegraphics[width=0.95\textwidth]{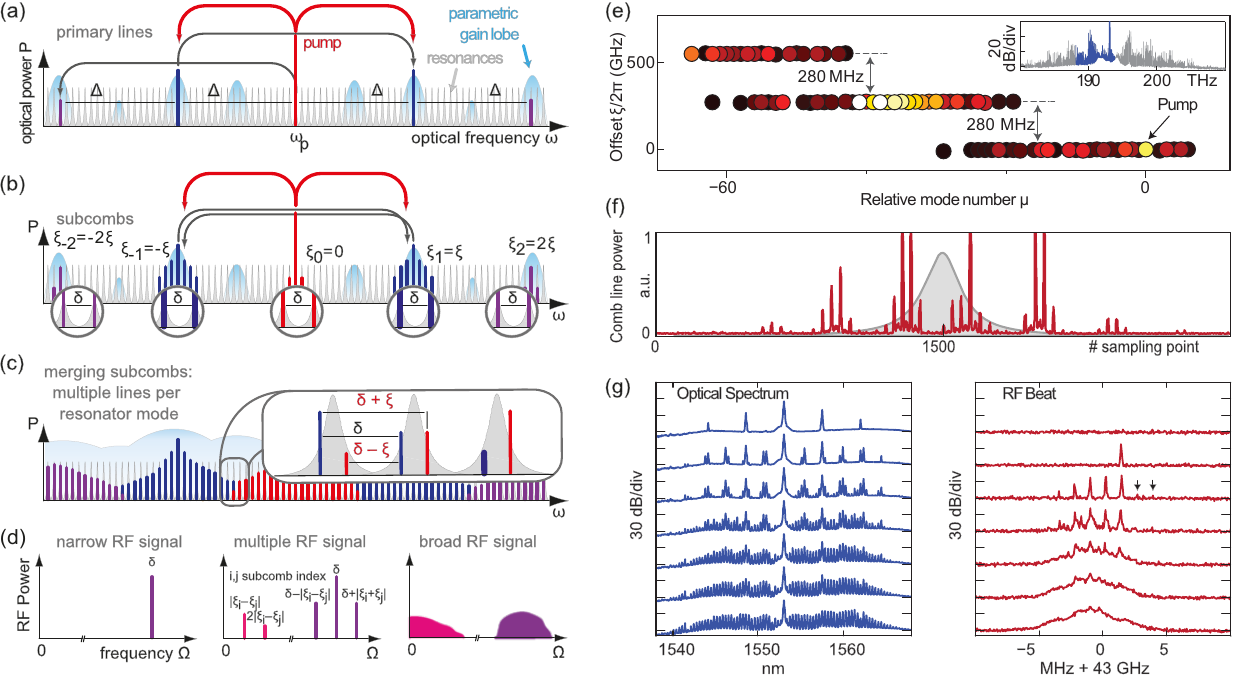}
\caption{\textbf{Kerr-comb dynamics.} (a) Primary comb lines and cascaded FWM. (b) Subcomb formation. (c) Merging of subcombs. (d) Progression of radio-frequency (RF) signal during comb formation. (e) Example of the optical spectrum and RF signal during comb formation (here in a MgF$_2$ whispering-gallery mode resonator). (f) Comb reconstruction based on a scanning laser, revealing subcomb with different offset. Inset: comb spectrum and reconstructed part marked in blue. (g) Power of a heterodyne beatnote signal between a continuous-wave laser and multiple optical lines in a single resonance of a Si$_3$N$_4$ microresonator. Peaks in the heterodyne signal (red) indicate the presence of a comb line. The total scan range of the laser is approximately 2~GHz and, for comparison, a resonance of 200~MHz width is indicated schematically (gray)~\cite{herr2012UniversalFormationDynamics}.}\label{fig:ch5:KerrCombsFormation}
\end{figure*}

In a frequency domain description, Turing rolls can be interpreted as a cascaded FWM process where the frequency spacing between the pump and first MI sidebands replicates itself exactly between all adjacent sidebands, as illustrated in Fig.~\ref{fig:ch5:KerrCombsFormation}a. 
Usually the spacing $\Delta$ between the sidebands in a Turing state usually corresponds to a multiple of the resonator's FSR. From Eq.~\ref{eq:first_MI_sideband} it follows that the spacing $\Delta$ between Turing roll lines will correspond to one FSR, only for low pump power and resonators with large FSR, high Q factor, or exceptionally strong dispersion, for which $D_2/\kappa=d_2/2\gtrsim1$\cite{herr2012UniversalFormationDynamics}.

\subsubsection{Chaotic states}
\label{sec:basics_kcomb:chaotic_states}
If in a Turing roll state, the detuning is reduced or the pump power is increased the system may enter into a chaotic state, which may appear on an optical spectrum analyzer as a dense frequency comb; see Fig.~\ref{fig:exp_comb_states} for an example.
Here the parametric gain exceeds threshold also for so far unpopulated resonator modes between the Turing roll sidebands. This results in new \textit{secondary} frequencies and introduces a new frequency spacing $\delta$ that is not generally an integer fraction of $\Delta$. Often this process manifests itself as the formation of $\delta$-spaced lines around the \textit{primary} Turing roll sidebands as illustrated in Fig.~\ref{fig:ch5:KerrCombsFormation}b. 
At this point, the spectrum does not represent a consistent frequency comb, but rather a set of multiple $\delta$-spaced \textit{subcombs} with frequencies 
\begin{equation}
    \omega_\mu^{(i)} = \omega_\mathrm{p} + \xi_i + \mu \delta
\end{equation}
where $i,j$ index the subcombs whose offset $\xi_{i,j}$ are not generally consistent, i.e., $\xi_j\neq\xi_i + n\delta, j\neq i, n\in\mathbb{N}$. As illustrated in Fig.~\ref{fig:ch5:KerrCombsFormation}d (left), a narrow repetition rate signal at frequency $\delta$ may be observed despite the inconsistency of the comb. As the detuning is reduced further (the pump power increased further), the subcombs grow and start overlapping, as observed via high resolution spectroscopy of the optical spectrum (Fig.~\ref{fig:ch5:KerrCombsFormation}e). All frequency components can interact and nonlinearly cascading into a large set of frequencies, including multiple frequency components per resonator mode as shown in Fig.~\ref{fig:ch5:KerrCombsFormation}f, and ultimately a continuum of frequencies within the linewidth of individual resonances. The spectral dynamics in the optical domain also manifest itself in the radio-/microwave domain at frequencies close (within several $\kappa$) to $\delta$ where first discrete frequency components and then a broad continuum may be observable. Importantly, as similar dynamics will be observed close to DC where intensity noise results as a direct consequence of the broadband optical/radio- and microwave signals; this may enable noise diagnostics with photo-detectors whose bandwidth does not allow for direct repetition rate detection (see Fig.~\ref{fig:ch5:KerrCombsFormation}d for illustration). A typical example of such dynamics is shown in Fig.~\ref{fig:ch5:KerrCombsFormation}g. 

While in some special cases, through careful adjustment of pump detuning and power, the different sub-combs may be aligned to result in a consistent frequency comb~\cite{herr2012UniversalFormationDynamics, delhaye2014SelfInjectionLockingPhaseLocked}, in general, the incommensurate comb lines will oscillate in a chaotic manner. The transition to chaos has been experimentally observed in both spectral and temporal domains~\cite{delhaye2011OctaveSpanningTunable, savchenkov2008TunableOpticalFrequency, ferdous2011SpectralLinebylinePulse, okawachi2011OctavespanningFrequencyComb} and further investigated in~\cite{herr2012UniversalFormationDynamics,coulibaly2019TurbulenceInducedRogueWaves}.

The presence of chaos in optical resonators has been predicted in early works~\cite{ikeda1979MultiplevaluedStationaryStatea,ikeda1980OpticalTurbulenceChaotic} and later studied in the context of RF-driven plasma governed by the driven dissipative NLSE~\cite{nozaki1986LowdimensionalChaosDriven} and extensively investigated after~\cite{coillet2014RoutesSpatiotemporalChaos,panajotov2017SpatiotemporalChaosTwodimensional,perinet2017EckhausInstabilityLugiatoLefever,coulibaly2019TurbulenceInducedRogueWaves,liu2017CharacterizationSpatiotemporalChaos,barashenkov1996ExistenceStabilityChart,leo2013DynamicsOnedimensionalKerr}.  Chaos can be observed both in mono- and bistable states of the microresonator and they can be divided into four categories: (i) chaotic MI in the mono-or bistable case (ii) spatiotemporal chaos, (iii) temporal chaos, and (iv) transient chaos. While the first two demonstrate similarity, temporal chaos remains localized in the cavity, and transient chaos dynamically decays into the stable CW state~\cite{leo2013DynamicsOnedimensionalKerr,anderson2016ObservationsSpatiotemporalInstabilities}.

The transition to the chaotic stage of MI can be described as follows: crossing the stability boundary of the Turing rolls, for sufficiently high pump power (f$>2$ for the parameters taken in Fig.~\ref{fig:ch5:cn_waves}), the dynamical system enters the chaotic state via the Hopf bifurcation which leads to temporal oscillations and collisions of those localized structures that were constituting the Turing roll. A distinct signature of the chaotic states is a positive-valued Lyapunov exponents~\cite{coulibaly2019TurbulenceInducedRogueWaves,liu2017CharacterizationSpatiotemporalChaos}, implying that two close trajectories in the phase space demonstrate exponential divergence~\cite{pikovsky2016LyapunovExponentsTool}.
Another remarkable feature of the chaotic states is the presence of rogue waves~\cite{coillet2014OpticalRogueWaves,coulibaly2019TurbulenceInducedRogueWaves} - extremely high amplitude events that appear more often than it is predicted by the linear theory~\cite{kharif2009RogueWavesOcean,onorato2016RogueShockWaves}.
Noteworthy, localized chaotic patterns called chimera states can occur when the pump laser or the resonator is modulated~\cite{nielsen2018InvitedArticleEmission, tusnin2020NonlinearStatesDynamics}.

\subsubsection{Dissipative Kerr solitons (DKS)}
\label{sec:basics_kcomb:DKS}

Temporal dissipative Kerr solitons, DKSs, are coherent pulse-like waveforms generated in driven Kerr-nonlinear microresonators, giving rise to broadband, low-noise frequency combs (see Fig.~\ref{fig:exp_comb_states}). They are an example of self-organization, arising from the double balance between anomalous group-velocity dispersion and nonlinearity, as well as cavity loss and parametric gain~\cite{akhmediev2005DissipativeSolitons,kippenberg2018DissipativeKerrSolitons,lugiato2018LugiatoLefeverEquation}. In contrast to conservative solitons, the additional balance between loss and gain uniquely defines its properties for a given driving power $f$ and detuning $\zeta_0$.  Both, temporal and spectral intensity profiles follow a square hyperbolic secant ($\sech^2$).

\begin{figure*}
\includegraphics[width=0.9\textwidth]{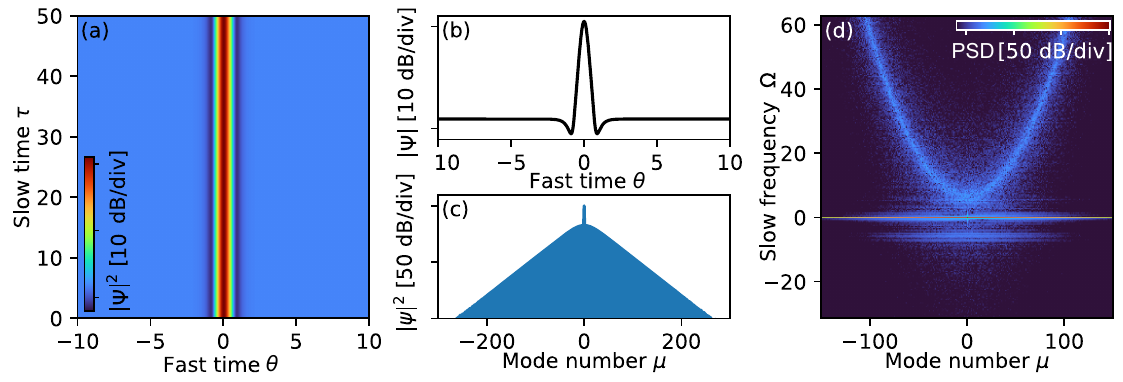}
\caption{\textbf{Dissipative Kerr soliton (DKS) in a microresonator.} (a) Spatiotemporal diagram demonstrating the evolution of the DKS. (b) A cross-section of the spatiotemporal diagram showing the fast time profile of the DKS. (c) Corresponding spectrum demonstrating the hyperbolic secant profile. (d) Nonlinear dispersion relation. Noise-induced dispersive waves highlight the dispersion parabola. DKS is represented by the line which indicates that the dispersion is exactly compensated by the cubic nonlinearity. } \label{fig:ch5:NDR_soliton} 
\end{figure*}

Numerical solution of the LLE corresponding to the DKS is presented in Fig.~\ref{fig:ch5:NDR_soliton}, panel (d) shows the so called nonlinear dispersion relation (NDR), indicating the spectral power in each mode $\mu$ measured offset from a $D_1$-spaced grid. The NDR reveals the equidistant $D_1$-spaced DKS comb lines as a pronounced, sharply defined horizontal line.
The stability of the DKS solution is identified by examining the eigenvalue spectrum of the corresponding Jacobian matrix. See Sec.~\ref{sec:basics_kcomb:numerical_methods} for details of numerical simulations and NDR technique.

While conservative solitons exhibit elastic collisions~\cite{novikov1984TheorySolitonsInverse}, this is not the case for DKS, and closely spaced DKSs with similar group velocities can decay upon collision~\cite{cole2017SolitonCrystalsKerr,wabnitz1993SuppressionInteractionsPhaselocked}. Another key difference is the presence of a stable background, which confines DKS existence mainly to the bistable region of the cavity.\footnote{In conservative systems, a similar background leads to breathing dynamics corresponding to the Kuznetsov–Ma breather~\cite{kuznetsov1977solitons,kibler2012ObservationKuznetsovMaSoliton,ma1979perturbed}.} 
These distinctions underlie the use of the terms 
temporal ``dissipative Kerr soliton'' or ``cavity soliton'' as adopted in the fiber-optics community.

\begin{figure}[h]
\includegraphics[width=0.99\columnwidth]{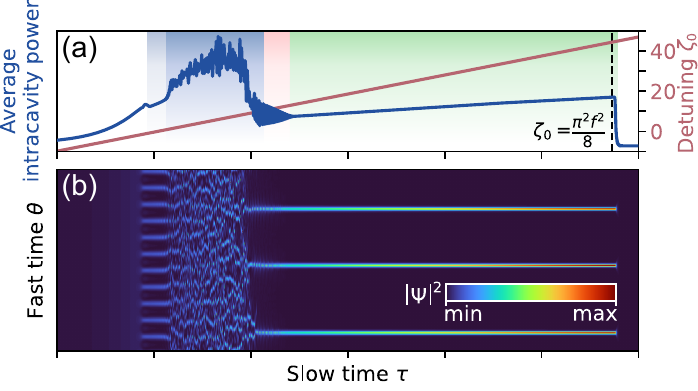}
\caption{\textbf{Tuning into the DKS state.} (a) Average intracavity power as a function of slow time is shown in blue. The laser detuning sweep is depicted in red. The black dashed line corresponds to the theoretical soliton existence range given by Eq.~\ref{eq:ch5:sol_ex_range}. Shaded areas depict distinct states outlined in Fig.~\ref{fig:ch5:phase_diagram}a, with colors preserved. (b) Spatiotemporal evolution of intracavity field represented by $|\Psi(\tau,\theta)|^2$. Normalized pump power $f=6$. }
\label{fig:ch5:Tuning_LLE}
\end{figure}

In contrast to fiber cavities, in high-finesse microresonators, DKS can form spontaneously from MI via a laser tuning procedure. For instance, changing the frequency of a pump laser with constant power from the blue-detuned side of the resonance to the red-detuned one, will first result in temporally modulated waveform, which then evolves into a DKS once the laser detuning reaches the
soliton existence range (Fig.~\ref{fig:ch5:phase_diagram}) on the red side of the resonance. The first generation of a DKS in a microresonator was achieved via this ``laser tuning method''. A numerical simulation of DKS formation as discussed above is presented in Fig.~\ref{fig:ch5:Tuning_LLE}, showing intracavity power and temporal intensity.

\begin{figure*}
\centering
\includegraphics[width=0.95\textwidth]{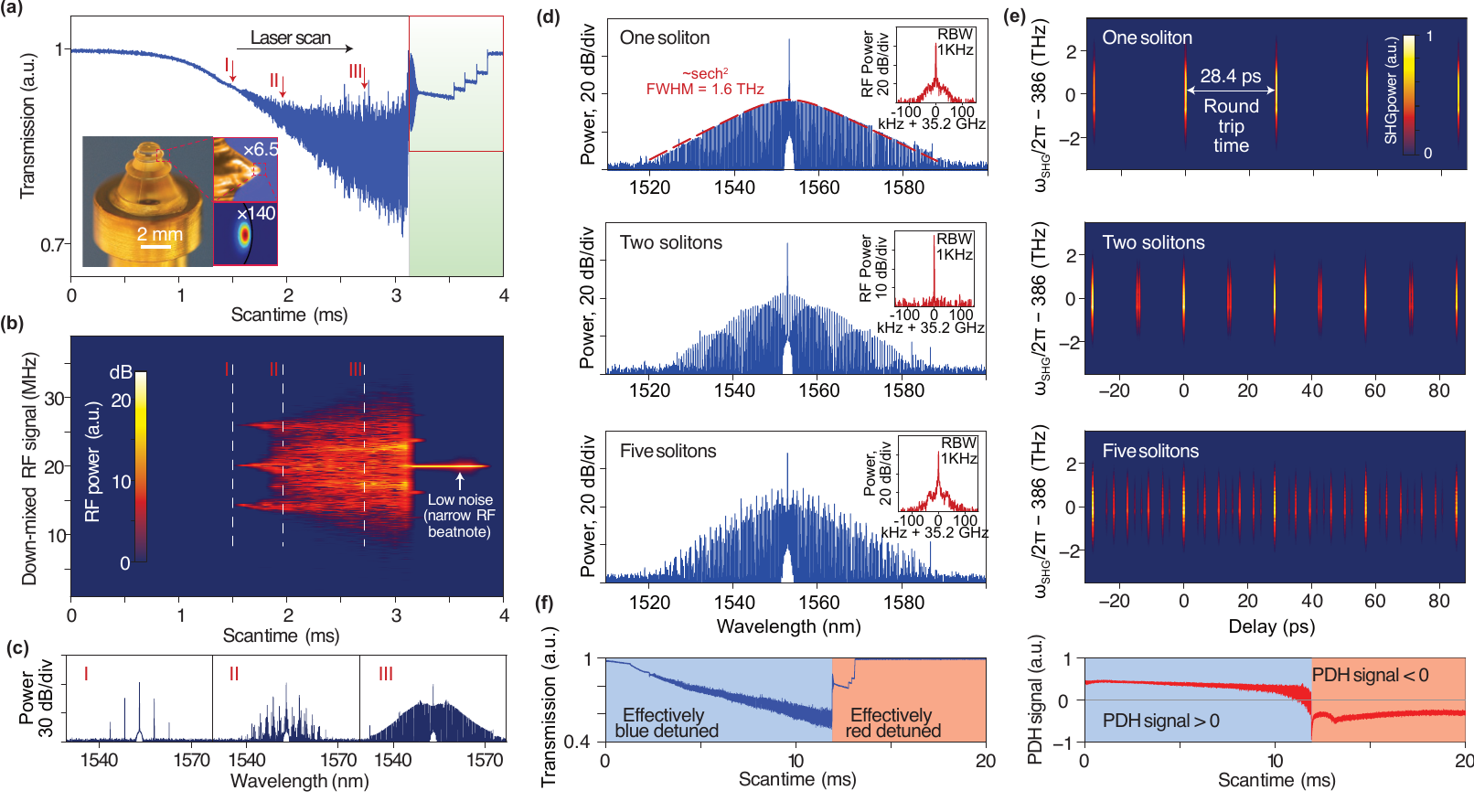}
\caption{\textbf{Experimental generation of DKS via laser tuning.} (a) Resonator transmission during a laser scan showing characteristic step-like transition features (``soliton steps''). The green-shaded region marks the DKS regime and shows discrete plateaus which are associated with different numbers of DKS. Inset: MgF$_2$ crystalline microresonator used in this experiment. (b) Repetition rate beatnote during the laser scan. (c) Characteristic spectra observed prior to entering the DKS regime.
(d) Examples of DKS spectra and (e) corresponding frequency resolved optical gating (FROG) traces. (f) Transmission and Pound-Drever-Hall (PDH) error signal. Effective blue and red detunings are shaded blue and red, respectively.  From~\cite{herr2014TemporalSolitonsOptical}. \label{fig:ch5:exp_steps}} 
\end{figure*}

In the tuning process, the system undergoes discrete phase transition, first from a noisy, chaotic state into a (usually) multi-DKS state, and then stepwise towards lower number of DKS, until eventually the laser may exit the DKS existence range. The discrete phase-transitions appear as discrete steps in the intracavity power, and experimentally, are observed characteristic steps in the coupling waveguide transmission signal as shown in Fig.~\ref{fig:ch5:exp_steps}(a)~\cite{herr2014TemporalSolitonsOptical}. 
Although the DKS existence range is sharply defined (see Fig.~\ref{fig:ch5:phase_diagram}), the gradual non-simultaneous decay can be explained by slight differences in the local environment of each DKS pulse \cite{luo2015SpontaneousCreationAnnihilation}, which may occur in real resonators for instance from input noise, Raman effects, and deviations from a purely quadratic dispersion (implying a modulation of the background field as well as coupling between DKS).
The step-like transition features are a hallmark of the DKS formation and interestingly resembles observations of soliton modes in nonlinear microwave resonators~\cite{gasch1984MultistabilitySolitonModes}.
The formation of DKS in the microresonator is also evidenced by recording the repetition rate signal during the laser scan (Fig.~\ref{fig:ch5:exp_steps}(b)), which upon transition to the soliton regime collapses into a narrow, low-noise beatnote. This transition to low noise manifests itself also in the drop of low-frequency noise close to DC (as can be seen from the Transmission trace in Fig.~\ref{fig:ch5:exp_steps}(a)), which provides complementary diagnostics in those cases where the repetition rate beatnote cannot be measured. 
Fig.~\ref{fig:ch5:exp_steps}(c) shows the optical spectra observed prior to DKS formation, and Fig.~\ref{fig:ch5:exp_steps}(d) shows examples of DKS spectra with different numbers of DKS, and their narrow repetition rate signals. 
The single DKS power spectrum shows a clear $\sech^2$-envelope, while states with multiple solitons (indexed by $j$) show characteristic interference 
pattern $\psi_\mu^\mathrm{multi} = \psi_\mu^\mathrm{single} \sum_j e^{i\mu \phi_j}$ determined by the resonator coordinates $\phi_j$ of the DKS, where $\psi_\mu^\mathrm{multi/single}$ 
denotes the single/multi DKS spectral field amplitudes. Frequency resolved optical gating (FROG)~\cite{trebino2000FrequencyresolvedOpticalGating} 
confirms the short femtosecond pulse nature (Fig.~\ref{fig:ch5:exp_steps}(e)) ~\cite{herr2014TemporalSolitonsOptical, herr2014ModeSpectrumTemporal}.

A defining feature of DKS is that they appear for an effectively red-detuned pump laser, i.e. a detuning $\zeta_0-\sum_\mu |\psi_\mu|^2>0$, as can be  experimentally confirmed by recording a Pound-Drever-Hall error signal (Fig.~\ref{fig:ch5:exp_steps}(f)). Prior to the observation of DKS, the effectively red-detuned regime had been considered non-favorable for microcombs due to its instability, or alternatively low-intracavity power. Moreover, the thermal dynamics of microresonators had long masked the DKS regime in experiments, as discussed in more detail in Sec.~\ref{sec:exp_DKS:Res_thermal}.

The exact analytical solution of the 1D NLSE in the form of a bright soliton is known in the conservative case~\cite{shabat1972ExactTheoryTwodimensional}.
An approximate DKS solution of the normalized LLE (Eq.~\ref{eq:ch5:LLE_norm}) can be expressed as a superposition of the conservative NLSE solitons and a CW background corresponding to the lower stable branch of the resonance. This solution can be analyzed with several approaches including the variational method~\cite{anderson1983VariationalApproachNonlinear}, the method of moments~\cite{maimistov1993evolution}, and the perturbation theory based on the inverse scattering transform~\cite{karpman1977PerturbationTheorySolitons}. The variational approach (further developed in~\cite{ chavezcerda1998VariationalApproachNonlinear, mertens2010NonlinearSchrodingerEquation,hasegawa2000SolitonbasedOpticalCommunications,malomed2002VariationalMethodsNonlinear})
is often used to obtain insights into the microresonator dynamics~\cite{grelu2016NonlinearOpticalCavity,yi2016TheoryMeasurementSoliton,tusnin2020NonlinearStatesDynamics,matsko2013TimingJitterMode}. Its application to the reduction of the LLE to a set of ODEs governing the DKS ansatz parameters will be described in the following.
 
The Lagrangian density that represents the conservative part of the LLE (the NLSE with a detuning term) is
 \begin{equation}
{\cal L}=\frac{i}{2}\left(\Psi^{*}\frac{\partial\Psi}{\partial\tau}-\Psi\frac{\partial\Psi^{*}}{\partial\tau}\right)-\frac{1}{2}\left|\frac{\partial\Psi}{\partial\theta}\right|^{2}+\frac{1}{2}|\Psi|^{4}-\zeta_{0}|\Psi|^{2},
\label{eq:ch5:lagr_dens}
\end{equation}
which is related to the Lagrangian as $L = \int{\cal L}d\theta$. The requirement of the variational derivative of the action over $\Psi^{*}$ to be zero leads to
\begin{equation}
\frac{\delta{\cal L}}{\delta\Psi^{*}}\equiv\frac{\partial{\cal L}}{\partial\Psi^{*}}-\frac{\partial}{\partial\tau}\frac{\partial{\cal L}}{\partial\Psi_{\tau}^{*}}-\frac{\partial}{\partial\theta}\frac{\partial{\cal L}}{\partial\Psi_{\theta}^{*}}=0,
\end{equation}

In order to account for the full LLE (Eq. \ref{eq:ch5:LLE_norm}), we
write the dissipative function as:
\begin{eqnarray}
\frac{\delta{\cal L}}{\delta\Psi^{*}} & = & {\cal R},\nonumber \\
{\cal R} & =& -i\Psi+if.
\label{eq:ch5:diss_func_var}
\end{eqnarray}
In this way, we obtain a reduced set of equations to be solved: 
\begin{equation}
\frac{\partial L}{\partial r_{i}}-\frac{d}{d\tau}\frac{\partial L}{\partial\dot{r}_{i}}  =  \int\left({\cal R}\frac{\partial\Psi^{*}}{\partial r_{i}}+{\cal R}^{*}\frac{\partial\Psi}{\partial r_{i}}\right)d\theta,
\label{eq:ch5:lagr_equation}
\end{equation}
where $r_{i}$ are different possibly time-dependent collective coordinates.
Using the ansatz of a stationary ($\frac{\partial\Psi}{\partial\tau}=0$)
soliton $\Psi = Be^{i\varphi_{0}}{\rm sech}(B\theta)$ which implies $r_{1}=B$ and $r_{2}=\varphi_{0}$. Substituting the soliton ansatz to Eq.~\ref{eq:ch5:lagr_dens} and integrating, we obtain: 
\begin{eqnarray}
L=-2B\frac{\partial\varphi_{0}}{\partial\tau}+\frac{1}{3}B^{3}-2B\zeta_0.
\end{eqnarray}

Finally, we obtain a finite-dimensional system of equations governing the evolution of the DKS parameters:
\begin{eqnarray}
\frac{dB}{d\tau}&=&-2B+\pi f\cos\varphi_{0} \nonumber,\\
\frac{d\varphi_{0}}{d\tau}&=&\frac{1}{2}B^{2}-\zeta_{0}.
\label{eq:ch5:FDDS}
\end{eqnarray}
This leads to the stationary parameters of the soliton attractor:
\begin{eqnarray}
B&=&\sqrt{2\zeta_{0}}, \label{eq:ch5:lagr_conclusion1}\\
\cos\varphi_{0}&=&\frac{2B}{\pi f}=\frac{\sqrt{8\zeta_{0}}}{\pi f}.\label{eq:ch5:lagr_conclusion2}
\end{eqnarray}
Equations~\ref{eq:ch5:lagr_conclusion1},\ref{eq:ch5:lagr_conclusion2} have two immediate consequences. The former expression, suggests that the amplitude of the soliton as well as its bandwidth increase with laser detuning and reaches the maximum at the end of the \textit{DKS existence range}.
The latter implies that the soliton existence range is limited to~\cite{wabnitz1993SuppressionInteractionsPhaselocked, barashenkov1996ExistenceStabilityChart}
\begin{equation}\label{eq:ch5:sol_ex_range}
\zeta_{0}^{\mathrm{max}}=\frac{\pi{}^{2}f{}^{2}}{8}.
\end{equation}  
Within the soliton existence range\footnote{A lower limit for the detuning $f^{2}<\frac{2}{27}\zeta_{0}^{\mathrm{min}}(\zeta_{0}^{\mathrm{min}\,2}+9)$ was derived in an asymptotic analysis~\cite{barashenkov1996ExistenceStabilityChart}.}, the lower branch of the CW solution is stable and can be found with the aid of Eq.~\ref{eq:ch4:lower_br_approx}. Thus, an approximate single soliton solution that accounts for a flat CW background is then given by 
\begin{eqnarray}
 & \Psi=\Psi_{0}+\Psi_{1}\simeq\Psi_{0}+Be^{i\varphi_{0}}{\rm sech}(B\theta).\label{eq:ch5:single_soliton}
\end{eqnarray}

From Eq.~\ref{eq:ch5:single_soliton}, Eq.~\ref{eq:ch5:sout}, and Eq.~\ref{eq:ch4:lower_br_approx} one can estimate the efficiency of the soliton microcomb generation, i.e. the pump photon \textit{conversion efficiency} to the comb lines given by $\eta=P_{\text {combs }}^{\text {out }} / P_{\mathrm{pump}}^{\text {in }}$. Depending on the pump power and detuning it has a value in the interval 1-5\%~\cite{jang2021conversion,bao2014NonlinearConversionEfficiency}. Outside the framework of the LLE, there are several ways to improve the conversion efficiency in microcombs~\cite{yang2024EfficientMicroresonatorFrequency}, such as operating in the normal dispersion regime~\cite{xue2017MicroresonatorKerrFrequency}, displacement of the pumped mode in coupled systems~\cite{helgason2023SurpassingNonlinearConversion}, or using synchronous pulse pumping scheme~\cite{obrzud2017TemporalSolitonsMicroresonators,anderson2021PhotonicChipbasedResonant,li2022EfficiencyPulsePumped} and these will be discussed in Sec.~\ref{sec:aspects_DKS}.

The spectral profile of the DKS is given by:
\begin{equation}
\begin{split}
    \psi_\mu &= \mathcal{FT}\left\{\sqrt{2 \zeta_0} \operatorname{sech}\left(\sqrt{\frac{\zeta_0 \kappa}{ D_2}} \phi\right)\right\} \\
&=\sqrt{D_2 / 2 \kappa} \operatorname{sech}\left(\frac{\pi \mu}{2} \sqrt{\frac{D_2}{\zeta_0 \kappa}}\right).
\end{split}\label{eq:ch5:soliton_FT}
\end{equation}

\begin{figure}
\includegraphics[width=\columnwidth]{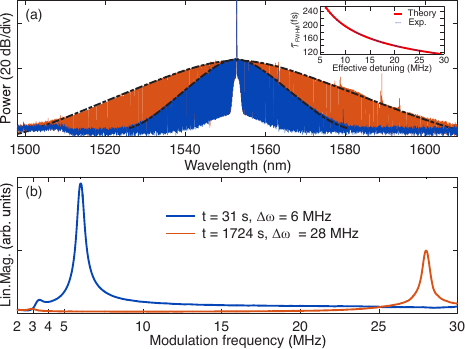}
\caption{\textbf{Detuning dependent properties of the DKS.} (a) Spectra of a DKS recorded at different detuning. The inset shows the dependence of the DKS duration as a function of the effective laser detuning from the (Kerr shifted) cavity resonance. (b) Vector network analyzer (phase modulation response) data showing cavity and soliton resonances at different effective detuning.  From~\cite{lucas2017DetuningdependentPropertiesDispersioninduced}.\label{fig:ch5:detuningdependence}} 
\end{figure}

It is often convenient to express the Eq.~\ref{eq:ch5:soliton_FT} in terms of real optical power:
\begin{equation}
P(\mu) \approx \frac{\kappa_{\mathrm{ex}} D_2 \hbar \omega_0}{4 g_\mathrm{K}} \operatorname{sech}^2\left(\frac{\pi \mu}{2} \sqrt{\frac{D_2}{2 (\omega_{0}-\omega_\mathrm{p})}}\right).
\label{eq:ch5:power_sp_dks}
\end{equation}
Figure~\ref{fig:ch5:detuningdependence}a shows experimentally recorded DKS optical spectra at different values of detuning (other parameters remain unchanged). The soliton spectral width significantly increases at higher detuning values in accordance with Eq.~\ref{eq:ch5:power_sp_dks}. The effective detuning value is measured with vector network analyzer~\cite{guo2017UniversalDynamicsDeterministica} by detecting the cavity and soliton resonances as shown Fig.~\ref{fig:ch5:detuningdependence}b.

Using the definition of the optical frequency $\omega=\omega_\mathrm{p}+\mu D_1$ and further transforming the azimuthal angle variable to actual intracavity time $t=\frac{\phi}{2 \pi} T_{\mathrm{R}}=\phi / D_1$, we obtain scaling laws for spectral envelope and pulse duration~\cite{herr2014TemporalSolitonsOptical,coen2013UniversalScalingLaws}:
\begin{equation}
\begin{split}
\psi\left(\omega-\omega_\mathrm{p}\right)=&\sqrt{D_2 / 2 \kappa} \operatorname{sech}\left[\left(\omega-\omega_\mathrm{p}\right) / \Delta \omega\right], \\
\text{with}\, \Delta \omega&=\frac{2 D_1}{\pi} \sqrt{\frac{\zeta_0 \kappa}{D_2}}
\end{split}\label{eq:ch5:soliton_spectrum}
\end{equation}
and
\begin{equation}
\Psi(t)=\sqrt{2 \zeta_0} \operatorname{sech}(t / \Delta t), \,\text {with } \Delta t=\frac{1}{D_1} \sqrt{\frac{D_2}{\zeta_0 \kappa}}.
\end{equation}

Using the expression for the soliton existence range  $\zeta_0^{\max }$ (Eq.~\ref{eq:ch5:sol_ex_range}), we can estimate the minimal achievable soliton duration and corresponding maximal spectral width in non-normalized units using the normalization in Eq.~\ref{eq:ch5:normalization} and the definition of the Kerr shift per photon Eq.~\ref{eq:ch4:g0}:
\begin{equation}
\Delta t_{\min }=\frac{1}{\pi D_1} \sqrt{\frac{\kappa D_2 n_0^2 V_{\mathrm{eff}}}{\eta P_{\mathrm{in}} \omega_0 c n_2}}\, \text{and} \, \Delta \omega_{\max} = \frac{2}{\pi \Delta t_{\min}}
\label{eq:ch5:sol_duration_and_width}
\end{equation}

\begin{figure}
\includegraphics[width=1.0\columnwidth]{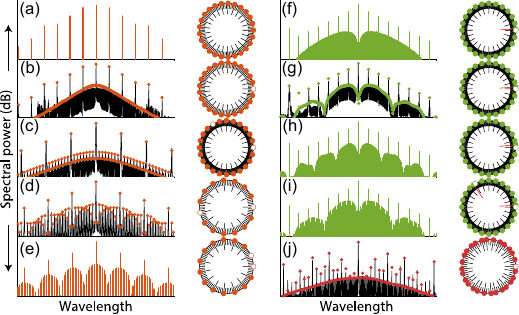}
\caption{\textbf{Soliton crystals in microresonators.}  Measured optical spectra are shown in black and simulations are in colour. Schematic depictions of the soliton distribution in the resonator are shown to the right of each spectrum. Major ticks in the schematic diagrams indicate the location or expected location of a soliton. Minor ticks indicate peaks of the extended background wave due to the mode crossing. (a) A perfect soliton crystal, consisting of 25 uniformly distributed solitons. (b-e) Soliton crystals exhibiting vacancies. (f-i) Soliton crystals exhibiting Frenkel defects. Shifted solitons still lie at peaks of the extended background wave. (j) A disordered crystal. From~\cite{cole2017SolitonCrystalsKerr}.
 \label{fig:ch7:SolitonCrystals}} 
\end{figure}

Examples of perfect soliton crystals as well as soliton crystals with defects are shown in Fig.~\ref{fig:ch7:SolitonCrystals}. Multi-soliton states and soliton crystals are usually stabilized via external pump modulations or AMX (see Sec.~\ref{sec:coupled_mode-families_and_resonators} for further details).

\subsubsection{Breathing solitons  }
\label{sec:basics_kcomb:Breathing}

In the conventional LLE, DKS can exhibit periodic oscillations~\cite{leo2013DynamicsOnedimensionalKerr, matsko2012ExcitationBreatherSolitons,parra-rivas2014DynamicsLocalizedPatterned} at the beginning of its existence range. This is observed in a variety of microresonator platforms~\cite{lucas2017BreathingDissipativeSolitonsa,afridi2022BreatherSolitonsAlN, yu2017BreatherSolitonDynamics} and is called breathing by analogy with a similar dynamics in single-pass optical fibers~\cite{dudley2014InstabilitiesBreathersRogue,copie2020PhysicsOnedimensionalNonlinear}. The 1D LLE breather exhibits similarities to the Kuznetsov-Ma solution of the 1D NLSE~\cite{kuznetsov1977solitons,kibler2012ObservationKuznetsovMaSoliton,ma1979perturbed}. The LLE breather has also been considered~\cite{bao2016ObservationFermiPastaUlamRecurrence} in the context of the Fermi-Pasta-Ulam-Tsingou recurrence~\cite{dauxois2008FermiPastaUlam, fermi1955StudiesNonlinearProblems}, the problem that inspired pioneering studies in soliton Physics~\cite{ablowitz1981SolitonsInverseScattering,zabusky1965InteractionSolitonsCollisionless}. Kuznetsov-Ma breather can be used as an ansatz for the Lagrangian perturbation approach~\cite{lucas2017BreathingDissipativeSolitonsa} described in the previous section. However, there are apparent differences that restrict this comparison, such as the periodic radiation of dispersive waves (DW) by the dissipative breathers.

\begin{figure*}
\includegraphics[width=0.9\textwidth]{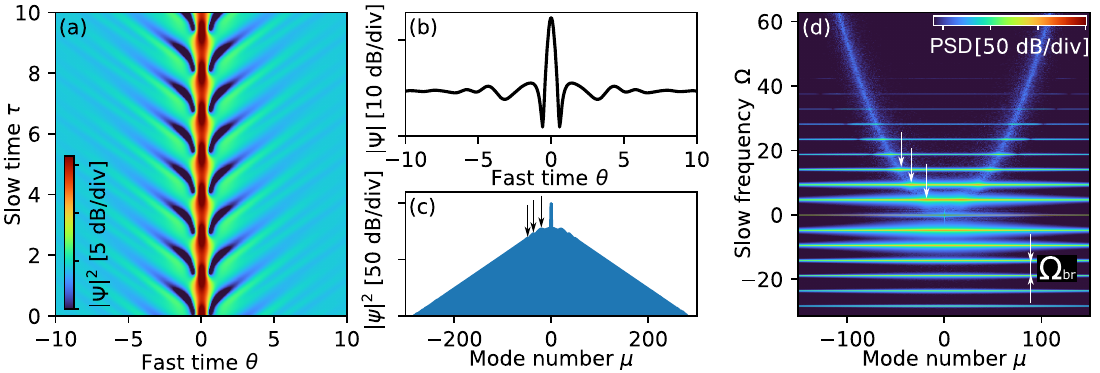}
\caption{\textbf{Driven-dissipative breather in a microresonator.}(a) Spatiotemporal diagram demonstrating the evolution of the breather. Diverging lines correspond to the dispersive waves radiating during propagation. (b) A cross-section of the spatiotemporal diagram showing the fast time profile of the breather. (c) Corresponding spectrum. (d) Nonlinear dispersion relation. The spacing between the horizontal lines corresponds to the breathing frequency $\Omega_\mathrm{br}$. The points where the ladder of lines crosses the dispersive parabola (depicted by white arrows) correspond to the enhancement of the comb lines shown with a black arrow in (c).  $\zeta_0 = 4.4$ $f$ = 3. \label{fig:ch5:NDR_breather} } 
\end{figure*}

The transition to the breathing state from the point of view of nonlinear dynamics corresponds to the Hopf bifurcation~\cite{parra-rivas2014DynamicsLocalizedPatterned,godey2014StabilityAnalysisSpatiotemporal,qi2019DissipativeCnoidalWaves}. 
By increasing the pump power, we cross the stability boundary of the DKS solution. This leads to the oscillatory dynamics. 
During the transition, two eigenvalues of the Jacobian matrix \{$\lambda_i$\} cross Re$(\lambda) = 0$ line, while the rest of the eigenvalue spectrum remains similar to the one, demonstrated in Fig.~\ref{fig:ch5:cn_waves}(d). 
Im($\lambda$) in this case, corresponds to the breathing frequency $\Omega_\mathrm{br}$~\cite{qi2019DissipativeCnoidalWaves}.

Periodic oscillations in slow time and radiation of the DW are shown in the spatiotemporal diagram - Fig.~\ref{fig:ch5:NDR_breather} (a). During every oscillation period, the breather radiates DW to the resonator which appear as diverging waves that gradually decay during the propagation. A cross-section shown in Fig.~\ref{fig:ch5:NDR_breather}(b) depicts the profile of the breather in the slow time, Fig.~\ref{fig:ch5:NDR_breather}(c) shows the corresponding power spectral density which exhibits an enhancement of certain comb lines depicted by black arrows. The NDR of the breather depicted in Fig.~\ref{fig:ch5:NDR_breather}(d) reveals the underlying mechanism of the sideband enhancement. The periodic nature of the breather has a distinct trace in the NDR representation. The periodic dynamics of a coherent structure results in a ladder of horizontal lines - in contrast to a single line corresponding to the stationary DKS. These lines correspond to the modulation sideband and the spacing between the lines corresponds to the breathing frequency $\Omega_\mathrm{br}$ which changes linearly with the pump laser detuning~\cite{lucas2017BreathingDissipativeSolitonsa}  
Spectral areas, where the ladder of lines crosses the dispersive parabola, correspond to the positions of the sidebands on the breather spectrum and are depicted with white arrows.

Breathers can be excited in the microresonators for instance by backward tuning. Numerical simulations include the thermal nonlinearity described in Sec.~\ref{sec:exp_DKS:Res_thermal} to capture the effects present in the experiments. As a result,
we observe that breathers are excited at the beginning of the soliton existence range\cite{leo2013DynamicsOnedimensionalKerr,lucas2017BreathingDissipativeSolitonsa} (which is different for different multisoliton solutions due to the thermal effects) that can be accessed by changing the direction of the laser frequency tuning after accessing the stable DKS state.

\subsection{Numerical techniques}
\label{sec:basics_kcomb:numerical_methods}
Numerical methods play an important role in studying the evolution of complex nonlinear waves~\cite{zabusky1965InteractionSolitonsCollisionless,fermi1955StudiesNonlinearProblems}.  In non-integrable nonlinear systems, such as nonlinear microresonators, where analytical methods do not provide sufficient insights, they are primary research tools~\cite{li2016ModelingFrequencyComb}. Below we introduce several important techniques:

\subsubsection{Time integration of nonlinear dynamics}
The dynamic evolution of intracavity states, including transient phenomena, can be simulated by integrating the LLE in time and frequency representations (Eqs.~\ref{eq:ch5:LLE_norm} or~\ref{eq:ch5:cme_normalized})~\cite{lamont2013RouteStabilizedUltrabroadband}. The state of the cavity field is described by a state vector $\boldsymbol x$ in frequency or time domain, which can be identified with either $\boldsymbol x=\{\psi_\mu\}$ with $\mu=-N/2, .. , N/2-1$, or $\boldsymbol x = \{\Psi(\phi_j)\}$  with $\phi_j = -\pi + j\frac{2\pi}{N}$ and $j=0, .. , N-1$. For computational efficiency we have assumed that the number of spectral components or discretizations points on the fast time axis is a power of 2. The state vectors can be propagated in time based on the derivatives 
\begin{equation}
    \frac{\partial}{\partial\tau}\boldsymbol x = \boldsymbol F(\boldsymbol x)
\end{equation}
where the vector functions $\boldsymbol F$ represents the RHS of the CMEs Eq.~\ref{eq:ch5:cme_normalized} or the LLE Eq.~\ref{eq:ch5:LLE_norm}. Any parameter defining $\boldsymbol F$, in particular detuning $\zeta_0$ or pump field $f$, can be made time dependent, but this is omitted here for clarity.
Numerically, higher-order schemes such as Runge-Kutta and adaptive step size control are often to used propagate the $\boldsymbol x$ along a discretized time axis $\tau_k$ with ($k=0,1,2, ..$);  these methods are conveniently available through standard numeric libraries.  
Both time and frequency versions are equivalent, an efficient computational implementation usually makes use of both~\cite{hansson2014NumericalSimulationKerr}, as it is well established in the field of nonlinear waveguide optics~\cite{agrawal2013NonlinearFiberOptics}. This approach leverages the fact that the spectral and temporal field amplitudes are efficiently connected numerically via the \textit{Fast Fourier Transform (FFT)}. Effects of dispersion, including complex or spectrally local features in the dispersion (such as AMX), can be modeled in the frequency domain, while nonlinear effects are most efficiently computed in the time domain. To avoid numeric artifacts, it is important to choose a sufficiently large $N$. Unwanted numeric aliasing from the nonlinear interaction and an unphysical back-folding of frequency components due to the cyclic nature of the FFT, can be avoided by temporarily increasing $N$ (here by a factor of $2^2$) in the computation of the nonlinear term~\cite{orszag1971EliminationAliasingFiniteDifference,derevyanko2008RuleDealiasingSplitStep,voumard2023SimulatingSupercontinuaMixed}.
Beyond standard numerical challenges in numeric integration of nonlinear optical wave interaction~\cite{dudley2006SupercontinuumGenerationPhotonic}, simulating microcombs requires careful implementation of noise sources, such as quantum shot noise (see Ref.~\cite{paschotta2004NoiseModelockedLasers} and Sec.~\ref{sec:ch6:noise:quantum}). Realistic noise levels are crucial to trigger the dynamics during adiabatic tuning through the MI regime.
There are several open-source examples of numerical solvers of the microresonator dynamics available that can readily be adapted to specific use cases~\cite{melchert2021PyGLLEPythonToolkit,moille2019PyLLEFastUser, wildi2023SidebandInjectionLocking, PyCOReFastTool}.

\subsubsection{Finding stable stationary solutions}
\label{sec:num_tool_stable_stationary}
Although analytical approximations exist in some cases, more exact stationary solutions of the CMEs or the LLE (Eqs.~\ref{eq:ch5:cme_normalized} and \ref{eq:ch5:LLE_norm}) have to be found numerically. Among other approaches, Newton's method is often employed~\cite{yang2010NonlinearWavesIntegrable} to find stationary solutions $\boldsymbol x: \, \boldsymbol F(\boldsymbol x)=0$ where $\boldsymbol x$ stands for $\boldsymbol \psi$ or $\boldsymbol \Psi$ and $\boldsymbol F(\boldsymbol x)$ is the right-hand side of Eqs.~\ref{eq:ch5:cme_normalized} or \ref{eq:ch5:LLE_norm}, respectively. 
With the definition of $\boldsymbol x_R = \operatorname{Re}\{\boldsymbol x\}$ $\boldsymbol x_I = \operatorname{Im}\{\boldsymbol x\}$, $\boldsymbol F_R (\boldsymbol x)= \operatorname{Re}\{\boldsymbol F(\boldsymbol x)\}$, and $\boldsymbol F_I (\boldsymbol x)= \operatorname{Im}\{\boldsymbol F(\boldsymbol x)\}$ the system's Jacobian is 
\begin{equation}
    \mathbb{J}_{\boldsymbol{F}}(\boldsymbol{x}) =
    \begin{pmatrix}
        \frac{\partial \boldsymbol{F}_R}{\partial \boldsymbol{x}_R} & \frac{\partial \boldsymbol{F}_R}{\partial \boldsymbol{x}_I} \\
        \frac{\partial \boldsymbol{F}_I}{\partial \boldsymbol{x}_R} & \frac{\partial \boldsymbol{F}_I}{\partial \boldsymbol{x}_I}
    \end{pmatrix}
\end{equation}
If $\boldsymbol F$ is holomorphic, a splitting into real and imaginary part is not needed (cf. Eq.~\ref{eq:mi_jacobian}). The stationary solutions can be iteratively found 
\begin{equation}
    \begin{pmatrix}
        \boldsymbol{x}_R \\
        \boldsymbol{x}_I
    \end{pmatrix}^{(k+1)}
    = 
    \begin{pmatrix}
        \boldsymbol{x}_R \\
        \boldsymbol{x}_I
    \end{pmatrix}^{(k)}
     - \mathbb{J}_{\boldsymbol{F}}^{-1}(\boldsymbol{x}^{(k)})\cdot 
    \begin{pmatrix}
        \boldsymbol F_R (\boldsymbol{x}^{(k)}) \\
        \boldsymbol F_I (\boldsymbol{x}^{(k)})
    \end{pmatrix}
\end{equation}
where $k$ is the iteration index and $k=0$ corresponds to an initial guess.
Newton's method is relatively straightforward to implement, although, in its simplest form, it has a limited area of convergence (requiring a reliable initial guess). Various modifications improving the performance have been proposed~\cite{kelley2003SolvingNonlinearEquations}. 
So far, we have assumed that the angular velocity of the co-moving time frame in which the waveform is stationary is $d_1$ (in the normalized time).
If this is not the case e.g., in the presence of higher order dispersion, Raman effects, coupled modes or multiple pump laser (Sec.~\ref{sec:aspects_DKS}) the temporal waveform will show an apparent drift with (normalized) angular velocity $\delta d_1$ in the co-moving time frame, and the spectral components will have a time dependent phase term $\exp(-i\mu (\delta d_1)\tau)$. To find stationary solution one can add $\delta d_1$ as a free parameter to the state vectors along with an additional condition e.g. $\frac{\partial}{\partial \tau}\sum_j \phi_j |\Psi(\phi_j)|^2 = 0$ (zero drift velocity), or $\sum_j \phi_j |\Psi(\phi_j)|^2 = 0$ (position fixed $\phi=0$) if the problem has a continuous time translation symmetry. To check for the stability of the solution, we consider the eigenvalues $\lambda_i$ of the Jacobian. If the real part of the eigenvalues 
does not exceed zero, then the solution can be considered stable. Note that any symmetry, such as translation symmetry, can cause eigenvalues close to zero, see discussion in Sec.~\ref{sec:AM_pumping}.

Noteworthy, these methods are often used in combination with continuation methods to compute the existence area of the numerical solutions and examine the bifurcation structure~\cite{kuznetsov2023ElementsAppliedBifurcation,seydel2010PracticalBifurcationStability,doedel1991NumericalAnalysisControl}. There are several established packages available, e.g.~\cite{doedel2007auto, veltz2020bifurcationkit,dhooge2003MATCONTMATLABPackage}.

\subsubsection{Comb line frequencies and beatnote between comb lines}
\label{sec:actualfreq_beatnote}
The comb lines are not always generated exactly at the frequencies $\omega_\mu$. To obtain the actual oscillation frequency of a comb line $\mu$,
$a_\mu$ can be Fourier-transformed along the \textit{slow} time axis $t$, which will provide for each mode $\mu$ its spectrum (power spectral density) around the frequency $\omega_\mathrm{p} + \mu D_1$:
\begin{equation}
\label{eq:actual_comb_spec}
    S^\mathrm{opt}_\mu(\omega) = \mathcal{F}[a_\mu](\omega-\omega_\mathrm{p} - \mu D_1),
\end{equation}
where we impose $\omega-\omega_\mathrm{p} - \mu D_1<<D_1$ to avoid interference with an adjacent line.
The (repetition rate) beatnote between the comb lines is given by
\begin{equation}
    \label{eq}
    S^\mathrm{rep}(\omega) = \sum_\mu \mathcal{F}[a_{\mu+1} a_\mu^*] (\omega-D_1),
\end{equation}
where we impose $\omega- D_1 \ll D_1$ to avoid interference with the beatnote resulting from next to nearest neighbor comb lines.

\subsubsection{Nonlinear dispersion relation}
\label{sec:ndr}

\begin{figure}
\includegraphics[width=\columnwidth]{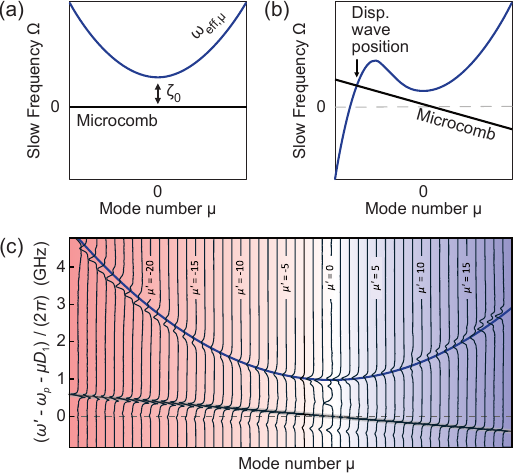}
\caption{\textbf{Nonlinear dispersion relation.} (a) Nonlinear dispersion relation showing the equidistant microcomb as a straight line. If white noise is injected the resonance spectrum $\omega_\mathrm{eff,\mu}$ can be revealed. (b) As in (a) but with third order dispersion. The tilt of the ``microcomb line'' indicates that the microcomb repetition rate deviates from $D_1$; the crossing of the $\omega_\mathrm{eff,\mu}$-line and the DKS marks the position of a dispersive wave. (c) Experimental measurement of the nonlinear dispersion by means of an auxiliary laser with frequency $\omega'$ that is coupled to the resonator and performing a calibrated laser scan. The $\omega_\mathrm{eff,\mu}$ appear as resonance features (indicated by the blue line) and the comb lines (indicated by the black line) show a characteristic interaction with the scanning laser~\cite{wildi2023SidebandInjectionLocking}; here the tilt of the black line arises from the Raman effect.} \label{fig:NDR_concept} 
\end{figure}

The dispersion of a microresonator can be conveniently represented via its integrated dispersion (Eq.~\ref{eq:ch3:dint} and Fig.~\ref{fig:ch3:dispersion}) as the deviation of the resonance frequencies from a $D_1$-spaced frequency grid centered on $\omega_0$. Relative to the same frequency grid, the power spectrum of the microcomb $S^\mathrm{opt}_\mu(\omega)= S^\mathrm{opt}_\mu(\Omega + \omega_0 + \mu D_1)$ (Section~\ref{sec:actualfreq_beatnote}) can be represented, where we introduce the \textit{slow frequency} $\Omega$; this is also referred to as the \textit{nonlinear dispersion relation (NDR)}~\cite{leisman2019EffectiveDispersionFocusing}. The NDR for a microcomb with a repetition rate $D_1/(2\pi)$ is schematically shown in Fig.~\ref{fig:NDR_concept}. It can be computed by Fourier-transforming the $\psi_\mu(t)$ and arranging the spectra as shown in Fig.~\ref{fig:NDR_concept}, or by taking the double Fourier transform of the spatiotemporal diagram. For the Fourier transform in time axis, a window function (e.g., a Hann window) can reduce the influence of nonperiodic boundaries. If white noise is numerically injected into the system, it will get enhanced at the frequencies that correspond to the effective resonance frequencies $\omega_\mathrm{eff,\mu}$ and appear when plotting $S^\mathrm{opt}_\mu$ as shown in Fig.~\ref{fig:NDR_concept}. The NDR reveals the presence of coherent structures, their amplitude and repetition rate, and the phase-matching between different resonances in the system, such as dispersive wave formation (Section~\ref{sec:sec:aspects_DKS:dispersion:HOD}) as exemplified in Fig.~\ref{fig:NDR_concept}b. The NDR has been applied for studying soliton Physics in hydrodynamics~\cite{tikan2022NonlinearDispersionRelation}, and characterization of the noise transfer mechanisms in soliton microcombs~\cite{anderson2021PhotonicChipbasedResonant}, complex FWM pathways in coupled~\cite{komagata2021DissipativeKerrSolitonsa,tikan2021EmergentNonlinearPhenomena} and modulated~\cite{anderson2023DissipativeSolitonsSwitchinga} microresonators, and served for the identification of the chimera state~\cite{tusnin2020NonlinearStatesDynamics}. Experimentally, the NDR can be measured by coherently probing the system with a narrow-linewidth scanning laser~\cite{herr2012UniversalFormationDynamics,tikan2021EmergentNonlinearPhenomena}, which can simultaneously reveal the resonance branch corresponding to the integrated dispersion~\cite{wildi2023SidebandInjectionLocking}.

\section{Generalized description of microcombs}
\label{sec:aspects_DKS}
The framework of the LLE as presented in Sec.~\ref{sec:basics_nln} captures key aspects of microcombs. However, distinct dynamics and phenomena can emerge through effects such as modified dispersion, modulation of the pump laser, Raman-nonlinear effects, and coupled resonator systems, necessitating a generalized description.

\begin{figure*}
\includegraphics[width=0.95\textwidth]{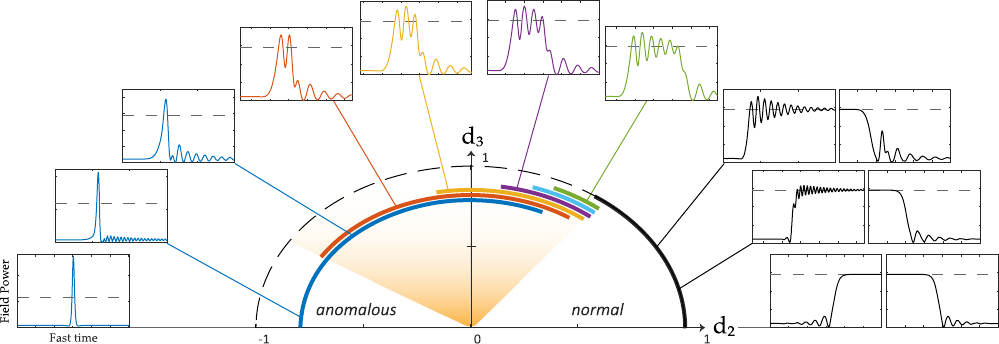}
\caption{\textbf{Localized dissipative structures in the second-/third-order dispersion (d$_2$-d$_3$) plane.} Clockwise from left: conventional DKS, dissipative solitons with dispersive-wave tails, zero-dispersion solitons with quantized periods (orange area), switching waves with dispersive-wave tails, conventional switching wave. The dashed gray line in the outer figures represents the CW high-state solution. Thick bands represent the existence range of structures in the circular path. From ~\cite{anderson2022ZeroDispersionKerr}.\label{fig:ch7:zds}} 
\end{figure*}

\subsection{Generalized dispersion}
\label{sec:aspects_DKS:dispersion}
In Section~\ref{sec:basics_kcomb:models_of_OFC}, we considered the ideal anomalous-dispersion case, where the resonance frequencies follow Eq.~\ref{eq:ch3:resonance_expansion} with $D_2>0$ and $D_i=0$ for all $i>2$. In practice, modification to the resonator mode structure can occur and strongly affect comb formation~\cite{herr2014ModeSpectrumTemporal}.

\subsubsection{Effects of higher order dispersion on DKS microcombs ($D_2>0$)}
\label{sec:sec:aspects_DKS:dispersion:HOD}
Considering higher order dispersion (HOD) terms, i.e. $D_j$, $j>2$ in Eq.~\ref{eq:ch3:resonance_expansion} becomes necessary for very broadband spectra or when $D_2$ is close to zero. HOD can be included in the CMEs~\ref{eq:ch5:CME_rf} by considering higher order terms in the definition of $D_\mathrm{int}$ or $\zeta_\mu$. The LLE is generalized by adding $i \sum_{j>2} \frac{D_j}{j !}\left(\frac{\partial}{i \partial \phi}\right)^j A$ to the right-hand side of Eq.~\ref{eq:ch5:LLE}. In normalized units (Eq.~\ref{eq:ch5:LLE_norm}), the coefficients in front of corresponding derivatives by $\phi$ are $d_j=D_j \frac{2}{\kappa j !}\left(\frac{\kappa}{2 |D_2|}\right)^{j / 2}$.

Different cases can be encountered~\cite{anderson2022ZeroDispersionKerr}: from conventional DKS in anomalous dispersion to zero dispersion soliton (ZDS) and switching-waves (SW) in the normal dispersion case. A `palette' of these states is shown in Fig.~\ref{fig:ch7:zds} as a function of the second and third-order dispersion (TOD) coefficients. Below, we discuss these states in detail.

\textit{Third order dispersion and dispersive waves.}
If third order dispersion ($D_3$) becomes relevant, the DKS dynamics is modified by the effect of the soliton dispersive wave (DW) formation (or ``Cherenkov radiation'')~\cite{akhmediev1995CherenkovRadiationEmitted, karpman1993RadiationSolitonsDue, wai1986NonlinearPulsePropagation}.
It arises when the DKS becomes directly phase matched to the resonances i.e., when the soliton line (Fig.\ref{fig:NDR_concept}b) crosses the dispersion profile, which can not occur in the pure anomalous second-order dispersion case.
The DKS spectrum is modified in two ways: (i) frequency combs at the crossing point demonstrate a local power enhancement~\cite{brasch2016PhotonicChipBased,jang2014ObservationDispersiveWavea} due to the improved phase-matching~\cite{milian2014SolitonFamiliesResonant,milian2015SolitonsFrequencyCombsa}, and (ii) the spectral center of the DKS spectrum is shifting away from the DW (``spectral recoil'). 

\begin{figure}[h]
\includegraphics[width=1.0\columnwidth]{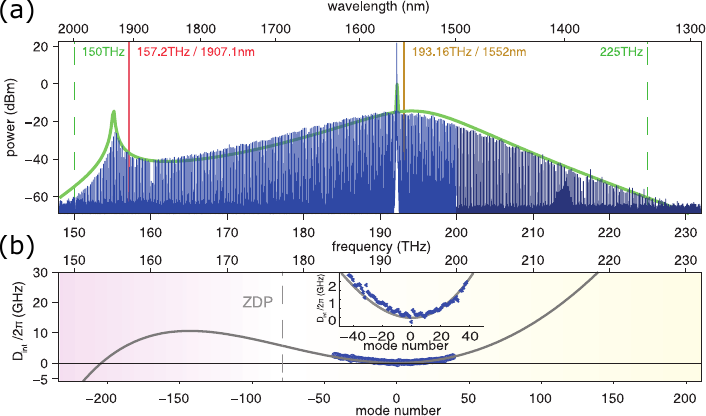}
\caption{\textbf{Frequency combs in the presence of the HOD and Cherenkov radiation.}(a) Optical spectrum shows the secant hyperbolic shape of a single soliton (with a 3-dB width of 10.8 THz) and the soliton Cherenkov radiation at 155 THz. The green dashed lines mark a span of two-thirds of an octave. The green solid line denotes the simulated spectral envelope. The different blue colors indicate measurements done with two different optical spectrum analyzers. (b)The integrated dispersion D$_\mathrm{int}$ from finite element method simulations for the measured resonator geometry (gray solid line). The gray dashed line indicates the zero dispersion point. The blue dots around 0 (the inset shows a zoom-in) are measured positions of around 80 resonances which show good agreement with the simulated dispersion. From~\cite{brasch2016PhotonicChipBased} \label{fig:ch7:Cherenkov}} 
\end{figure}

An experimental example of DW formation is shown in Fig.~\ref{fig:ch7:Cherenkov}, where however the `recoil effect' is compensated by the Raman effect (Sec.~\ref{sec:srs}).

In the time domain, the DW manifests itself as oscillatory tail as shown in Fig.~\ref{fig:ch7:zds}. The DW oscillatory tail and the DKS are co-propagating and stationary in the same co-moving frame, in contrast to DWs in the non-resonant generalized NLS describing supercontinua in single-pass waveguides~\cite{dudley2006SupercontinuumGenerationPhotonic,skryabin2010ColloquiumLookingSoliton}.

It has been numerically demonstrated in~\cite{parra-rivas2014ThirdorderChromaticDispersion}, that the presence of the third-order dispersion modifies the bifurcation structure of the DKS, reducing its dynamical instability region. Moreover, DW tail can trap neighboring DKS and create a complex soliton bound state~\cite{weng2019PolychromaticCherenkovRadiation,vladimirov2018EffectCherenkovRadiation,wang2017UniversalMechanismBinding} (see~Sec.\ref{subsubsec:interaction_dks_diss_patterns}).

Employing the method of moments, often seen as an alternative~\cite{ankiewicz2008ComparisonLagrangianApproach} to the Lagrangian approach described in Sec.~\ref{sec:basics_kcomb:DKS}, the modification of the DKS parameters due to the HOD effects can be estimated.
This approach investigates the evolution of integrals of motion of the NLSE, e.g. those corresponding to energy and momentum:\footnote{Often the integration limits are set to infinity since the microresonator roundtrip time is much long than the DKS duration.}
\begin{align}
    \mathcal{Q}&= \int_{-\pi}^\pi \left|\Psi\right|^2 d \phi \\
     \mathcal{P}&= -\frac{i}{2} \int_{-\pi}^\pi  \left(\Psi^* \frac{\partial \Psi}{\partial \phi}-\Psi \frac{\partial \Psi^*}{\partial \phi}\right)d \phi,
     \label{eq:ch6:integrals_NLS}
\end{align}
and their moments in the presence of a perturbation $\mathcal{R}[\Psi]$ given in this case by the LLE and its generalizations. For the general expressions with an arbitrary perturbation see~\cite{maimistov1993evolution}. The quantities defined by Eq.~\ref{eq:ch6:integrals_NLS} are no longer conserved in the presence of the perturbation. With a suitable choice of ansatz, their evolution equations can be reduced to a finite-dimensional system analogous to Eq.~\ref{eq:ch5:FDDS}.

The treatment of the case of TOD effect on DKS is presented  in~\cite{cherenkov2017DissipativeKerrSolitons}.
The DKS solution ansatz suitable for the LLE with the TOD is given by (in normalized units Eq.~\ref{eq:ch5:LLE_norm2}):
\begin{equation}
    \Psi=\sqrt{2 \zeta_{0}} \operatorname{sech}\left(\sqrt{2 \zeta_{0}} \tilde{\theta}\right) e^{i d_{3} g(\tilde{\theta})+i \arccos \left(\sqrt{8 \zeta_{0}} / \pi f\right)}-i \frac{f}{\zeta_{0}},
\end{equation}
where $\tilde{\theta}=\theta-2 d_{3} \zeta_{0} \tau$ is the normalized fast time that accounts for the TOD-induced group velocity, $g(\tilde{\theta})=\left[4 \zeta_{0} \tilde{\theta}-3 \sqrt{2 \zeta_{0}} \tanh \left(\sqrt{2 \zeta_{0}} \tilde{\theta}\right)\right]$. 
Thus, the DKS in the presence of the TOD acquires additional group velocity that can be expressed as $v_g=2 d_3 \zeta_0$. Associated change of the DKS repetition rate is given by 
\begin{equation}
\label{eq:ch6:dw_recoil}
    \Omega_\mathrm{DW}=\zeta_0 \frac{\kappa D_3}{6 D_2}.
\end{equation}
The position and the amplitude of the DKS maximum are shifted by
\begin{equation}
    \mu_r=\zeta_0 \frac{D_3 \kappa}{3 D_2^2}; \, \left|a_{\mu_r}\right|^2 \approx \frac{D_2}{2 \kappa}.
\end{equation}
The position of the DW is estimated as real part of
\begin{equation}
\mu_{d}=-\frac{3 D_{2}}{D_{3}}-\frac{D_{3} \kappa}{3 D_{2}^{2}}\left[2 \zeta_{0}-2\left(f / \zeta_{0}\right)^{2}-i\right],
\end{equation}
suggesting that the DW is not emerging exactly at the point where the dispersion profile crosses zero D$_\mathrm{int}$ point given by $\mu_{d0} = - 3 D_{2}/D_{3}$ but with an additional shift caused by the tilting of the DKS line (cf. Fig.~\ref{fig:NDR_concept}). The expressions above are derived for large enough detuning $\left(\zeta_0 \gg 1\right)$ and low CW background $\left(f / \zeta_0 \ll 1\right)$.

\begin{figure}[h]
\includegraphics[width=1.0\columnwidth]{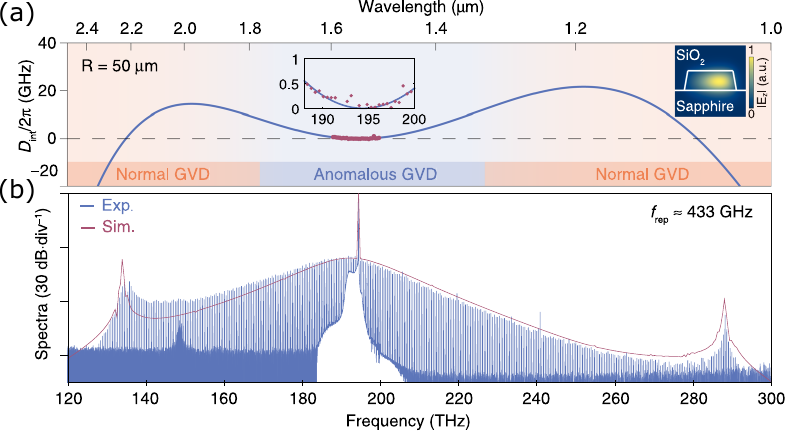}
\caption{\textbf{Octave-spanning Kerr frequency comb.} (a) D$_\mathrm{int}$ of a 50$\mu$m-radius AlN resonator
(cross section: 1.0 x 2.3 $\mu$m$^2$), where the anomalous and normal GVD regimes are indicated by solid light blue and orange
colors in the bottom. Insets: measured (red dots) and simulated (blue curve) D$_\mathrm{int}$, and electric field abs(E$_\mathrm{z}$) of the
TM00 mode. (b) Microcomb spectra from the experiment (Exp., blue curve) and simulation
(Sim., red curve) at an on-chip pump power of 390 mW. The resonator Q$_\mathrm{int}$ is 1.6 million and the repetition rate (f$_\mathrm{rep}$) is estimated to be around 433 GHz. From~\cite{liu2021AluminumNitrideNanophotonics}. \label{fig:ch7:octave}} 
\end{figure}

\textit{Octave-spanning frequency combs.}
The parametric gain in microresonators can support ultra-broadband and octave spanning spectra~\cite{delhaye2011OctaveSpanningTunable, okawachi2011OctavespanningFrequencyComb}, where higher order dispersion beyond TOD becomes relevant~\cite{coen2013ModelingOctavespanningKerra, wang2014BroadbandKerrFrequency,zhang2013GenerationTwocyclePulses}. DKS microcombs have been demonstrated in systems where negative even order dispersion terms, such as $D_4$, reduce the integrated dispersion $D_\mathrm{int}$ in the far-out wings of the combs resulting in spectral flattening or even pronounced dual DW formation~\cite{li2017StablyAccessingOctavespanninga,pfeiffer2017OctavespanningDissipativeKerr, liu2021AluminumNitrideNanophotonics,weng2022DualmodeMicroresonatorsStraightforwarda, briles2021HybridInPSiN}. Moreover, spectral extension via a second pump laser~\cite{zhang2020SpectralExtensionSynchronization} to beyond an octave  has been demonstrated~\cite{moille2021UltrabroadbandKerrMicrocomb}. Noteworthy, it has been observed that the higher order dispersion that gives rise to ultra-broadband combs can lead to the generation of coherent sub-combs with different spectral offset frequency, especially in high-repetition rate system ~\cite{puzyrev2022FrequencyCombsMultiple}.

\subsubsection{Normal dispersion frequency combs and switching waves ($D_2<0$)}
\label{sec:switching_waves}

\begin{figure*}
\includegraphics[width=0.9\textwidth]{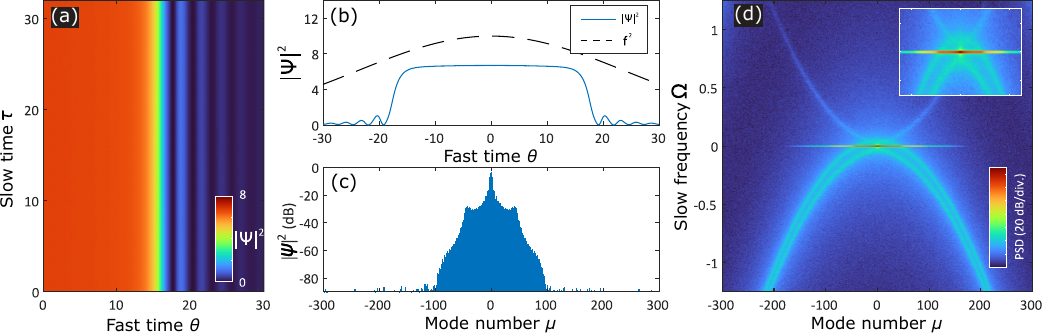} 
\caption{\textbf{Switching wave solution in a normal dispersion microresonator}. (a) Spatiotemporal diagram showing SW field in cavity fast time propagating over resonator slow time. (b) Fast time-domain field snapshot. The blue line shows the SW profile while dashed black the pump pulse profile  (c) Power spectral density of the SW showing the characteristic flat top spectrum. (d) NDR of SW obtained by taking dual Fourier transform of the plot (a). }  \label{fig:ch7:platicon}
\end{figure*}

\begin{figure}
\includegraphics[width=1.0\columnwidth]{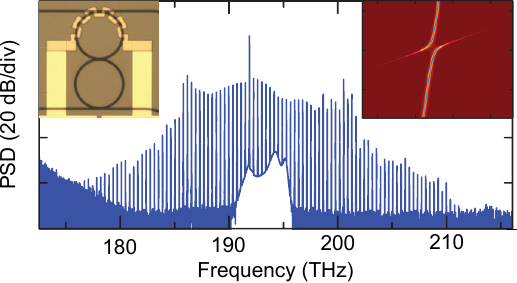}
\caption{ \textbf{Experimental observation of a switching wave microcomb.} SW microcomb generated in coupled normal dispersion resonators (inset on the left) used for controlling the AMX position (inset on the right). Optical spectrum demonstrates characteristic flat-top profile of SW comb. 
From~\cite{xue2015NormaldispersionMicrocombsEnabled}. \label{fig:ch6:nd_exp}} 
\end{figure}

Many resonator platforms exhibit normal dispersion especially towards shorter near-infrared and visible wavelengths~\cite{xue2016NormaldispersionMicroresonatorKerr}, which is generally unfavorable for conventional bright DKS formation, requiring more sophisticated microresonator arrangements~\cite{yuan2023SolitonPulsePairs}.

As in the conservative NLSE, the normal dispersion LLE has a \textit{dark soliton} solution approximated by hyperbolic tangent~\cite{kivshar1998DarkOpticalSolitons,weiner1988ExperimentalObservationFundamental}. The dissipative dark soliton, which constitutes a pathway to microcombs in the normal dispersion regime, is a member of a more general class of coherent structures that exist in microresonators with normal GVD, namely SWs or \textit{platicons}~\cite{xue2016NormaldispersionMicroresonatorKerr,lobanov2015FrequencyCombsPlaticons}.  Their counterpart in the normal dispersion NLSE are called shock waves that are approximated by the modulated train of cnoidal waves inherent to the normal dispersion case~\cite{el2016DispersiveShockWaves}.
In microcombs, the SWs connect the upper and lower homogenous intensity states given by Eq.~\ref{eq:ch4:cubic} with rising and falling fronts~\cite{coen1999ConvectionDispersionOptical}. Due to second order dispersion, $D_2<0$ a spatially decaying oscillatory pattern is present at the lower side of the front, while the top of these structures remains flat if the influence of $D_3$ can be neglected (see Fig.~\ref{fig:ch7:zds}). Except at the so-called Maxwell point, the SW fronts drift (have nonzero speed in the rotating frame), which would result in either the upper or lower intensity state to take over, destroying the SW. However, away from the Maxwell point, the oscillatory patterns at the lower end of the fronts can interact and through interlocking create stable states where the lower intensity ends of the fronts are relatively close to each other, effectively creating a `dark' pulse. The stabilization mechanism is similar to that underpinning bound soliton states as discussed in Sec.~\ref{subsubsec:interaction_dks_diss_patterns}. Generally, SW can be classified by the number of oscillations~\cite{nazemosadat2021SwitchingDynamicsDarkpulse}. In the presence of TOD oscillatory patterns also emerge at the high intensity end of the front, creating stability also for `bright' structures as well as structures in the zero dispersion regime (see Sec.~\ref{sec:zero_disp_combs})~\cite{parra-rivas2016OriginStabilityDark,garbin2017ExperimentalNumericalInvestigations,anderson2022ZeroDispersionKerr}. Stable SW exist in a band within the $f-\zeta_0$-plane where $\zeta_0>0$ where their plurality is reflected by a characteristic snaking structure in the intracavity power vs. pump power plane.~\cite{parra-rivas2016OriginStabilityDark}. Their existence is highly dependent on the resonator dispersion and, in particular, might be affected by the method that is used to excite them, as discussed below.

The formation of DKS relies on MI occurring when differential SPM and XPM balance dispersion and a small detuning. 
In the normal dispersion regime -- as an analysis similar to the one in Sec.~\ref{sec:basics_kcomb:MI} shows -- the MI induced emergence of primary sidebands and Turing patterns is possible ~\cite{haelterman1992DissipativeModulationInstability,hansson2013DynamicsModulationalInstability} if the pump laser is red-detuned (effectively emulating anomalous dispersion)~\cite{ godey2014StabilityAnalysisSpatiotemporal}; this however necessitating larger pump power, and can thermally destabilize the system through red-detuning (see Sec.~\ref{sec:exp_DKS:Res_thermal}).
Thus, alternative methods for triggering sideband formation and then combs in the normal dispersion regime are pursued. 
The first discovery of the normal dispersion combs fiber resonators reports their excitation via the periodic modulation of the intracavity field caused by the boundary conditions and described by the Ikeda map formalism~\cite{coen1997ModulationalInstabilityInduced}. See details in Sec.\ref{sec:FI_combs}.
\begin{figure}
\includegraphics[width=1.0\columnwidth]{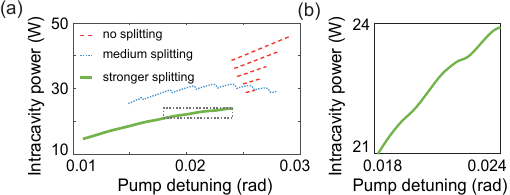}
\caption{ \textbf{Intracavity power vs detuning in a switching wave comb.} (a) Simulation of average intracavity power for different mode-splitting strength. (b) Magnified view of showing smooth transitions between SW states. Adapted from 
From~\cite{nazemosadat2021SwitchingDynamicsDarkpulse}. \label{fig:ch6:nd_steps}} 
\end{figure}
Another approach leverages a displaced resonator modes, which can result from a randomly occurring AMX (Sec.~\ref{sec:basics_lin_AMX}), or from an engineered mode-splitting from coupling to another resonator mode (Sec.~\ref{sec:coupled_mode-families_and_resonators:AMX}). In both cases, the red-shifted modes can enable the phase matching, resulting low-power threshold MI~\cite{helgason2021DissipativeSolitonsPhotonic,kim2019TurnkeyHighefficiencyKerr,xue2015ModelockedDarkPulse,xue2015NormaldispersionMicrocombsEnabled,yu2022ContinuumBrightDarkpulse}, and thermally stable blue-detuned pumping (cf. Sec.~\ref{sec:exp_DKS:Res_thermal}. As Fig.~\ref{fig:ch6:nd_steps} shows, the mode-splitting can drastically affect the SW solutions and experimental access to them.
Alternatively, self-injection locking (SIL) may be utilized without the need for AMX or mode-shifts, achieving stable-access to the SW microcombs on the red-detuned side ~\cite{lihachev2022PlaticonMicrocombGeneration,wang2022SelfregulatingSolitonSwitching}. Finally, pulsed, i.e. poly-chromatic pumping may be used to excite normal dispersion combs~\cite{anderson2023DissipativeSolitonsSwitchinga,xu2021FrequencyCombGeneration}.

The particular spectral profile of a SW microcomb can be understood by reconstructing the NDR (see Sec.~\ref{sec:ndr}). Fig.~\ref{fig:ch7:platicon} shows the NDR for a pulse-pumped normal dispersion microresonator. Pulse pumping is used to excite frequency combs and stabilize them by locking the SW fronts to the pump power value of the pulse that corresponds to the Maxwell point~\cite{anderson2022ZeroDispersionKerr}. The NDR reveals three dispersion parabolas. The parabolas with down-ward opening correspond to the noise of the upper and lower intensity state of the, with a differential SPM-shift between both. The parabola with up-ward opening results from FWM between the (high-intensity-related) down-ward parabola and the pump~\cite{milian2014SolitonFamiliesResonant}. Coherent SW combs constitute a line on the NDR as in the case of DKS, however, in the normal dispersion case it crosses the upper down-ward opening parabola, resulting in enhanced intensity in the wing of the comb (see Fig.~\ref{fig:ch7:platicon}(c,d) and Fig.~\ref{fig:ch6:nd_exp}). While SW combs do generally not achieve the wide spectral span characteristic of DKS, their pump-to-comb conversion efficiency is often much higher, as the effective pump detuning in SW combs is usually smaller in comparison to DKS (especially when the resonance is red-shifted) ~\cite{xue2017MicroresonatorKerrFrequency,fulop2018HighorderCoherentCommunications}. 

\subsubsection{Microcombs at zero second-order dispersion ($D_2\approx0$)} 
\label{sec:zero_disp_combs}
Conservative soliton in the vicinity of zero dispersion have first been studied in single-pass fiber optics~\cite{wai1986NonlinearPulsePropagation}. 

Zero-dispersion solitons (ZDS) in microresonators, including their bifurcation diagrams have been analyzed in~\cite{parra-rivas2017CoexistenceStableDark,tallambe2017ExistenceSwitchingBehaviora}, and subsequently observed in experiments~\cite{anderson2022ZeroDispersionKerr,li2020ExperimentalObservationsBrighta,li2021ObservationsExistenceInstability, zhang2023QuinticDispersionSoliton}. 
In the vicinity of the zero $D_2$, the oscillating tail associated with the dispersive wave becomes more pronounced. A splitting of the single-peak DKS with a DW into a multi-peak structure as highlighted in Fig.~\ref{fig:ch7:zds} by red and yellow marks. The existence range of the ZDS with a different number of peaks can overlap, allowing for the direct switching between the number of peaks. The analysis of the stability of stationary solutions reveals that this switching has a snaking bifurcation structure and bright ZDS solutions can co-exist with the dark once~\cite{parra-rivas2017CoexistenceStableDark,tallambe2020CoexistenceBrightDark}. Moreover, as in the case of DKS in the presence of TOD, the breathing dynamics is suppressed extending the stability range of ZDS~\cite{parra-rivas2017CoexistenceStableDark}.

\begin{figure}
\includegraphics[width=1\columnwidth]{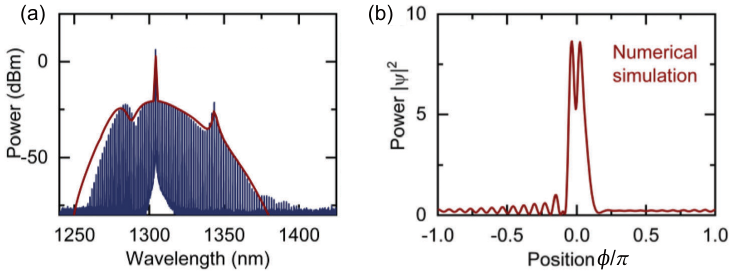}
\caption{\textbf{Experimental demonstration of zero-dispersion soliton formation.} (a) Optical spectrum with simulated spectral envelop. (b) Simulated temporal waveform. Adapted from~\cite{zhang2023QuinticDispersionSoliton}.
\label{fig:ch7:zds_exp}} 
\end{figure}

The peak power level of the ZDS is lower compared to the conventional DKS and corresponds to the upper CW state of the cavity given by Eq.~\ref{eq:ch4:cubic}. The multi-peak structure is reflected by the oscillations in the frequency domain as suggested by simulation as well as experimental data~\cite{anderson2022ZeroDispersionKerr}. An experimental spectrum of a ZDS is shown in Fig.~\ref{fig:ch7:zds_exp}.

\begin{figure}
\includegraphics[width=1.0\columnwidth]{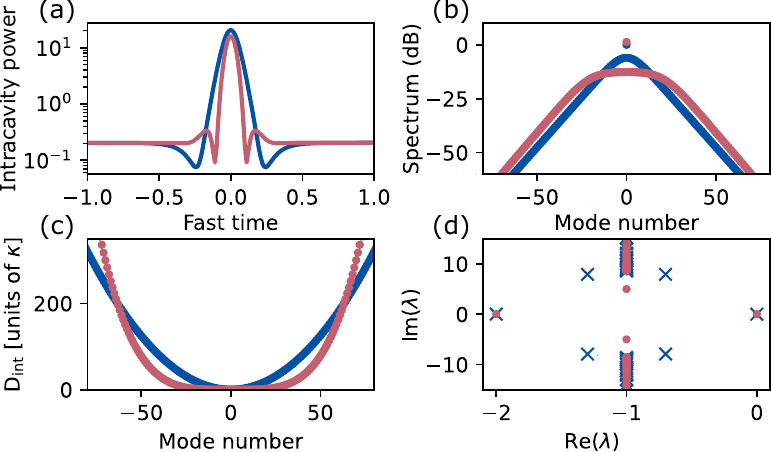}
\caption{Comparison of (a) PQS [red line, D$_4$ = 3x10$^{-4}\kappa$/2] and DKS [blue line, D$_2$ = 0.1$\kappa$/2] intracavity power profile $|\Psi|^2$, (b) the same comparison of power spectral density (c) Integrated dispersion profile presented in units of $\kappa/2$, (d) corresponding Jacobian eigenvalues. Normalized pump power and detuning are the same for both simulations.
Similar plot has been presented in \cite{taheri2019QuarticDissipativeSolitons}. \label{fig:ch7:quart_solitons}} 
\end{figure}

\textit{Pure quartic solitons. } The influence of fourth-order dispersion on solitons has been first investigated in fiber optics~\cite{hook1993UltrashortSolitonsMinimumdispersion}, where solitons of the modified NLS decay during the propagation due to the radiation of DW. However, if second and third order dispersion are negligible, pure quartic solitons (PQS) with Gaussian profile have been observed in a single-pass photonic crystal waveguide~\cite{blanco-redondo2016PurequarticSolitons}. Also in microresonators, dissipative PQS can exist, as shown in Fig.~\ref{fig:ch7:quart_solitons}. Their spectra have a flat profile over a large number of modes~\cite{yao2021PureQuarticSolitons}.

The shape of the DKS changes in the case of quartic dispersion. To obtain an exact expression, a combination of the Lagrangian approach (described in Sec~\ref{sec:basics_kcomb:DKS}) and the direct numerical simulation has been used~\cite{taheri2019QuarticDissipativeSolitons}. A Gaussian function as an ansatz yields
\begin{equation}
A  = A_0 \exp[i \vartheta - i h (\phi-\phi_0) - (1+iC)(\phi - \phi_0)^2 /2w^2],
\end{equation}
where $A_0$ - pulse amplitude, $\vartheta$ - phase, $\phi_0$ - delay, $w$ - width, $h$ - frequency, and $C$ - chirp. 
Setting $C$ and $h$ to zero as suggested by direct numerical simulations, one obtains the following expressions for the stationary PQS amplitude, width, and phase:
\begin{equation}
A_0=\sqrt{\frac{8 \sqrt{2}}{7}  \frac{\omega_0-\omega_\mathrm{p}}{g_\mathrm{K}}}, \,w=\frac{1}{2} \sqrt[4]{\frac{7}{2}  \frac{D_{4}}{\omega_0-\omega_\mathrm{p}}}
\end{equation}
and $\vartheta=\cos ^{-1}[(\kappa / \sqrt{\kappa_\mathrm{ex}} s_\mathrm{in}) \sqrt{(\sqrt{2} / 7) ( (\omega_0-\omega_\mathrm{p}) / g_\mathrm{K})}]$. As for the conventional DKS, the phase term defines the PQS existence range which is comparable with the quadratic case. Bifurcation structure analysis has been performed in~\cite{parra-rivas2022QuarticKerrCavitya}, for both anomalous and normal dispersion cases.

\subsubsection{Periodic modulation of the dispersion and Faraday Instability}
\label{sec:FI_combs}
Introducing dispersion modulation along the resonator~\cite{kordts2016HigherOrderMode,li2020RealtimeTransitionDynamics}, or any other kind of periodic modulation of system parameters (such as Kerr coefficient~\cite{staliunas2013ParametricPatternsOptical}) in CW-driven systems can result in the Faraday Instability (FI)~\cite{mussot2018ModulationInstabilityDispersion} leading to the creation of sidebands similar to the case of MI. Such systems can be described by the Floquet theory~\cite{nayfeh2004NonlinearOscillations,conforti2014ModulationalInstabilityDispersion,conforti2016ParametricInstabilitiesModulated}. In the simplest case of a small step-like modulation of the dispersion, the frequency of FI-induced primary comb lines can be analytically  approximated as~\cite{conforti2014ModulationalInstabilityDispersion}:
\begin{equation}
\mu_{\mathrm{max}}=\sqrt{\left\{\frac{2}{d_2^{av}}\left( \xi_0 -2 \rho \right)\right\} \pm \left[\frac{2}{d_2^{av}} \sqrt{\left(\frac{m \pi}{T}\right)^2+ \rho^2}\right]}
\label{eq:ch7:FImodes}
\end{equation}
where $m$ is the order of parametric resonance, $T$ is the modulation period in normalized slow time, $d_2^{av}$ is the average normalized dispersion. We note that the first term under the square root (curly brackets) represents the conventional MI and can be obtained by differentiating the Eq.~\ref{eq:ch5:mi_gain} over $\mu$ and equating the result to zero, while the effect of the periodic perturbation is represented by the second term (square brackets).

Experimental studies dedicated to the optical FI have mostly been conducted in the context of fiber-based devices operating in the quasi-CW~\cite{ mussot2018ModulationInstabilityDispersion}  and the DKS~\cite{luo2015ResonantRadiationSynchronously,nielsen2018InvitedArticleEmission} regimes. This regime is characterized by period-doubling dynamics~\cite{Bessin2019Real} and competition between Turing (modulation) and Faraday instabilities~\cite{copie2016CompetingTuringFaraday,yang2020CoherentSatellitesMultispectral}. In microresonators, primary combs formation due to the FI~\cite{huang2017QuasiphasematchedMultispectralKerra} and the peak-like enhancement in microcomb spectra in both normal and anomalous dispersion regimes have been observed~\cite{anderson2023DissipativeSolitonsSwitchinga}.

\subsubsection{Spectrally dependent cavity loss}

A spectrally dependent loss profile in the cavity can, via the nonlinear gain-through-loss effect \cite{bessin2019GainthroughfilteringEnablesTuneable, perego2018GainLossesNonlinear, perego2021TheoryFilterinducedModulation}, give rise to new nonlinear dynamics and enable the formation of filter-driven solitons \cite{xue2023DispersionlessKerrSolitons, turitsyn2020SolitonsincOpticalPulses}.
To date, such structures have been demonstrated predominantly in fiber-based cavities. A comprehensive analysis of these effects lies beyond the scope of this review.

\subsection{Generalized Nonlinearity}
\subsubsection{Effect of quadratic optical nonlinearity }
Besides leveraging the Kerr-nonlinearity, frequency combs can be generated based on the second-order (quadratic) optical nonlinearity~\cite{boyd2020NonlinearOptics} potentially extending the accessible frequency range and reducing the power requirements~\cite{buryak2002OpticalSolitonsDue}. This is made possible by emerging integrated photonics platforms such as  LiNbO$_3$~\cite{boes2023LithiumNiobatePhotonics,qi2020IntegratedLithiumNiobate, zhu2021IntegratedPhotonicsThinfilm}, LiTaO$_3$~ \cite{wang2024LithiumTantalatePhotonic}, BaTiO$_3$\cite{abel2019LargePockelsEffect}, and several III--V materials (e.g., AlGaAs~\cite{mobini2022AlGaAsNonlinearIntegrated}, AlN~\cite{li2021AluminiumNitrideIntegrated,liu2023AluminumNitridePhotonic}, GaN~\cite{gromovyi2022LowlossGaNoninsulatorPlatform, zheng2022IntegratedGalliumNitridea}, GaP~\cite{kuznetsov2025UltrabroadbandPhotonicchipbasedParametric, nardi2024IntegratedChirpedPhotoniccrystal, wilson2020IntegratedGalliumPhosphide}, see Ref.~\cite{liu2023H2NonlinearPhotonics} for comparison). Additionally, a second-order nonlinearity can be induced via the photogalvanic effect in materials that lack a $\chi^{(2)}$ response such as Si$_3$N$_4$~\cite{lu2021EfficientPhotoinducedSecondharmonic,billat2017LargeSecondHarmonic, porcel2017PhotoinducedSecondorderNonlinearity,li2023HighcoherenceHybridintegrated780}. 
Microcomb generation approaches can be broadly classified into two categories: (i) microcombs based $\chi^{(2)}$ nonlinearity and (ii) microcombs utilizing the electro-optic effect 

\begin{figure}
\includegraphics[width=0.7\columnwidth]{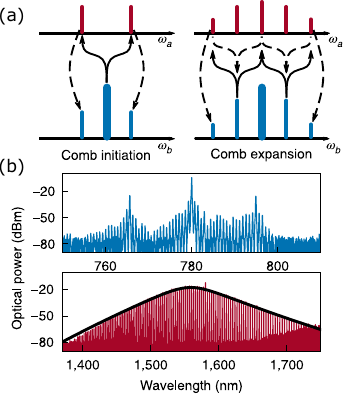}
\caption{\textbf{Quadratic comb generation from a CW pump in AlN microring system} (a) The principle of the comb formation: a near-visible pump produces infrared frequencies, while simultaneous SHG phase-matching produces near-visible frequencies. (b) Generated frequency combs. In near-visible (top) and infrared (bottom) frequency ranges.
From \cite{bruch2021PockelsSolitonMicrocomb}. \label{fig:ch6:pockels_soliton}} 
\end{figure}

\textit{(i) $\chi^{(2)}$ microcombs.}
Pure $\chi^{(2)}$ microcombs have been studied in free-space cavities~\cite{ricciardi2020OpticalFrequencyCombs}, WGM resonators \cite{breunig2016ThreewaveMixingWhispering,smirnov2023SelfStartingSolitonComb}, and, more recently, in integrated microresonators~\cite{bruch2021PockelsSolitonMicrocomb,lu2023TwocolourDissipativeSolitons}, as well as in theoretical works~\cite{ricciardi2020OpticalFrequencyCombs,smirnov2023SelfStartingSolitonComb,nie2022PhotonicFrequencyMicrocombs}.
Physically, $\chi^{(2)}$-based combs rely on a nonlinear coupling and  cascaded bi-directional frequency conversion between a fundamental pump and either its second harmonic (SHG)~\cite{lu2023TwocolourDissipativeSolitons}, or its half-harmonic (OPO) ~\cite{bruch2021PockelsSolitonMicrocomb}, as depicted in Fig.~\ref{fig:ch6:pockels_soliton}(a). 
Both regimes are fundamentally different due to the threshold nature of the latter one.
The nonlinear dynamics in $\chi^{(2)}$ microresonator, akin to the $\chi^{(3)}$-case, can be described by the Ikeda map~\cite{ricciardi2020OpticalFrequencyCombs}, coupled modes equations~\cite{skryabin2020CoupledmodeTheoryMicroresonators}, and mean-field PDEs~\cite{parra-rivas2019LocalizedStructuresDispersive, villois2019FrequencyCombsMicroring}.
For instance, the temporal domain coupled mean field equations for fundamental $A_f$ and half-harmonic $A_h$ envelopes in the doubly resonant degenerate OPO case can be written as:
\begin{equation}
\begin{aligned}
 \partial_t A_h=& - \left(\frac{\kappa_h}{2}+i(\delta\omega_f/2 + \epsilon)\right)A_h +\Delta D_{1} \frac{\partial A_h}{\partial \phi }  \\
& +i\frac{D_{2 h}}{2} \frac{\partial^2A_h}{\partial \phi^2}+ig_{2 h} A_h^* A_f  ,  \\
  \partial_t A_f=&- \left(\frac{\kappa_f}{2}+i\delta\omega_f\right)A_f  +\sqrt{\kappa_{\mathrm{ex,f}}}s_{\mathrm{in}} \\
& +i\frac{D_{2 f}}{2} \frac{\partial^2A_f}{\partial \phi^2}+ig_{2 f} A_h^2,
\label{eq:ch6:pockels_lle}
\end{aligned}
\end{equation}
where $g_{2 f,h}$ - is the second order nonlinear coupling coefficient, $\delta\omega_f = (\omega_{0,f}-\omega_{p})$, $\epsilon = \omega_{0,h}-\omega_{0,f}/2$ describes the frequency mismatch between fundamental and harmonic mode, $\Delta D_1$ represents the temporal walk-off.
The last two variables have a significant impact on $\chi^{(2)}$ combs~\cite{smirnov2023SelfStartingSolitonComb} and careful engineering of resonator dispersion, along with phase-matching techniques~\cite{jankowski2024UltrafastSecondorderNonlineara} is required. 

For a broad range of parameters, the coupled mean field equations can be reduced to a nonlocal version of the LLE~\cite{nikolov2003QuadraticSolitonsNonlocal, leo2016FrequencycombFormationDoubly,nie2022PhotonicFrequencyMicrocombs}. 
When the phase mismatch is large, cascaded second-order nonlinearity effectively mimics the Kerr effect~\cite{ulvila2013FrequencyCombGeneration,szabados2020FrequencyCombGeneration}. In this case, Eq.~\ref{eq:ch6:pockels_lle} can be further simplified to the single LLE-like equation which supports sech-shaped (amplitude) soliton solutions.
In contrast, phase-matched frequency combs result in the sech$^2$-shaped amplitude profiles~\cite{skryabin2021SechsquaredPockelsSolitons}.
Frequency combs in the microresonator with second-order nonlinearity can be further connected to physical concepts like photon-photon polaritons and dressed states~\cite{puzyrev2021BrightsolitonFrequencyCombs,skryabin2021PhotonphotonPolaritonsChi2}.

Generally, Kerr nonlinearity is always present when the $\chi^{(2)}$ soliton microcombs are formed. The intensity-dependent refractive index shifts the dissipative soliton existence range, affects its dynamics, and competes with the cascading three-wave mixing~\cite{cui2022SituControlEffective}. Therefore, realistic simulations of microcomb generation must include both self-phase and cross-phase modulation effects~\cite{ding2024TheoreticalAnalysisMicrocavity,xue2016SecondharmonicassistedFourwaveMixing}. An example of the corresponding dissipative soliton spectrum is shown in Fig.~\ref{fig:ch6:pockels_soliton}(b).

In the case, when Kerr nonlinearity drives the dissipative soliton formation, the conventional DKS can be spectrally translated via the three-wave-mixing~\cite{xue2016SecondharmonicassistedFourwaveMixing,he2019SelfstartingBichromaticLiNbO3, herr2018FrequencyCombDownconversion}, or powered by a $\chi^{(2)}$-process~\cite{englebert2021ParametricallyDrivenKerr}.

\textit{(ii) Electro-optic microcombs.} 
Electro-optical phase-modulation in a microresonator leads to $\chi^{(2)}$-nonlinear wave mixing between the optical fields and the modulating microwave. This induces a coupling between neighboring modes, leading to the formation of frequency combs~\cite{parriaux2020ElectroopticFrequencyCombs}. Created by an external microwave driving signal, electro-optic microcombs do not rely on self-organization, and hence do not require a specific detuning for start-up or operation. Microcombs based on this principle became possible through advances in nanofabrication with integrated electro-optic modulators operating at CMOS-compatible voltage levels~\cite{hu2022HighefficiencyBroadbandOnchip, shams-ansari2022ThinfilmLithiumniobateElectrooptic, wang2018IntegratedLithiumNiobate, zhang2019BroadbandElectroopticFrequency, zhang2025UltrabroadbandIntegratedElectrooptic}. 
The corresponding input-output relation is described by~\cite{ho1993OpticalFrequencyComb,zhang2019BroadbandElectroopticFrequency}:
\begin{equation}
\begin{aligned}
   & E_{out} (t)  = \\ &\left(\sqrt{(1-\gamma)(1-k)} - k \sqrt{\frac{1-\gamma}{1-k}}\frac{r e^{-i\beta \sin\omega_\mathrm{m} t}}{1 - r e^{-i\beta \sin \omega_\mathrm{m} t}}\right) E_{in} (t),
\end{aligned}
\end{equation}
where $\beta = V_{\mathrm{p}} / V_\pi$ is the single-pass peak phase shift of the phase modulator where $V_{\mathrm{p}}$ is the microwave drive peak amplitude, $V_\pi$ is the half-wave voltage of the phase modulator; $\omega_\mathrm{m}$ is the modulation frequency; $r$ denotes the roundtrip transmission coefficient for the circulating field; $\gamma$ and $k$ refer to the coupler insertion loss and the power coupling parameter, respectively; and E$_{in}$(t) and E$_{out}$(t) are the temporal envelopes of the incoming and outgoing optical fields.

When combined with Kerr nonlinearity, the resulting effective generalized LLE takes the form of the driven-dissipative Gross–Pitaevskii-type equation~\cite{tusnin2020NonlinearStatesDynamics}:
\begin{equation}
\begin{aligned}
        \frac{\partial A}{\partial t} = &-\Big(\frac{\kappa}{2} + i(\omega_{0}-\omega_{p})\Big)A +\frac{i D_2}{2}\frac{\partial^2 A}{\partial \phi^{2}}\\& + 2i J_s \cos\big(s \phi + \theta_\mathrm{op}\big)A + ig_\mathrm{K} |A|^2 A + \sqrt{\kappa_{\mathrm{ex}}}s_{\mathrm{in}},
\end{aligned}
\end{equation}
where $J_s$ is the mode coupling rate proportional to $\beta$, $\theta_\mathrm{op}$ is the offset phase. The effective potential leads to the intracavity angle-dependent dynamics and new nonlinear states facilitating single DKS generation.

Intracavity phase modulation has also been utilized to create microcombs of different shape~\cite{tusnin2020NonlinearStatesDynamics,englebert2023BlochOscillationsCoherently}. A more general perspective on this process is provided by the synthetic frequency dimension approach~\cite{yuan2021SyntheticFrequencyDimensions,yang2025ControllingFrequencycombGeneration}, in which each microresonator mode is treated as a bosonic mode, and electro-optic modulation acts as a coupling mechanism between them. 

Several combined approaches have demonstrated advantages in microcomb generation, including frequency-modulated OPO~\cite{stokowski2024IntegratedFrequencymodulatedOptical} and doubly-resonant schemes~\cite{rueda2019ResonantElectroopticFrequency, zhang2025UltrabroadbandIntegratedElectrooptic}.

\subsubsection{Stimulated Raman scattering}
\label{sec:srs}
Raman scattering is an inelastic scattering process where photons
of the pump field $\omega$ are scattered off of optical phonons $\Omega_R$ (molecular vibration)
inside the material~\cite{boyd2020NonlinearOptics}. 
The Raman shift $\Omega_R$ in dielectrica is typically on the order of 10 THz. In stimulated Raman scattering (SRS) the scattered light field of
frequency $\omega_\mathrm{R} = \omega-\Omega_\mathrm{R}$ stimulates further frequency conversion from $\omega$ to $\omega_\mathrm{R}$ and creation of phonons of frequency $\Omega_\mathrm{R}$. As such, SRS is a non-parametric process that does not conserve the photonic energy.

Microresonator-based Raman-lasing has been demonstrated~\cite{spillane2002UltralowthresholdRamanLasera, grudinin2007UltralowthresholdRamanLasing}
and a transition between FWM and SRS regimes as function of pump laser detuning
has been reported~\cite{kippenberg2004KerrNonlinearityOpticalParametrica}. Particularly when pumping in the normal
dispersion regime, where the initial degenerate FWM process is suppressed SRS can
be observed. 

When seeded by parametrically generated comb, SRS may serve as an additional gain
mechanism transferring energy from the pump to comb lines as it has been observed in both normal~\cite{cherenkov2017RamanKerrFrequencyCombs} and anomalous~\cite{karpov2016RamanSelfFrequencyShift,yi2016TheoryMeasurementSoliton,milian2015SolitonsFrequencyCombsa} dispersion regimes.

\begin{figure}[h]
\includegraphics[width=0.99\columnwidth]{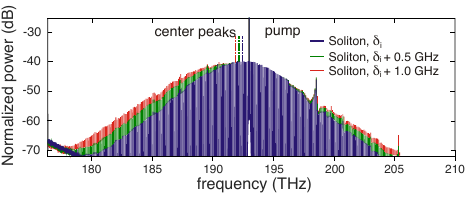}
\caption{ \textbf{Effect of the stimulated Raman scattering on DKS.} 
Compared to an initial detuning $\delta_i$, increasing the detuning leads to wider DKS spectra (shorter pulses) and larger Raman-shift, as here observed in a $100-\mathrm{GHz}$ $ \mathrm{Si}_3 \mathrm{N}_4$. The dashed vertical lines indicate the center of mass of the soliton spectra. From~\cite{karpov2016RamanSelfFrequencyShift}\label{fig:ch7:raman_specr}} 
\end{figure}

Generally SRS will lead to a self-frequency-shift of the spectrum towards lower frequency as it is well known in nonlinear fiber optics~\cite{dudley2006SupercontinuumGenerationPhotonic,agrawal2013NonlinearFiberOptics}. 
In a microresonator, the value of the spectral shift is fixed and depends on the temporal waveform inside the resonators and thus on the operation point on the parameters space (Fig.~\ref{fig:ch5:phase_diagram}). An example of a redshift of a DKS in dependence on the pump laser detuning is shown in Fig.~\ref{fig:ch7:raman_specr}; such a shift is not observed in chaotic MI states.

To quantitatively describe the value of the Raman scattering induces self-frequency shift of DKS, the following modification of the nonlinear term in the LLE (Eq.~\ref{eq:ch5:LLE}) can be made:   

\begin{equation}
ig_\mathrm{K} |A|^2A \rightarrow ig_\mathrm{K} A\left[R(t, \phi) * |A(t, \phi)|^2\right],
\label{eq:ch7:Raman_conv}
\end{equation}
where $R(t, \phi) = \left(1-f_{\mathrm{R}}\right) \delta(t, \phi)+f_{\mathrm{R}} h_{\mathrm{R}}(t, \phi)$ - delayed nonlinear response function. $f_{\mathrm{R}}$ denotes the fractional contribution of the vibrational Raman part in the response, $h_{\mathrm{R}}(t, \phi)$ - is the Raman response function. $*$ indicates the convolution. 

For DKS duration longer than the damping time of molecular vibrations, it is possible to simplify the convolution term~\cite{karpov2016RamanSelfFrequencyShift,cherenkov2017RamanKerrFrequencyCombs}: 
\begin{equation}
ig_\mathrm{K} |A|^2A \rightarrow i g_\mathrm{K} \left(|A|^2 A-f_{\mathrm{R}} \phi_\mathrm{R} A \frac{\partial|A|^2}{\partial \phi}\right),
\label{eq:ch7:Raman_simple}
\end{equation}
where we denoted $\phi_{\mathrm{R}}=\int_{-\pi}^\pi h_{\mathrm{R}}\left(t, \phi^{\prime}\right) \phi^{\prime} d \phi^{\prime}$, and $f_{\mathrm{R}} \phi_{\mathrm{R}}$ = $D_1 \tau_\mathrm{R}$, where $\tau_\mathrm{R}$ is the Raman shock time.

The DKS self-frequency shift can be estimated using the Lagrangian approach~\cite{yi2016TheoryMeasurementSoliton} described for the case of the unperturbed LLE in Sec.~\ref{sec:basics_kcomb:DKS} (alternative derivation using the method of moments can be found in~\cite{karpov2016RamanSelfFrequencyShift}). 

The soliton ansatz is modified to account for the self-frequency shift:
\begin{equation}
    A (t, \phi)= B \operatorname{sech}\left[\left(\phi -\phi_0 \right) / \phi_s\right] e^{-i D_1 \Omega\left(\phi -\phi_0\right)} e^{i \varphi_0},
\end{equation}
which is represented by the term $\Omega$. $\phi_0$ is the DKS position. As in the case of the LLE, DKS amplitude $B$ and duration $\phi_s$ are related, so we can eliminate the latter one. The dissipative function $\mathcal{R}$ is modified to account for the Raman term (Eq.~\ref{eq:ch7:Raman_simple}). Going through the same procedure as described in Sec.~\ref{sec:basics_kcomb:DKS}, we obtain a finite-dimensional set of equations for the amplitude, phase, position, and self-frequency shift. 
Considering steady-state solutions of this system, we obtain the Raman self-frequency shift:
\begin{equation}
\label{eq:ch6:raman_recoil}
    \Omega_{\text{Raman}}=-\frac{8 D_2 \tau_R D_1^2}{15 \kappa \phi_s^4},
\end{equation}
where $\phi_s$ is the DKS pulse width that depends on the pump laser detuning.
The DKS existence range and DKS duration are limited in the presence of the Raman scattering~\cite{wang2018StimulatedRamanScattering}.
Importantly, the Raman self-frequency shift can be compensated by a competing process of the DW generation as shown in Fig.~\ref{fig:qp1}. 

\begin{figure}[h]
\includegraphics[width=1.0\columnwidth]{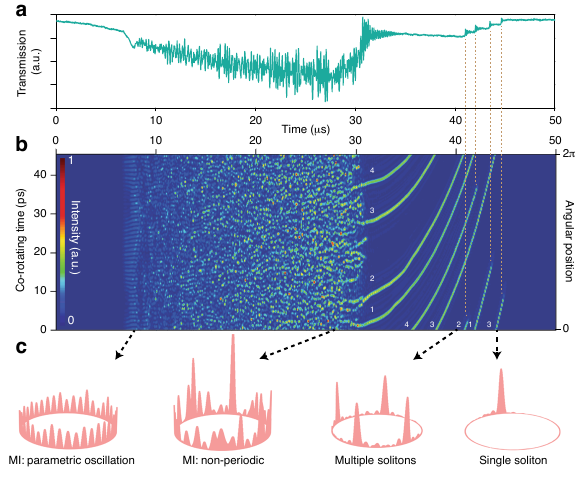}
\caption{  \textbf{Experimental observation of the nonlinear microresonator dynamics.} (a) Microresonator pump power transmission when the pump laser frequency scans from higher to lower frequency. Multiple steps indicate the formation of solitons. (b) Imaging of soliton formation corresponding to the scan in c. The x-axis is time and the y-axis is time in a frame that rotates with the solitons (full scale is one round-trip time). The right vertical axis is scaled in radians around the microcavity. Four soliton trajectories are labeled and folded back into the cavity coordinate system. The color bar gives their signal intensity. (c) Soliton intensity patterns measured at four moments in time are projected onto the microcavity coordinate frame. The patterns correspond to initial parametric oscillation in the MI regime, non-periodic behavior (MI regime), four soliton, and single soliton states. From~\cite{yi2018ImagingSolitonDynamics}.\label{fig:ch7:raman_img}}
\end{figure}

Imaging of the real microresonator dynamics clearly shows the influence of the Raman effect on the DKS dynamics~\cite{yi2018ImagingSolitonDynamics}. As displayed in Fig.~\ref{fig:ch7:raman_img}, increasing the pump laser detuning leads to a larger relative drift of the DKS which increases while the pump laser is tuned to the low frequency (red) part.

Noteworthy, two co-existing DKSs can be generated in a microresonator using the Raman effect. The mode that has the strongest Raman gain can be populated through the scattering which effectively acts as an additional pump source generation the second Raman DKS~\cite{yang2017StokesSolitonsOptical}; also see~\cite{lin2017NonlinearPhotonicsHighQ} for further details.

\subsubsection{Stimulated Brillouin scattering}
Stimulated Brillouin scattering (SBS) \cite{boyd2020NonlinearOptics} is similar to SRS; however, in SBS, the scattering involves lower-frequency acoustic phonons of frequency $\Omega_\mathrm{B} \ll \Omega_\mathrm{R}$, typically on the order of 10~GHz.
Compared to FWM and SRS, the SBS gain is usually significantly larger ($\sim100\times$) while its gain bandwidth is relatively narrow ($<100$~MHz). In integrated waveguides, the geometry and material composition can have noticeable impact on the Brillouin gain spectrum~\cite{gyger2020ObservationStimulatedBrillouin}. For SBS to occur in high-Q microresonators, the resonance spacing must match $\Omega_\mathrm{B}$ enabling low-noise SBS lasers~\cite{grudinin2009BrillouinLasingCaF, tomes2009PhotonicMicroElectromechanicalSystems, lee2012ChemicallyEtchedUltrahighQ}, which can, in turn, be leveraged as a low-noise pump laser for microcombs~\cite{bai2021BrillouinKerrSolitonFrequency}. Moreover, by carefully tailored resonance detuning, Kerr and SBS effects can be balanced, providing a pathway to comb initiation in the normal dispersion regime~\cite{bunel2025BrillouinInducedKerrFrequency}.

\subsection{Generalized pump}
\label{sec:generalized_pump}
In this section, we discuss generalized resonator-pumping schemes that go beyond the use of a single continuous-wave laser. These schemes can trigger the formation of microcombs, reduce the required pump power, and provide access to specific comb states. Furthermore, synchronization between the microcomb and the pump laser enables all-optical control of the comb’s repetition rate. This synchronization can be interpreted as an injection-locking phenomenon described by the Adler equation~\cite{adler1946StudyLockingPhenomena} and, more generally, within the Kuramoto model~\cite{strogatz2000KuramotoCrawfordExploring}.
Below, we discuss bi-chromatic pumping with two CW lasers. Phase modulated, amplitude modulated and pulsed pumping, may be viewed as generalizations. We will also describe how injection of one microcomb into another can lead to synchronization.

\subsubsection{Bi-chromatic pumping}
\label{sec:bichromatic_pump}
Bichromatic pumping usually involves a main pump laser $f$ driving the mode $\mu=0$ that is sufficiently powerful to generate a comb and a secondary often weaker pump laser $f'$ of frequency  $\omega_\mathrm{p}'$ driving the mode $\mu'\neq0$ in the wing of the comb. 

\textit{Threshold reduction and access to specific comb states.} Bi-chromatic pumping with two pump lasers of comparable power has been shown to lower the parametric threshold~\cite{strekalov2009GenerationOpticalCombs} and to support DKSs~\cite{hansson2014BichromaticallyPumpedMicroresonator}. If two pump lasers are driving a mode, offset from the DKS center frequency, where $D_\mathrm{int}=0$, spectrally symmetric DKS can be generated with high efficiency~\cite{matsko2023LowThresholdKerr}, and even when the two pumps have negligible spectral overlap with the DKS~\cite{moille2024ParametricallyDrivenPureKerra}. Importantly, bi-chromatic pumping can trigger the formation of combs in the normal dispersion regime~\cite{xu2021FrequencyCombGeneration,anderson2022ZeroDispersionKerr}, in cases where monochromatic driving fails.
Moreover, bi-chromatic pumping can synthesize soliton crystals \cite{lu2021SynthesizedSolitonCrystals}.
A powerful secondary laser positioned in the wing of the DKS can lead to spectral extension~\cite{zhang2020SpectralExtensionSynchronization, moille2021UltrabroadbandKerrMicrocomb}, and dark-bright bound states if the secondary laser falls into the normal dispersion regime~\cite{zhang2022DarkBrightSolitonBound}.

\begin{figure}[h]
\includegraphics[width=1.0\columnwidth]{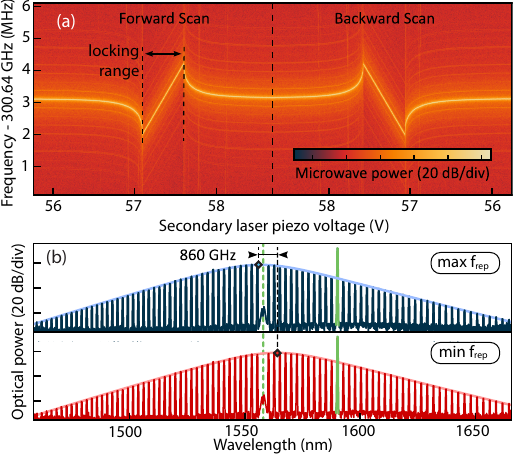}
  \caption{\textbf{Sideband injection locking.} (a) DKS repetition rate monitored during a scan of the secondary (sideband injection) laser across one of the comb lines. (b) A secondary pump laser (green) defines jointly with a main pump laser (here filtered out, dashed green) the repetition rate of the microcomb. Shown are the spectra at the minimally and maximally observable repetition rate, which imply a spectral shift between both spectra. Adapted from~\cite{wildi2023SidebandInjectionLocking}. \label{fig:sideband_injection}} 
\end{figure}

\textit{Synchronization between the microcomb and the pump lasers.} If the secondary laser is spectrally close to a comb line, the microcomb will lock to the secondary pump, so that the frequency spacing between both pump lasers is a multiple of the comb's repetition rate (\textit{sideband injection locking}). 
This permits optical control over all comb frequencies and in particular the repetition rate (Fig.~\ref{fig:sideband_injection}a); it enables optical frequency division~\cite{taheri2017OpticalLatticeTrap, wildi2023SidebandInjectionLocking} for low noise microwave generation~\cite{kudelin2024PhotonicChipbasedLownoise, sun2024IntegratedOpticalFrequency, zhao2024AllopticalFrequencyDivision} and works with any comb line, including the DW of DKS and normal dispersion combs~\cite{moille2023KerrinducedSynchronizationCavity, wildi2023SidebandInjectionLocking}. 
It implies that the temporal waveform of the frequency comb moves at the same group velocity as the temporal beating pattern between the two pump lasers. 
The underlying mechanism can be understood both in time or frequency domains as momentum conservation. 

\textit{Time-domain description. }
The differential group velocity of the waveform inside a co-moving frame with angular velocity $d_1$ is  
\begin{equation}
    \label{eq:GV_momentum_spectralCOG}
    \dot{\phi}=d_2 \frac{\mathcal{P}}{N}=d_2 \bar{\mu}
\end{equation}
where $N=\frac{\mathcal{Q}}{2\pi}= \frac{1}{2 \pi} \int_0^{2 \pi}|\Psi|^2 \mathrm{~d} \phi = \sum_\mu\left|\psi_\mu\right|^2$ is proportional to the number of photons in the cavity and $\bar{\mu}=\sum_\mu \mu \left|\psi_\mu\right|^2 / \sum_\mu\left|\psi_\mu\right|^2$ is the `photonic center of mass'. A drifting waveform in the co-moving frame can be identified as one that has a non-zero momentum $\mathcal{P}$ or whose spectral center of mass $\bar{\mu}$ is shifted \cite{obrzud2017TemporalSolitonsMicroresonators, wildi2023SidebandInjectionLocking}, implying a modified group velocity (assuming non-zero GVD $d_2\neq0$); the spectral shift is shown in Fig.~\ref{fig:sideband_injection}b.

The LLE (Eq.~\ref{eq:ch5:LLE_norm}) can be modified to include a pump term that depends on the spatial coordinate in the resonator $\phi$ and the slow time $\tau$~\cite{taheri2017OpticalLatticeTrap}
\begin{equation}
\tilde{f}(\phi, \tau)=f + f' \, e^{i\mu'  \phi} \, e^{-i\tfrac{2}{\kappa}\left( \omega_\mathrm{p}' - \omega_\mathrm{p} -D_1 \mu' \right)\tau} ,
\label{eq:ch7:dual_pump}
\end{equation}
Using the method of moments, the derivative of the momentum $\mathcal{P}$ of Eq.~\ref{eq:ch6:integrals_NLS} can be expressed as~\cite{taheri2017OpticalLatticeTrap, wildi2023SidebandInjectionLocking}:
\begin{equation}
    \label{eq:momentum_force}
    \begin{aligned}
    \frac{d \mathcal{P}}{d \tau}
    & =-2 \mathcal{P}-i \int_{-\pi}^\pi \left(\Psi^* \frac{\partial \tilde{f}}{\partial \phi}-\Psi \frac{\partial \tilde{f}^*}{\partial \phi}\right)d \phi \\
    & =-2 \mathcal{P} + 2\mu' |f'| | \psi_{\mu'} \sin(\Phi) \\
    \end{aligned}
\end{equation}
where $\Phi=\omega_\mathrm{p}' - \omega_\mathrm{p} -D_1 \mu'\tau +\angle\psi_{\mu'} - \angle f'$ is the phase angle between the secondary laser and the comb line $\mu'$. The secondary pump acts as a force driving the otherwise damped momentum. 

Specifically for the DKS, it can be shown that the force is spatially periodic with an angular period $2\pi/\mu'$, corresponding to the periodicity of the beating pattern between the two lasers. The resulting periodic potential can trap multiple solitons and stabilize soliton crystals \cite{taheri2017OpticalLatticeTrap, lu2021SynthesizedSolitonCrystals}.
Eq.~\ref{eq:momentum_force} describes the injection of an oscillation of frequency $|\omega_\mathrm{p} - \omega_\mathrm{p}'|$ to which the microcomb's repetition rate can (sub-)harmonically lock. Solving for the steady state of the momentum 
\begin{equation}
	\mathcal{P} = \mu' |f'| | \psi_{\mu'} | \sin(\Phi)
\end{equation}
with $\Phi \in [-\pi, \pi]$ and $2\pi \delta f_\mathrm{rep} = D_2 \frac{\mathcal{P}}{N}$. The full width of the injection locking range for the secondary laser is
\begin{equation}
\label{eq:injection_locking_range}
    \begin{aligned}
        \Delta \omega_\mathrm{p}' 
         = 2\mu'^2 D_2 \frac{\left|f^{\prime}\right|\left|a_{\mu^{\prime}}\right|}{N} 
         \approx 8\pi \mu'^2 \eta D_2 \frac{\sqrt{P^{\prime} P_{\mu^{\prime}}}}{\sum_\mu P_\mu},
    \end{aligned}
\end{equation}
where $P'$ and $P_{\mu}$ are the power levels of the secondary pump and the comb lines measured in transmission of the bus waveguide.
The intracavity temporal waveform cannot react instantaneously to a change of the secondary pump laser frequency implying a bandwidth limitation of the locking mechanism. While this limits the maximal actuation bandwidth, it suppresses high-frequency noise even within the cavity linewidth, which was observed in the related cases of phase and amplitude modulated pumping ~\cite{weng2019SpectralPurificationMicrowave,brasch2019NonlinearFilteringOptical}. Considering in a bi-chromatically pumped resonator the dynamic evolution of $\mathcal{P}$ in the small signal limit for frequencies that are well within the linewidth of the cavity the locking bandwidth can be estimated to $\Delta_\mathrm{BW}=\tfrac{1}{2}\Delta \omega_\mathrm{p}'\cos{(\bar{\Phi})}$, where $\bar{\Phi}$ is the temporal mean of $\Phi$~\cite{sun2025MicrocavityKerrOptical}.

\textit{Frequency domain description. }
Replacing in the CMEs~\ref{eq:ch5:cme_normalized} $D_1$ with $D_1'=|\omega_\mathrm{p} - \omega_\mathrm{p}'|$, i.e. transforming into a rotating frame corresponding to the locked state the pump term is given by $\delta_{0\mu} f + \delta_{\mu'\mu} f'$.
Based on the momentum (mode number) conservation in the parametric Kerr processes, the spectral center of gravity $\bar{\mu}=\mathcal{P}/N$ in the steady state can be directly inferred from the mean momentum injected by the two pump lasers \cite{wildi2023SidebandInjectionLocking} 
\begin{align}
	\begin{split}
        \bar{\mu}=
        \mu^{\prime} \frac{\left|a_{\mu^{\prime}}\right|\left|f^{\prime}\right| \sin \left(\Phi\right)}{N}
	\end{split}
\end{align}
which then leads again to Eq.~\ref{eq:injection_locking_range}. In principle, this equation also for non-DKS microcombs, however, additional effects, such as spectral shape changes, may modify the locking range.

\subsubsection{Phase modulated pumping}
\label{sec:PM_pumping}
\begin{figure}[h]
\includegraphics[width=1.0\columnwidth]{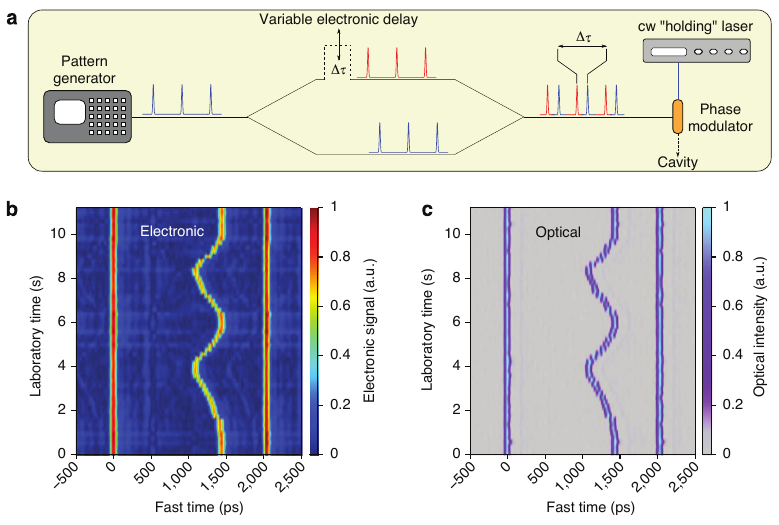}
  \caption{\textbf{Soliton tweezing via pump phase modulation.}(a) Configuration used to generate an electronic signal consisting of two interleaved periodic sets of 90 ps quasi-Gaussian pulses whose relative delay can be continuously varied. (b) Typical real-time manipulation of the electronic signal and, (c) corresponding identical temporal motion of picosecond optical temporal CSs trapped to the phase peaks. From~\cite{jang2015TemporalTweezingLight}. \label{fig:ch7:tweezing}} 
\end{figure}

Phase modulation imprints a phase profile on the pump laser
\begin{equation}
    \tilde{f} (\phi) = f \exp[i \Gamma_m (\phi)],
    \label{eq:ch7:pm_pump}
\end{equation}
For DKS it causes a drift velocity $\dot{\phi}$ of dissipative solitons proportional to the phase gradient $\partial\Gamma(\phi)/\partial \phi$~\cite{firth1996OpticalBulletHoles}, towards $\max{(\Gamma)}$, enabling control over the positions of DKS \cite{jang2015TemporalTweezingLight} (Fig.~\ref{fig:ch7:tweezing}) and deterministic generation of single DKS by providing only one stable position within the resonator~\cite{taheri2015SolitonFormationWhisperingGalleryMode,lobanov2016HarmonizationChaosSoliton, cole2018KerrmicroresonatorSolitonsChirped}.

The same time or frequency domain approaches introduced in the context of bi-chromatic pumping in Section~\ref{sec:bichromatic_pump} can also be used here. From eq.~\ref{eq:momentum_force} it follows that a phase gradient creates a force modifying the momentum and with it the soliton drift velocity. In a frequency domain description, phase modulation of a CW pump laser with modulation frequency $\Omega\approx\mu' D_1$ creates sidebands of frequencies $\omega_\mathrm{p} + n\Omega$ ($n=\pm1, \pm2, .. $). For sinusoidal phase modulation $\Gamma(\phi)=\beta \sin{(\mu' \phi)}$ with modulation index $\beta$, the sideband amplitudes are described by Bessel functions of the first kind of order $n$. For weak modulation, it is sufficient to consider only the two sidebands $n=\pm1$ adjacent to the pump with amplitude $\pm\beta f/2$. Following the same approach as in Section~\ref{sec:bichromatic_pump} the locking range is now twice (two injection sidebands) that of Eq.~\ref{eq:injection_locking_range} with $P'=\beta^2 P_\mathrm{p}/4$ where $P_\mathrm{p}$ is the pump power.

\begin{figure}[h]
\includegraphics[width=1.0\columnwidth]{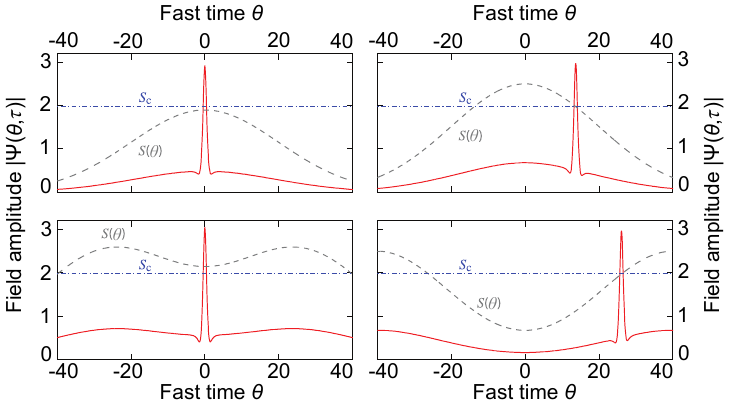}
\caption{ \textbf{Amplitude modulated pumping.}
Steady-state DKS profiles corresponding to various driving field distributions (gray dashed line) that can be below, above or crossing the critical value (blue dash-dotted line). Modified from ~\cite{erkintalo2022PhaseIntensityControl} and \cite{hendry2018SpontaneousSymmetryBreaking}. \label{fig:ch7:pulse_pump1}} 
\end{figure}

\subsubsection{Amplitude modulated and pulsed pumping}
\label{sec:AM_pumping}
Amplitude modulation is phenomenologically similar to phase modulation and may also be interpreted as a polychromatic pump. Both time and frequency domain approaches, following those introduced in Section~\ref{sec:bichromatic_pump} may be used for the description and the locking range. Synchronization of the microcomb waveform to the modulated background ~\cite{obrzud2017TemporalSolitonsMicroresonators,brasch2019NonlinearFilteringOptical,anderson2021PhotonicChipbasedResonant,hendry2019ImpactDesynchronizationDrift,parra-rivas2014EffectsInhomogeneitiesDrift,xu2021FrequencyCombGeneration,anderson2023DissipativeSolitonsSwitchinga} including rational pumping~\cite{xu2020HarmonicRationalHarmonic} is possible. In contrast to phase modulation, locking can also occur at temporal positions where the intracavity pump field amplitude gradient is different from zero \cite{erkintalo2022PhaseIntensityControl}. 
To estimate the pump driving level at which a DKS will be stationary, a semi-analytical approach based on the neutral (or Goldstone) mode theory~\cite{maggipinto2000CavitySolitonsSemiconductor} can be used to yield the DKS drift velocity associated with the pump amplitude modulation:
\begin{equation}
   \dot{\phi}=a \left(\zeta_0, \tilde{f}\left(\phi_\mathrm{DKS}\right)\right) \frac{\partial \tilde{f}\left(\phi_\mathrm{DKS}\right)}{\partial \phi}, 
   \label{eq:ch7:am_grup_vel}
\end{equation}
where the coefficient $a$ can be estimated as a projection of the perturbation to the LLE on the neutral mode~\cite{hendry2018SpontaneousSymmetryBreaking,erkintalo2022PhaseIntensityControl}.  
There are two cases when $\dot{\phi}=0$: (i) the derivative of the pump modulation profile is zero, or (ii) $a = 0$, both occurring at least twice along the resonator. 
The Jacobian spectrum, neutral mode profile, and hence the coefficient $a$ can be estimated numerically for example with the Newton method described in Sec.~\ref{sec:basics_kcomb:numerical_methods}.
Depending on whether the pump field falls below, above or crosses a critical level, the DKS can be trapped at the maximum, (non-zero) minimum or slope of the modulated pump.

Thus, the system can exhibit a spontaneous symmetry breaking when the driving power is increased~\cite{xu2014ExperimentalObservationSpontaneous, hendry2018SpontaneousSymmetryBreaking}. 

\textit{Synchronous pulsed driving.} A particular form of amplitude modulated is driving the microcomb with a pulsed pump laser \cite{obrzud2017TemporalSolitonsMicroresonators, malinowski2017OpticalFrequencyComb, weng2021GainswitchedSemiconductorLaser, lilienfein2019TemporalSolitonsFreespace, brasch2019NonlinearFilteringOptical}. Assuming Gaussian pulses~\cite{erkintalo2022PhaseIntensityControl} the pump term is:
\begin{equation}
    \tilde{f}=f \exp \left(-\frac{\phi^2}{2 \Delta\phi^2}\right),
\end{equation}
where $\Delta\phi$ is the duration of the pump impulse. 
\begin{figure}[h]
\includegraphics[width=1.0\columnwidth]{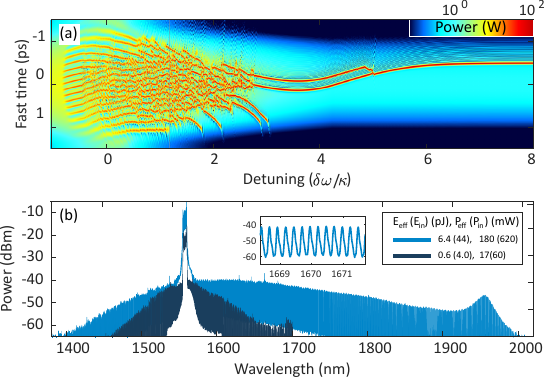}
\caption{ \textbf{Pulse pumping.} (a) Numerical simulation of the intracavity power evolution as a function of the pump laser detuning. (b) Experimentally recorded optical spectra. From~\cite{anderson2021PhotonicChipbasedResonant} \label{fig:ch7:pulse_pump0}} 
\end{figure}
An example of the pulse-driven frequency comb generation is given in Fig.~\ref{fig:ch7:pulse_pump0}. Pulsed pumping localizes the dynamics inside the cavity, thereby restricting the area of the chaotic state spreading (see Fig.~\ref{fig:ch7:pulse_pump0}(a)). By concentrating the pump power temporally, it substantially improves the pump efficiency~\cite{obrzud2017TemporalSolitonsMicroresonators,li2022EfficiencyPulsePumped} enables broadband spectra~\cite{anderson2021PhotonicChipbasedResonant}, and can mitigate thermal effects as discussed in Sec.~\ref{sec:exp_DKS:Res_thermal}.
Isolated pulses may be used to write and erase DKS inside the cavity~\cite{wang2018AddressingTemporalKerr}. Moreover, pulsed driving of a $\chi^{(2)}$-nonlinear microresonator has also been used for spectral up-and down conversion of the driving pulses~\cite{herr2018FrequencyCombDownconversion}.

\subsubsection{Asymmetric synchronization between microcombs}
\begin{figure}[h]
\includegraphics[width=1.0\columnwidth]{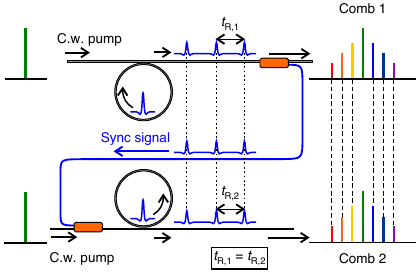}
\caption{ \textbf{Synchronization of coupled microcombs} Both microcombs share the same pump laser and, by injecting comb 1 into comb 2, their repetition rates (or temporal pulse intervals $t_\mathrm{R,1/2}$) synchronize. From~\cite{jang2018SynchronizationCoupledOptical}. \label{fig:sync_DKS}} 
\end{figure}

When two microcombs, whose repetition rates are integer multiples/fractions of the other, are pumped by the same pump laser, synchronization between both combs may be achieved by asymmetric injection of one comb into the other, as illustrated in Fig.~\ref{fig:sync_DKS}. This has been demonstrated for DKS and SW microcombs~\cite{jang2018SynchronizationCoupledOptical,jang2019ObservationArnoldTongues,kim2021SynchronizationNonsolitonicKerr}, as well as for, OPO injection into a DKS microcomb~\cite{zhao2024AllopticalFrequencyDivision}. The underlying mechanism is similar to the one discussed in Sec.~\ref{sec:bichromatic_pump} and \ref{sec:AM_pumping}. In accordance with Eq.~\ref{eq:momentum_force}, for the case shown in Fig.~\ref{fig:sync_DKS}, the synchronization can be described by an Adler-type equation for the waveform position $\tau_2$ in the second microresonator~\cite{jang2018SynchronizationCoupledOptical}:
\begin{equation}
\frac{\mathrm{d} \tau_{2}}{\mathrm{~d} t}=\Delta \tau-k \sin \left(2 \pi \frac{\tau_{2}}{t_{\mathrm{R}, 1}}\right),
\label{eq:ch7:adler}
\end{equation}
where $\Delta \tau$ stands for the relative drift rate per roundtrip in uncoupled microresonators, $k$ is a coupling constant, and $t_{\mathrm{R}, 1}$ is the roundtrip time of the master microresonator. Eq.~\ref{eq:ch7:adler} predicts that the synchronization can be achieved if the coupling strength exceeds the natural drift rate $\Delta \tau$. 

\subsubsection{Interaction between multiple solitons and dissipative patterns}
\label{subsubsec:interaction_dks_diss_patterns}
Even in the absence of poly-chromatic or modulated pump laser, the background field inside the cavity can be modulated through the presence of another DKS or dissipative pattern. A narrowband perturbation of the resonance spectrum (e.g. AMX) can cause a CW-like feature (increased or decreased intensity of a comb line), which has a similar effect as an secondary CW pump laser in bi-chromatic pumping (Sec.~\ref{sec:bichromatic_pump}). It creates a spatially extended oscillatory feature in the waveform inside the cavity that can lead to a long range interaction between solitons, including a locking of their temporal separation~\cite{wang2017UniversalMechanismBinding}. On some level, this effect is always present in experimental systems due to small imperfections, and explains the long term stability of the relative temporal of pulses in multi-solitons states, including soliton crystals~\cite{cole2017SolitonCrystalsKerr, karpov2019DynamicsSolitonCrystals}. 
Moreover, there can be a direct attractive or repulsive interaction of DKS in close proximity~\cite{mitschke1987ExperimentalObservationInteraction}, where the intensity slope of one DKS pulse can act as as a modulated background field, stabilizing heteronuclear molecules~\cite{weng2020HeteronuclearSolitonMolecules} and composite states~\cite{weng2020FormationCollisionMultistabilityEnabled}. The formalism introduced in Section~\ref{sec:bichromatic_pump} can be modified to describe the interaction between DKS and other dissipative patterns.

\subsubsection{Multiplexed DKS with multiple pump lasers}
\label{subsubsec:multiplexed_solitons}
\begin{figure}[h]
\includegraphics[width=1.0\columnwidth]{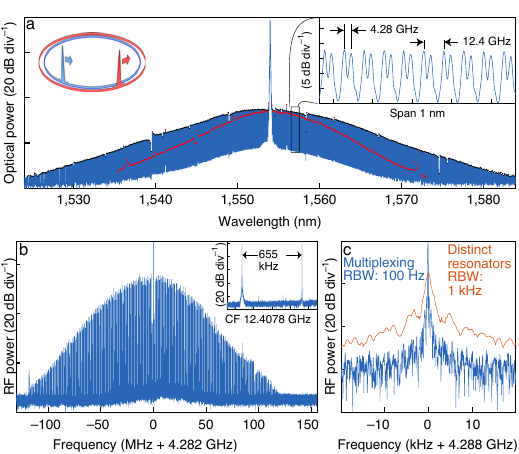}
\caption{ \textbf{Dual-comb spectroscopy with multiplexed solitons.} (a) Generated dual-comb optical spectrum from two multiplexed co-propagating DKS in a single resonator generated by two distinct pump lasers. The combs are interleaved and offset from each other. (b) Resulting dual-comb heterodyne beatnotes (Resolution bandwidth RBW: 3~kHz); CF: center frequency. (c) Focus on one line of the heterodyne signal. Adapted from~\cite{lucas2018SpatialMultiplexingSoliton}. \label{fig:multi_DKS_DCS}} 
\end{figure}

Multiple pump lasers may also be utilized to generate independent DKS within a single resonator in different mode families with the same~\cite{lucas2018SpatialMultiplexingSoliton} or different polarizations~\cite{xu2021SpontaneousSymmetryBreaking}, that can propagate in the same or opposite directions~\cite{yang2017CounterpropagatingSolitonsMicroresonators,yang2019VernierSpectrometerUsing,lucas2018SpatialMultiplexingSoliton,bao2021QuantumDiffusionMicrocavity,joshi2018CounterrotatingCavitySolitons}. As the DKS are in the same resonator, any noise resonator induced noise is highly correlated, enabling implementation of dual-comb spectroscopy with intrinsically high mutual coherence as how in Fig.~\ref{fig:multi_DKS_DCS}. Noteworthy, multiplexed DKS and SW propagating in orthogonally polarized linearly uncoupled modes can appear in single-mode fiber-based resonators. The study of this system reveals the importance of the spontaneous symmetry breaking~\cite{averlant2017CoexistenceCavitySolitons,garbin2021DissipativePolarizationDomain,xu2022BreathingDynamicsSymmetrybrokena}, and the possibility of the deterministic switching between two symmetry-broken soliton states~\cite{xu2021SpontaneousSymmetryBreaking}.

\begin{figure}[h]
\includegraphics[width=1.0\columnwidth]{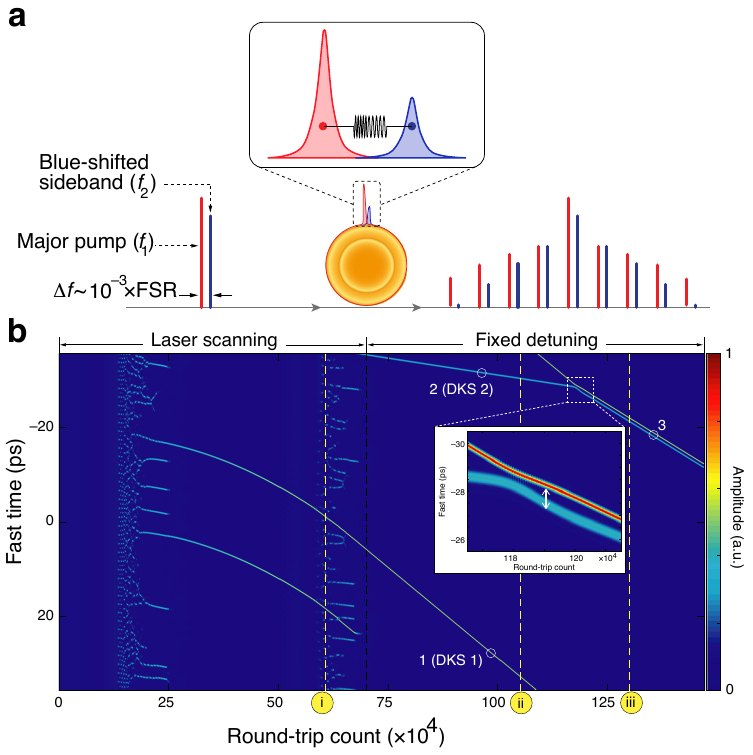}
\caption{\textbf{Generation of heteronuclear soliton molecules. } (a) Principle of the discrete pumping scheme. Closely bound dissimilar solitons are generated by pumping a single resonance with two laser fields of different frequencies simultaneously. (b) Simulated intracavity field evolution showing the formation of soliton molecules. An enlargement of the soliton binding is shown in the inset. Owing to the periodicity of the fast time axis, the solitons moving out of the map from the bottom will reappear from the top. Adapted from ~\cite{weng2020HeteronuclearSolitonMolecules}.\label{fig:ch7:sol_mol}} 
\end{figure}

\subsection{Coupled mode-families and coupled resonators}
\label{sec:coupled_mode-families_and_resonators}

Linear coupling between different mode families -- within a resonator or between resonators -- can substantially modify microcombs dynamics, enabling new functionality and phenomena. Many of the phenomena can be understood as emerging from resonance frequency shift caused by AMX (Sec.~\ref{sec:basics_lin_AMX}), however, a full model of the coupled system is needed for an accurate description. Moreover, if the coupled modes have considerable mode-overlap (e.g. they are in the same resonator), XPM between the mode-families must be considered~\cite{daguanno2016NonlinearModeCoupling,matsko2016OpticalCherenkovRadiation}. Depending on the nature of the coupled system, a time- or frequency domain description may be advantageous.

Generally, the coupling between the modes can be described by a coupling matrix. Similar to the conventional treatment of coupled oscillators, it is often convenient to introduce an orthonormal basis of normal modes or supermodes~\cite{morin2012IntroductionClassicalMechanics}, that diagonalize the coupling matrix as it has been done in Sec.~\ref{sec:basics_lin_AMX}.
Below, we consider several important examples of coupled systems that can be extended to other situations.

\subsubsection{Coupling between mode families.} 
Usually microresonators do not only support a single fundamental family of longitudinal modes but also higher order transverse mode families. If some modes of the coupled mode families are approximately frequency degenerate they can couple as discussed in Sec.~\ref{sec:basics_lin_AMX}.
Assuming that the coupling rate between the transverse modes does not depend on frequency the governing equation for two transverse mode families can be conveniently written in the time domain as~\cite{yi2017SinglemodeDispersiveWaves,guo2017IntermodeBreatherSolitonsb,yang2021DispersivewaveInducedNoise}:

\begin{equation}
\begin{aligned}
\frac{\partial A_1}{\partial t} =&  -\left(\frac{\kappa_1}{2} + i(\omega_{0}-\omega_{p}) \right)A_1  + i\frac{1}{2}D_{2,1}\frac{\partial^{2}A_1}{\partial\phi^{2}}  \\ & + i(g_{0,1}|A_1|^{2}A_1+g_{0,12}|A_2|^{2}A_1)  +i J A_2 +\sqrt{\kappa_\mathrm{ex,1}}s_\mathrm{in}
\\
\frac{\partial A_2}{\partial t} =&  -\left(\frac{\kappa_2}{2} + i(\omega_{0}-\omega_{p}) + i \delta\right)A_2  + i\frac{1}{2}D_{2,2}\frac{\partial^{2}A_2}{\partial\phi^{2}}  \\ & - \delta D_1 \frac{\partial A_2}{\partial\phi} + i(g_{0,2}|A_2|^{2}A_2+g_{0,12}|A_1|^{2}A_2)  +i J A_1,
\end{aligned}
\label{eq:ch7:coupled_LLE_AMX}
\end{equation}
where indexes 1(2) correspond to the main (auxiliary) mode families. $\delta$ - detuning between the mode families, $\delta D_1 = D_{1,2} - D_{1,1}$, $g_{0,12}$ characterizes the XPM effect and is inversely proportional to effective mode overlapped volume\footnote{The XPM effect only modulates phase but does not lead to an energy exchange between the coupled modes.}. The remaining coefficients are the same as in Eq.~\ref{eq:ch5:LLE} with corresponding indexes. 
As described in Sec.~\ref{sec:basics_lin_AMX}, the 
coupling between modes of different mode families leads to mode hybridization and mode-splitting. As the involved mode families usually differ in FSR by an amount $\Delta \mathrm{FSR}$, AMXs are usually periodic in the mode number $\mu$ and repeat every $\Delta\mu\approx\Delta \mathrm{FSR}/\mathrm{FSR}$, where $\mathrm{FSR}$ is the approximate FSR of the mode families. Moreover, as the coupling higher order mode may have a lower $Q$-factor, this may also reduce the $Q$-factor in the comb generating mode family, as describe by Eq.~\ref{eq:ch3:amx}. The probability of observing coupling between transverse modes depends strongly on the spectral density of resonances and its occurrence is reduced in resonators with larger FSR, higher-Q factor and fewer transverse modes.

\begin{figure} 
\includegraphics[width=0.85\columnwidth]{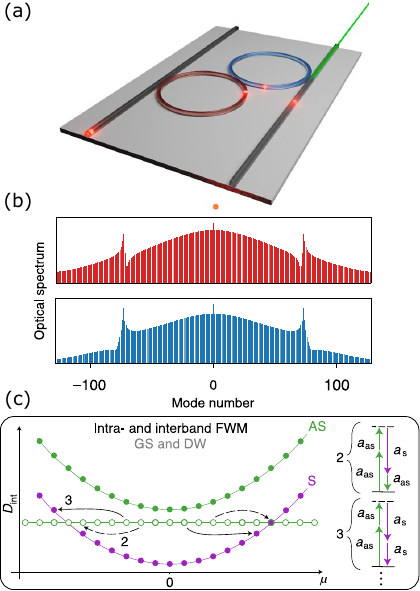}
\caption{\textbf{DKS in a photonic dimer.} (a) Schematic image of two coupled microresonators constituting the photonic dimer. (b) The modified shape of the soliton microcomb generated in the AS supermode of the dimer, displaying two symmetrically spaced sidebands, corresponding to the intersection of soliton with the lower (S supermode) parabola. (c) schematic representation of the NDR and one of the emerging FWM pathways. From ~\cite{tikan2021EmergentNonlinearPhenomena}. \label{fig:ch7:coupled_res}} 
\end{figure}

\subsubsection{Coupling between resonators. }
The configuration where two different microresonators are coupled to each other (see Fig.~\ref{fig:ch7:coupled_res}a for an example) can be described by Eqs.~\ref{eq:ch7:coupled_LLE_AMX} without XPM effects $g_{0,12}=0$ (as the mode families have almost no overlap). The detuning between the coupled mode families $\delta$ can be freely tuned by differential actuation (e.g., heating) of the resonators. Coupled resonators have also been termed `photonic molecules' to reflect their analogy with the level structure in molecules~\cite{Zhang2019,boriskina2010PhotonicMoleculesSpectral} and they have played a role in the field of topological photonics that extends this analogy to the case of solid-like arrangements with non-trivial topological properties~\cite{ozawa2019TopologicalPhotonics,mittal2021TopologicalFrequencyCombsb,tusnin2023NonlinearDynamicsKerr}. One can distinguish two cases:

\textit{Coupling between resonators of different FSR.} If resonators with different FSRs are coupled (heteronuclear photonic molecules)~\cite{helgason2021DissipativeSolitonsPhotonic,ji2023EngineeredZerodispersionMicrocombsa,yuan2023SolitonPulsePairs,xue2015NormaldispersionMicrocombsEnabled,helgason2023SurpassingNonlinearConversion}, this is formally equivalent to the case of coupling between higher-order transverse longitudinal modes without the effect of XPM and Eq.~\ref{eq:ch7:coupled_LLE_AMX} with ($g_{0,12}=0$) may be used. By modifying the ratio of the two FSRs, the mode-hybridization may be controlled~\cite{ji2023EngineeredZerodispersionMicrocombsa,pidgayko2023VoltagetunableOpticalParametric}. In the so-called Vernier configuration~\cite{gomes2021OpticalVernierEffect,boeck2010SeriescoupledSiliconRacetrack}, mode hybridization can also be arranged to affect only one mode (e.g., the pump mode), which can offer several advantages (Sec.~\ref{subsubsection:pump_mode_hybridization}).

\textit{Coupling between resonators of identical FSR.} If identical resonators are coupled (homonuclear photonic molecule or photonic dimer)\footnote{Coupled-resonator optical waveguide (CROW) has been the subject of intense investigation in the context of optical filters and delay lines~\cite{van2016OpticalMicroringResonators,morichetti2012FirstDecadeCoupled,Yariv1999,Poon2004}.}
~\cite{tikan2021EmergentNonlinearPhenomena,tikan2022ProtectedGenerationDissipative,mittal2021TopologicalFrequencyCombsb,tusnin2020NonlinearStatesDynamics}, we can further set $\kappa_{1} = \kappa_{2}$, $D_{2,1} = D_{2,2}$, $\delta D_1 = 0$.  All longitudinal modes are hybridized according to Eq.~\ref{eq:ch3:amx} with $\omega_\mathrm{diff}=\delta= \sqrt{4J^2 + \delta^2}$.

\subsubsection{Coupling between counter-propagating modes. } 

\begin{figure}[h]
\includegraphics[width=1.0\columnwidth]{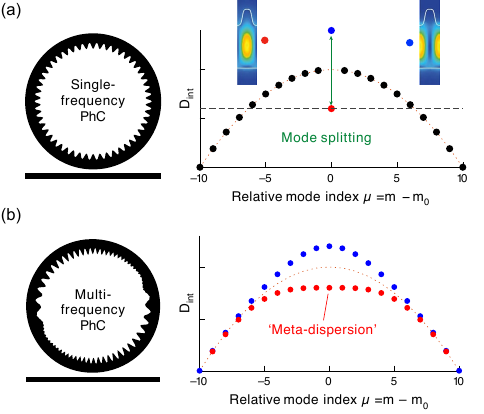}
  \caption{\textbf{Tailored dispersion in photonic crystal ring resonators.} (a) A single periodic corrugation in a photonic crystal ring resonator (PhCR) can lead to targeted mode-hybridization and mode splitting. (b) Multiple spatial corrugation frequencies can be superimposed to implement complex dispersion modification. Adapted from~\cite{lucas2023TailoringMicrocombsInversedesigned}. \label{fig:meta_disp}} 
\end{figure}

In ring (and other traveling wave) resonators, frequency degenerate counter-propagating modes exist\footnote{Coupling between counter-propagating waves is absent in standing wave Fabry-P\'erot resonators~\cite{wildi2023DissipativeKerrSolitons}.}. In the presence of frequency independent coupling $J$, the system is described by Eqs.~\ref{eq:ch7:coupled_LLE_AMX} with $\delta D_1=0$ and $g_{0,12}=2g_{0,1}=2g_{0,2}$. In contrast, a frequency domain description readily allows to include frequency dependent coupling $J_\mu$:
\begin{eqnarray}
    \begin{aligned}
        \partial_t a_\mu= & -\left(1+i \zeta_\mu\right) a_\mu+i \sum_{\nu, \eta} a_\nu a_\eta a^* _{\nu+\eta-\mu} \\ & + 2 i a_\mu \sum_\eta\left|b_\eta\right|^2  +i \frac{2}{\kappa}J_\mu b_\mu+\delta_{\mu0}f_a \\ 
        \partial_t b_\mu= & -\left(1+i \zeta_\mu\right) b_\mu+i \sum_{\nu, \eta} b_\nu b_\eta b^*_{\nu+\eta-\mu} \\ & +2 i b_\mu \sum_\eta\left|a_\eta\right|^2 +i \frac{2 }{\kappa}J_\mu a_\mu
    \end{aligned}
\end{eqnarray}
where we have assumed here that only the mode $a_0$ is driven. Frequency dependent coupling can be realized for instance in \textit{Photonic crystal ring resonators (PhCRs)}~\cite{arbabi2011RealizationNarrowbandSingle,yu2021SpontaneousPulseFormation, black2022OpticalparametricOscillationPhotoniccrystal, lu2022HighQSlowLight, lucas2023TailoringMicrocombsInversedesigned, ulanov2024SyntheticReflectionSelfinjectionlocked}. In these resonators the refractive index is modulated along the resonator length e.g., via a corrugated, width-modulated waveguide. The spatial period of this modulation phase-matches the frequency degenerate counter-propagating modes (Bragg condition). In a ring resonator coupling the counter-propagating modes with index $m$ implies a modulation with angular period of $\Lambda_\phi = \pi/m$ along the entire ring. The coupling between opposite direction modes will lead to a set of hybrid modes whose resonance frequencies are split and symmetrically arranged at $\omega_m \pm J_m$, where $J_m$ is the coupling rate. Multiple corrugation periods may be superimposed to create complex spectral coupling patterns~\cite{lucas2023TailoringMicrocombsInversedesigned}, as illustrated in Fig.~\ref{fig:meta_disp}.

\subsubsection{Effects of mode hybridization across the comb spectrum}
\label{sec:coupled_mode-families_and_resonators:AMX}
Mode hybridization from linear coupling modifies the system's dynamics and opportunities to engineer new functionality. Below we discuss several important effects:\\

\textit{(i) Dispersive waves and suppression of DKS.} Mode-hybridization and the associated resonance shifts modify the phase-matching condition in nonlinear frequency conversion. While it does not capture the full dynamics of the hybridization mode structure, a simplified model that approximates the mode frequency shifts of a single comb generating mode family  
\begin{equation}
    \omega_\mu=\omega_0+D_1 \mu+\frac{1}{2} D_2 \mu^2+\frac{a / 2}{\mu-b-0.5}.
\end{equation}
can capture some of its essential features ~\cite{herr2014ModeSpectrumTemporal}. 

\begin{figure}
\includegraphics[width=1.0\columnwidth]{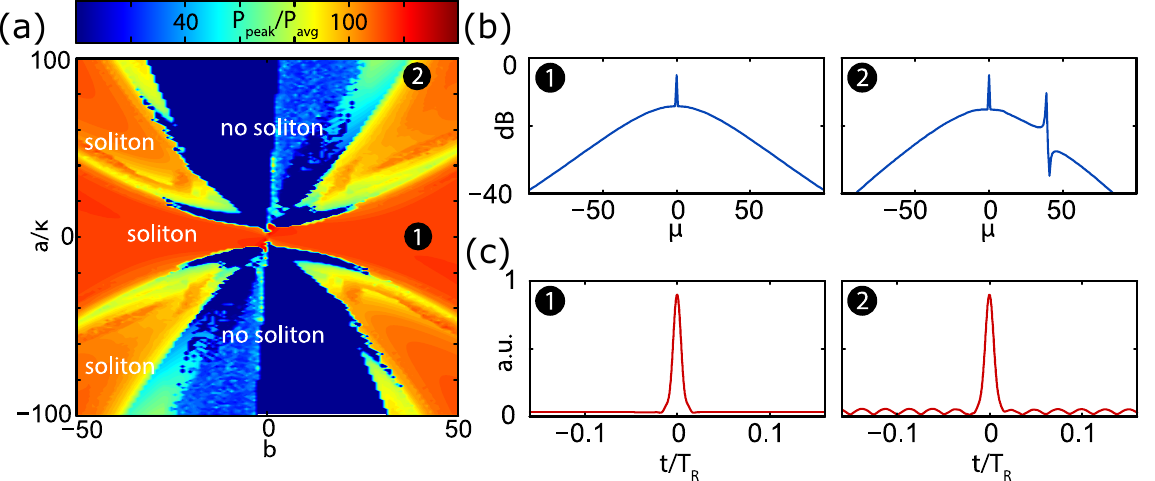}
\caption{ \textbf{Numerical investigation of soliton formation in different dispersion scenarios.} (a) The high ratio of peak to average power is used as an indicator of soliton formation for different situations characterized by an AMX that is parametrized by magnitude $a$ and distance $b$ from the pump laser. (b) Optical spectra for different simulation parameters in panel (a) (1, 2) show the characteristic``up-dow'' feature induced by an avoided mode crossing. (c) Temporal field envelope inside the microresonator corresponding to panel (b). The pulses have the same duration but oscillatory features in the background field appear in the presence of avoided mode crossings. From ~\cite{herr2014ModeSpectrumTemporal}.\label{fig:ch7:amx_solitons}} 
\end{figure}

\begin{figure}[h]
\includegraphics[width=0.99\columnwidth]{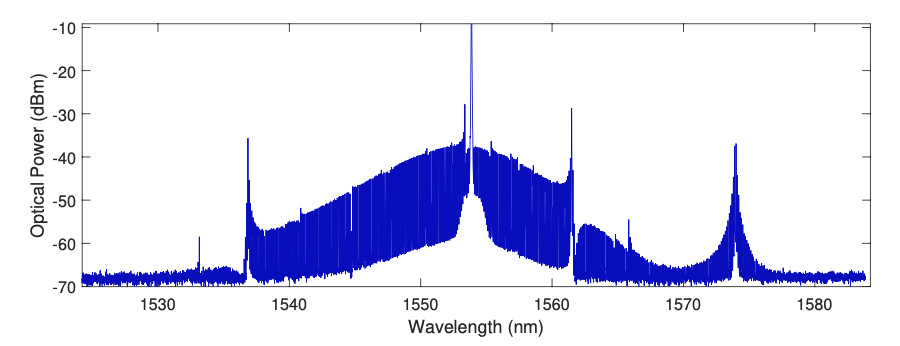}
\caption{ \textbf{AMX influence on DKS generation.} Example of an experimentally generated DKS in the presence of two avoided mode crossings. From~\cite{karpov2016RamanSelfFrequencyShift}.\label{fig:ch7:amx_experiment}} 
\end{figure}

Based on this model, Fig.~\ref{fig:ch7:amx_solitons} reveals qualitatively the characteristic modification of the soliton spectral envelope by AMX and the time domain waveform; it also shows that DKS may be suppressed if pronounced AMX occur close to the pump laser. In particular AMX, can modify the integrated dispersion $D_\mathrm{int}$ in the spectral wing of a comb, resulting in dispersive-wave like features(Sec.~\ref{sec:sec:aspects_DKS:dispersion:HOD}). In contrast to the DWs emerging from higher-order dispersion, DWs from AMXs are usually narrow (even single mode)~\cite{yi2017SinglemodeDispersiveWaves}, corresponding to extended spatially oscillatory feature; an experimetal example is shown in Fig.~\ref{fig:ch7:amx_experiment}.

Effectively a DW can play the role akin to that of a second pump laser and stabilize the temporal spacing in multi-soliton and soliton crystal states, as discussed in Sec.~\ref{subsubsec:interaction_dks_diss_patterns}). Moreover, the modified phase matching condition can also determine the mode number of the first oscillating sideband and impact the formation of soliton crystals~\cite{cole2017SolitonCrystalsKerr,he2020PerfectSolitonCrystals,karpov2019DynamicsSolitonCrystals}. By controlling for instance coupled resonators, DW can be tuned in position and strength~\cite{yang2016SpatialmodeinteractioninducedDispersiveWaves,okawachi2022ActiveTuningDispersivea}.
A full description of the hybrid modes reveals the spectral recoil of the microcomb spectrum which can be leveraged for reduced reduced noise microcombs~\cite{yi2017SinglemodeDispersiveWaves} (see Sec.~\ref{sec:ch6:noise:soliton}).

\begin{figure}[h]
\includegraphics[width=0.99\columnwidth]{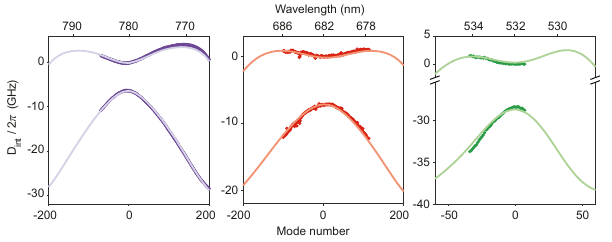}
\caption{ \textbf{Broadband dispersion modification in coupled resonators.} Measured integrated dispersion of coupled microresonators showing anomalous dispersion in one of the supermodes at wavelength of 780~nm, 682~nm, and 532~nm. Adapted from~\cite{liu2025NearvisibleIntegratedSoliton}.\label{fig:near_vis_dks}} 
\end{figure}

\textit{(ii) Broadband mode hybridization}
Mode hybridization may be utilized for broadband dispersion engineering (cf. Section~\ref{sec:basics_lin_disp_eng}). This includes the creation of anomalous dispersion in resonators with normal dispersion via coupling to higher order modes~\cite{wang2020DiracSolitonsOptical}, coupled between resonators~\cite{ji2023EngineeredZerodispersionMicrocombsa,yuan2023SolitonPulsePairs,liu2025NearvisibleIntegratedSoliton} as show in Fig.~\ref{fig:near_vis_dks}, and broadband mode-by-mode tailoring of mode-splitting in photonic crystals ring resonators~\cite{lucas2023TailoringMicrocombsInversedesigned} as shown in Fig.~\ref{fig:meta_disp}.

In a pair of identical coupled resonators (photonic dimer), DKSs can be generated in both symmetric and anti-symmetric supermodes~\cite{tikan2021EmergentNonlinearPhenomena} corresponding to two synchronously circulated DKS in both rings as shown in Fig.~\ref{fig:ch7:coupled_res}. The interaction of the 
DKS line with the symmetric supermode parabola~\cite{komagata2021DissipativeKerrSolitonsa} as schematically depicted in Fig.~\ref{fig:ch7:coupled_res}(c) yields a hyperbolic secant spectral shape perturbed by two Fano-like structures caused by the DW generation in the lower parabola. In the supermode basis, linear DW can be separated from the unperturbed DKS spectrum and AMXs exhibit an abnormal behavior~\cite{churaev2023HeterogeneouslyIntegratedLithium} due to the complex transverse mode families interaction~\cite{tikan2022ProtectedGenerationDissipative}.
More complex multi-resonator topologies have been investigated~\cite{mittal2021TopologicalFrequencyCombsb,tusnin2023NonlinearDynamicsKerr,flower2024ObservationTopologicalFrequency}).

\textit{(iii) Intermode breathers. }
The DKS line on the NDR can cross a resonance of another mode family establishing phase matching between this mode and the microcomb. The mode interaction can lead to intermode breather solitons~\cite{guo2017IntermodeBreatherSolitonsb}, where energy is periodically exchanged between the DKS and the auxiliary phase-matched mode. This effect has to be considered when designing coupled resonator systems. 

\begin{figure}[h]
\includegraphics[width=1.0\columnwidth]{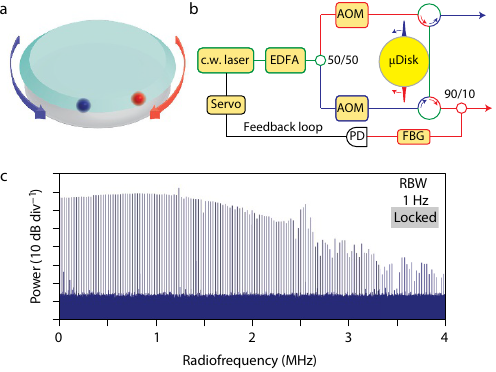}
\caption{\textbf{Counter-propagating DKS in microresonators.} (a) Schematics of a microresonator with counter-propagating solitons. (b) Experimental setup. FBG, fibre-Bragg grating; PD, photodetector; EDFA, erbium-doped-fibre-amplifier; AOMs, acousto-optic modulators. (c) High-mutual coherence dual-comb signal observed when both DKS are pumped with frequency offset lasers and two lines, one from each comb, in the wing of the combs lock together. Adapted from~\cite{yang2017CounterpropagatingSolitonsMicroresonators}. \label{fig:ch7:cw_ccw}}
\end{figure}

\textit{(iv) Synchronization between symmetrically coupled microcombs} In weakly coupled systems that each support microcomb generation on their own, synchronization between the microcombs can occur through their symmetric coupling.
For instance, this is the case for DKS counter-propagating in the same ring-type resonator and driven by a shared and bidirectionally coupled pump laser, as shown in Fig.~\ref{fig:ch7:cw_ccw}a,b; here, a coupling between both propagation direction emerges from back-scattering, resulting in a locking of the repetition rate~\cite{yang2017CounterpropagatingSolitonsMicroresonators}. A related example are counter-propagating solitons with different pump lasers and different repetition rate, where a mutual coupling and injection locking (cf. Sec.~\ref{sec:bichromatic_pump}) between two comb lines in the wing can occur, resulting in a high-mutual coherence dual-comb heterodyne signal~\cite{yang2017CounterpropagatingSolitonsMicroresonators, yang2019VernierSpectrometerUsinga} (Fig.~\ref{fig:ch7:cw_ccw}c).

The independent control over both pump lasers in bi-directionally pumped microresonators provides extended control over the DKS generation and reveals the effect of soliton blockade~\cite{fan2021ControllingMicroresonatorSolitons,fan2020SolitonBlockadeBidirectional}.

\subsubsection{Effects of pump mode hybridization}
\label{subsubsection:pump_mode_hybridization}
Hybridization of only the pump mode and the associated resonance shift have been utilized to address several shortcoming of DKS and SW microcombs. While most effects can be intuitively understood in a simplified picture of a frequency shifted pump mode, capturing the full dynamics requires solving the coupled equations.

\begin{figure} 
\includegraphics[width=1\columnwidth]{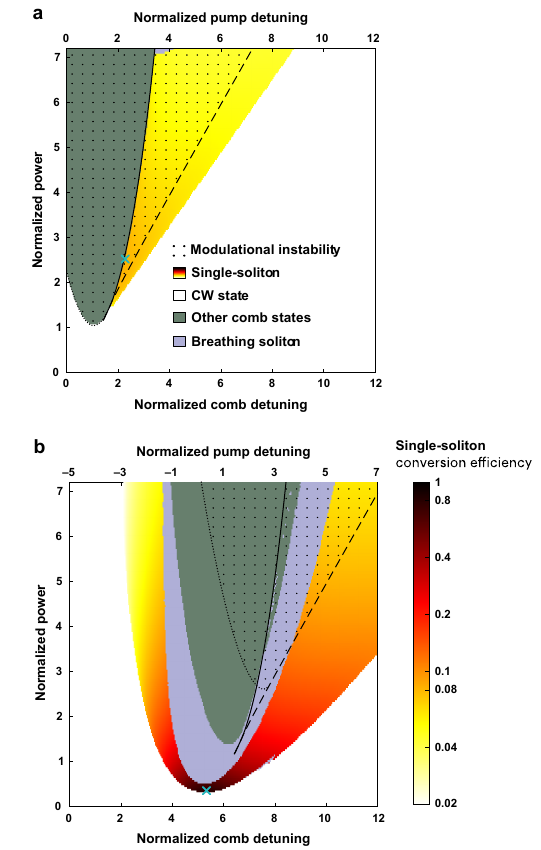}
\caption{\textbf{Efficiency enhancement in microresonators with a displaced pump mode.} (a) Single microresonator case. (b) Coupled microresonators arranged in the Vernier configuration that leads to the pump mode displacement. From ~\cite{helgason2023SurpassingNonlinearConversion}. \label{fig:ch7:coupled_res_efficiency}} 
\end{figure}

\textit{(i) High conversion efficiency DKS}
The conversion efficiency from pump laser to DKS sidebands is usually small  1-5\%~\cite{jang2021conversion,bao2014NonlinearConversionEfficiency}, due to the far detuned pump laser, which limits pump to resonator coupling~\cite{wang2016IntracavityCharacterizationMicrocomb}. Hybridization of the pump mode permits to shift one of the hybrid modes resonance frequencies closer to the pump laser position or even into an effectively blue-detuned pumping regime. This drastically increases the conversion efficiency and, for instance, values of 50\% have been observed in a 100 GHz photonic molecule~\cite{helgason2023SurpassingNonlinearConversion}. Figure~\ref{fig:ch7:coupled_res_efficiency} demonstrates the theoretical prediction of the conversion efficiency for a conventional microring resonator and the resonator with a displaced pump mode. 

\textit{(ii) Suppression of non-solitonic states and deterministic single-soliton generation}
When the pumped mode is hybridized and the pump laser is scanned across the red hybrid resonance, the parametric threshold is effectively reached at a higher detuning (with regard to the unshifted mode, and the set of all other modes which define the dynamics). If this higher detuning falls within the soliton existence range a transient MI will directly evolve into the DKS attractor and lead to DKS pulse formation without entering MI or chaotic states~\cite{yu2021SpontaneousPulseFormation,bao2017SpatialModeinteractionInduced}. The mechanism of spontaneous pulse formation can greatly reduce the challenge associated with thermal effects when tuning into a DKS state, however, Kerr-nonlinear or thermal shadowing effects (Sec.~\ref{sec:ch6:exp_DKS:thermal_shadowing}) from the blue-shifted hybrid mode may complicate access to the red-shifted hybrid resonance. This challenge can be overcome via SIL~\cite{ulanov2024SyntheticReflectionSelfinjectionlocked} (Sec.~\ref{sec:exp_DKS:low_noise_ucombs:SIL}).
The red-shifted hybrid mode corresponds to strong anomalous dispersion. If strong enough it will force the first oscillating MI sideband to be adjacent to the pump \cite{herr2012UniversalFormationDynamics}, seeding deterministically a \textit{single} DKS pulse. This is the case when
\begin{align}
    \frac{\gamma}{\kappa} > \frac{f^2}{8}
\end{align}
Due to the pump mode hybridization, not all power is available in the soliton generating mode family. This implies an increase of the threshold power, which can be estimated as
\begin{align}
    f^2_\mathrm{th} = 4\frac{\gamma}{\kappa} + \frac{\kappa}{\gamma}
\end{align}
where $\gamma > \kappa/2$ was assumed \cite{ulanov2024SyntheticReflectionSelfinjectionlocked}.

\textit{(iii) Initiation of normal dispersion combs. }
A sufficiently split hybridized pump mode provides locally anomalous dispersion. This permits in a deterministic manner the local anomalous dispersion needed to initiate MI and SW microcombs in the otherwise normal dispersion regime requires; it also enables (stable) effectively blue detuned operation, as discussed in Sec.~\ref{sec:switching_waves}~\cite{helgason2021DissipativeSolitonsPhotonic,kim2019TurnkeyHighefficiencyKerr,xue2015NormaldispersionMicrocombsEnabled,yu2022ContinuumBrightDarkpulse}.

\subsection{Coupled pump system and nonlinear microresonator}

In most demonstrations, the pump system is an external laser that is isolated from the microresonator. The schemes discussed below are in stark contrast to this paradigm and a bi-directional coupling between an actively lasing resonator and the nonlinear microresonator is essential.

\subsubsection{Laser self-injection locking}
\label{sec:exp_DKS:low_noise_ucombs:SIL}

\begin{figure*}
\includegraphics[width=1.0\textwidth]{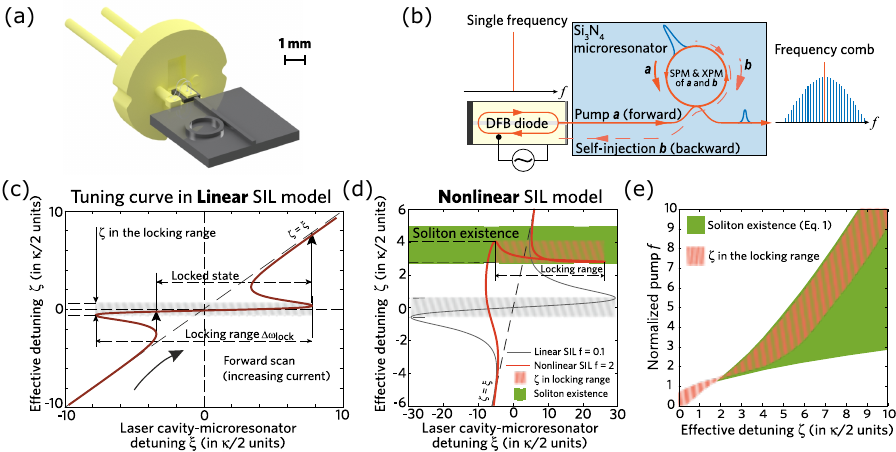}
\caption{\textbf{Self-injection locking of DKS.} (a) Illustration of the soliton microcomb device via direct butt coupling of a laser diode to the $\mathrm{Si}_3 \mathrm{~N}_4$ chip. (b) Principle of laser self-injection locking. The DFB laser diode is self-injection locked to a high- $Q$ resonance via Rayleigh backscattering and simultaneously pumps the nonlinear microresonator to generate a soliton microcomb. (c) Schematic of the self-injection locking dynamics without taking into account the microresonator nonlinearity, i.e., linear SIL model. (d) Nonlinear SIL model coincides with the linear one for low pump powers f<1, but the tuning curve changes significantly at higher pump power f >1 and shifts up. (e) The model predicts that attainable $\xi$ values in the SIL regime are red-detuned and located inside the soliton existence range. From~\cite{voloshin2021DynamicsSolitonSelfinjection}.\label{fig:sil1}}
\end{figure*}

Another, example of the extension of the microcombs dynamics is the SIL (self-injection locking) regime~\cite{kondratiev2023RecentAdvancesLaser} - coupling of a laser diode pump cavity and a high-Q Kerr microresonator\footnote{We note that another approach of Kerr comb generation using an external active media known as laser cavity soliton microcomb has been actively developed~\cite{baoLaserCavitysolitonMicrocombs2019}.} as depicted in Fig.~\ref{fig:sil1}(a,b). Several factors make the microresonator pumped by SIL laser an exceptional experimental technique for the microcomb generation that include pump laser linewidth reduction~\cite{kondratiev2017SelfinjectionLockingLaser,vasilev1996HighcoherenceDiodeLaser,liang2010WhisperinggallerymoderesonatorbasedUltranarrowLinewidth, pavlov2018NarrowlinewidthLasingSolitona}, frequency pulling~\cite{lihachev2022LownoiseFrequencyagilePhotonic,snigirev2023UltrafastTunableLasers,corato-zanarella2023WidelyTunableNarrowlinewidth}, and facilitated access to the coherent frequency comb state in both normal~\cite{wang2022SelfregulatingSolitonSwitching,lihachev2022PlaticonMicrocombGeneration} and anomalous dispersion cases~\cite{voloshin2021DynamicsSolitonSelfinjection,shen2020IntegratedTurnkeySoliton,xiang2021LaserSolitonMicrocombs,ulanov2024SyntheticReflectionSelfinjectionlocked}, including octave-spanning frequency combs~\cite{briles2021HybridInPSiNa}.

In the presence of the feedback, the emission frequency of the pump diode $\omega_\mathrm{p}$ (that follows the resonance frequency of the laser cavity $\omega_\mathrm{L}$ linearly in the absence of the feedback) is locked to the resonance frequency of the microresonator $\omega_0$. Thus, the effect of the laser cavity noise and fluctuations on the emitted light frequency is significantly reduced. 

The full description of the SIL requires considering a set of equations that include the coupled LLEs for the forward and backward propagating modes~\cite{yang2017CounterpropagatingSolitonsMicroresonators,kondratiev2020ModulationalInstabilityFrequency,skryabin2020HierarchyCoupledMode} and a laser diode equations often taken in the form of the Lang-Kobayashi equation with a resonant feedback~\cite{lang1980ExternalOpticalFeedback}.
Laser and microresonator equations are coupled via the following input-output relations:
\begin{equation}
    \begin{aligned}
s_{\mathrm{in}}&=i \sqrt{\gamma} \sqrt{T} \exp \left(i \varphi_{\mathrm{B}}\right) A_{\mathrm{L}} \\
s_{\mathrm{L, in}}&=i \sqrt{\kappa_{\mathrm{ex}}} \sqrt{T} \exp \left(i \varphi_{\mathrm{B}}\right) \overline{A_{\mathrm{B}}},
\label{eq:ch7:SIL_inout}
\end{aligned}
\end{equation}
here $A_{\mathrm{B}}$ and $A_\mathrm{L}$ are the slowly-varying field envelops in the backward propagating microresonator modes and the laser respectively, $\varphi_\mathrm{B}$ is the phase of the backscattered signal, $T$ is the power transmission coefficient on the feedback waveguide, $\gamma$ is the laser cavity loss rate considered to be mainly given by the output mirror of the diode.

However, in most cases, the SIL equations can be substantially simplified. The key assumptions are weak backscattering (i.e., no resonance splitting) and the sufficiently rapid relaxation dynamics of the laser, allowing the laser power to follow the external input resulting from backscattering. 
These assumptions lead to a simplified system that includes the conventional LLE (Eq.~\ref{eq:ch5:LLE}) with the pump term given by Eq.~\ref{eq:ch7:SIL_inout} for the forward propagating field envelope in the microresonator, together with coupled pump mode backward propagating equation and laser field phase equation~\cite{wang2022SelfregulatingSolitonSwitching}:
\begin{equation}
    \begin{aligned}
\frac{\partial A_{\mathrm{F}}}{\partial t} & =-\left(\frac{\kappa}{2}+i \delta \omega\right) A_{\mathrm{F}}+i \frac{D_2}{2} \frac{\partial^2 A_{\mathrm{F}}}{\partial \phi^2} +i g_\mathrm{K}\left|A_{\mathrm{F}}\right|^2 A_{\mathrm{F}}\\
& -\sqrt{\kappa_{\mathrm{ex}} \gamma T} A_{\mathrm{L}} e^{  i \varphi_{\mathrm{B}}} \\
\frac{d \overline{A_{\mathrm{B}}}}{d t} &=  -\left(\frac{\kappa}{2}+i \delta \omega\right) \overline{A_{\mathrm{B}}}+2 i g_\mathrm{K} \overline{A_{\mathrm{B}}} \,  \overline{\left|A_{\mathrm{F}}\right|^2} +i \overline{\Gamma}_0 \, \overline{A_{\mathrm{F}}}  \\
\delta \omega & =\delta \omega_{\mathrm{L}} -\operatorname{Im}\left[\left(1-i \alpha_{\mathrm{g}}\right) \sqrt{\kappa_{\mathrm{ex}} \gamma T} e^{i \varphi_{\mathrm{B}}} \frac{\overline{A_{\mathrm{B}}}}{A_{\mathrm{L}}}\right],
\end{aligned}
\end{equation}
where $\delta\omega =  \omega_0-\omega_\mathrm{p}$, $\delta\omega_\mathrm{L} = \omega_0-\omega_\mathrm{L}$. $|\overline{\Gamma}_0|$ is the averaged scattering coefficient, $\alpha_g$ is the phase-amplitude coupling (Henry) factor.

In the case when the microresonator is considered as a linear low-loss reflector and the reflection is due to a weak backscattering, the corresponding diode laser \textit{tuning curve} is shown in Fig.~\ref{fig:sil1}(c). Here the normalized detunings are defined as  $\zeta_0 = 2\delta\omega/\kappa$ and $\xi_0 = 2\delta\omega_\mathrm{L}/\kappa$.
This effect occurs over a range of frequencies called the \textit{locking range} that can be estimated as~\cite{kondratiev2017SelfinjectionLockingLaser}:
\begin{equation}
    \frac{\Delta \omega_{\text {lock }}}{\omega_0} \approx \sqrt{1+\alpha_g^2} \frac{\beta}{Q_d}.
    \label{eq:ch7:SIL_locking_bw}
\end{equation}
The linewidth reduction factor in this case is expressed as:
\begin{equation}
    \frac{\delta \omega_{\text{lock}}}{\delta \omega_{\text {free }}} \approx \frac{Q_d^2}{Q_0^2} \frac{1}{16 \beta^2\left(1+\alpha_g^2\right)}.
    \label{eq:ch7:SIL_lw_reduction}
\end{equation}
Here $\Delta \omega_{\text{lock}}$ is the locking bandwidth, $Q_d$ and  $Q_0$ are quality factors of the laser cavity and a microresonator, respectively. $\beta = 2|\overline{\Gamma}_0|/\kappa \ll1$ is the normalized coupling coefficient,  $\delta \omega_{\text{lock}}$ and $ \delta \omega_{\text {free}}$ are linewidth in the SIL and the free-running regimes, respectively.
The linewidth narrowing in the SIL regime can exceed 3~orders of magnitude. For this reason, SIL is widely used for the fabrication of compact hybrid and heterogeneously chip-integrated narrow-linewidth lasers used for the frequency comb generation~\cite{xiang2021LaserSolitonMicrocombs}, some of them reaching sub-Hz regime~\cite{jin2021HertzlinewidthSemiconductorLasersa}.

It has been shown that the tuning curve (see Fig.~\ref{fig:sil1}(c)) is given by~\cite{voloshin2021DynamicsSolitonSelfinjection,kondratiev2017SelfinjectionLockingLaser} 
\begin{align}
    \begin{split}
\label{ch6:eq:SIL_ring_tuningcurve_0}
        \xi = \zeta + \frac{K_0}{2} \frac{2\zeta \cos{\varphi_\mathrm{lock}}+(1-\zeta^2)\sin{\varphi_\mathrm{lock}}}{(1+\zeta^2)^2},
    \end{split}
\end{align}
where $K_0 = 8 \eta \beta\gamma T \sqrt{1+\alpha_g^2}/ \kappa$ is the SIL stabilization coefficient and $\varphi_\mathrm{lock} = 2\varphi_\mathrm{B} - \mathrm{arctan}(\alpha_g)$ is the SIL phase. Factor 2 accounts for the roundtrip phase. Often a constant phase is added to $\varphi_\mathrm{lock}$.

The nonlinear tuning curve (Fig.~\ref{fig:sil1}(d), red) follows an expression similar to Eq.~\ref{ch6:eq:SIL_ring_tuningcurve_0} with $\xi$, $\zeta$, $\varphi_\mathrm{lock}$, and $\beta$ replaced by their nonlinear counterparts~\cite{voloshin2021DynamicsSolitonSelfinjection}.
Figure~\ref{fig:sil1}(d) demonstrates an important difference between the linear and nonlinear tuning curves. The nonlinear SIL locking range has a significant overlap with the DKS existence range (DKS range) (see Sec.~
\ref{sec:basics_kcomb:DKS} for details) in laser detuning, as shown in Fig.~\ref{fig:sil1}(e). 

\begin{figure}
\includegraphics[width=1.0\columnwidth]{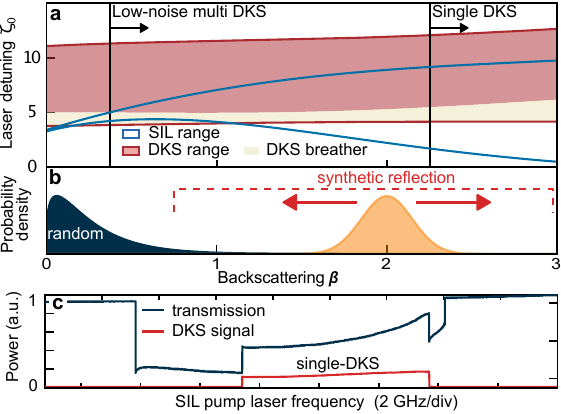}
\caption{\textbf{SIL with controlled backscattering strength.}  (a) Simulated SIL and DKS range indicating that a minimal backscattering $\beta$ is required to access the DKS range. Sufficiently large $\beta$ results in single DKS formation. (b) Random imperfection based backscattering can be insufficient to provide access to DKS. Through a deliberate synthetic back-reflection, reliable access to the DKS regime may be achieved. (c) Characteristic SIL resonance shape, observed while increasing the pump diode laser current (here $\beta=2.67$). Blue: coupling waveguide transmission; Red: transmission with pump laser filtered out, indicating single DKS formation (confirmed via microwave beatnote detection; not shown). Adapted from~\cite{ulanov2024SyntheticReflectionSelfinjectionlocked}.}\label{fig:synth_reflection_SIL}
\end{figure}

Accessing a single DKS on demand requires achieving a large positive detuning ($\zeta_0$). When the backscattering coefficient is sufficiently large, it is possible to achieve an overlap between the detuning range of the DKS and the SIL detuning range, as shown in Fig.~\ref{fig:synth_reflection_SIL}a. The extent of the locking range can be estimated by numerically solving Eq.~\ref{ch6:eq:SIL_ring_tuningcurve_0}.
For this purpose, the laser diode-microresonator coupling, which occurs naturally through weak resonant feedback caused by a back-scattering inside the microresonator from material defects, etching roughness, inhomogeneities of the refractive index, etc.)~\cite{gorodetsky2000RayleighScatteringHighQ}, can be enhanced by adding a loop mirror ~\cite{siddharth2022UltravioletPhotonicIntegrated,corato-zanarella2023WidelyTunableNarrowlinewidth} or by using a corrugated microresonator (Fig.~\ref{fig:synth_reflection_SIL}b)~\cite{ulanov2024SyntheticReflectionSelfinjectionlocked}. A characteristic SIL resonance shape is shown in Fig.~\ref{fig:synth_reflection_SIL}c.

\subsubsection{Microresonators with broadband active gain media}
\textit{Microresonator nested in broadband gain loop.} Microresonators have been nested in longer gain cavities.
Here they act both as filter and as a nonlinear device. This configuration has been explored to generate microcombs without a CW pump laser~\cite{peccianti2012DemonstrationStableUltrafast, pasquazi2012StableDualMode, johnson2014MicroresonatorbasedCombGeneration}. When the roundtrip time of temporal structures in the microresonator and the gain loop is matched by an adjustable delay-line, then stable Turing patterns~\cite{bao2020TuringPatternsFiber} as well as self-starting \textit{laser cavity solitons}~\cite{baoLaserCavitysolitonMicrocombs2019, rowley2022SelfemergenceRobustSolitons} can emerge. The system is described by an LLE equation for the microresonator driven, where the pump term consists of a set of supermodes that experience gain in the fiber section. 

\textit{Microresonators with broadband intracavity gain. } Broadband active gain media may also be included in the nonlinear cavity, blurring the boundaries between coherently driven passive and active system. Operating below the lasing threshold and driven by an external CW laser, such systems can give rise to \textit{active cavity solitons}.  Benefiting from the additional gain mechanism, they reach high peak power in the presence of a comparatively weak CW background. A generalized LLE with gain also describes the nonlinear dynamics in coherently driven quantum-cascade lasers below threshold~\cite{columbo2021UnifyingFrequencyCombs}, including the observation of soliton pulses~\cite{kazakov2025DrivenBrightSolitons}.

\section{Thermal effects and noise}
\label{sec:exp_DKS}
This section discusses thermal effects and noise that are relevant to the experimental generation of microresonator frequency combs. 

\subsection{Thermal effects in microresonators}
\label{sec:exp_DKS:Res_thermal}
Thermal effects may be used to tune the resonance frequencies of microresonators e.g., via electric heaters~\cite{joshi2016ThermallyControlledComb}, and  they also enable a self-stabilizing ``thermal lock'' of the resonator to the pump laser (Section~\ref{sec:thermal_bistab_lock}). However, they can also complicate the generation of microcombs as detailed in Section~\ref{sec:ch6:exp_DKS:thermal_shadowing}. Even for weakly-absorbing low-loss resonators, the high-circulating power levels can increase the resonator temperature. Here, we 
show how the description of Section~\ref{sec:basics_kcomb:models_of_OFC} can be extended to include thermal effects. The important topic of thermo-refractive noise will be described in Section~\ref{sec:ch6:noise:trn}.

\subsubsection{Temperature dependence of the resonance frequencies}
\label{sec:thermal_res_shift}
A temperature change $\Delta T$ of the microresonator from a base temperature induces a shift $\Delta \omega_\mu$ of the \textit{cold} resonance frequency $\omega_\mu$ via thermo-refractivity and thermal expansion~\cite{jiang2020OptothermalDynamicsWhisperinggallery}
\begin{equation}
    \label{eq:thermal_res_shift}
    \Delta \omega_\mu = -\omega_\mu \left(\frac{1}{n}\frac{\mathrm{d}n}{\mathrm{d}T}  + \frac{1}{L}\frac{\mathrm{d}L}{\mathrm{d}T}\right) \,  \Delta T
\end{equation}
so that the \textit{hot} resonance frequency is $\omega^{T}_\mu = \omega_\mu + \Delta \omega_\mu$.
Here, $1/n\cdot\mathrm{d}n/\mathrm{d}T$ is the thermo-refractive coefficient and $1/L\cdot\mathrm{d}L/\mathrm{d}T$ the coefficient of thermal expansion, for which we have neglected any chromatic dispersion.
Already small relative shifts of the resonance frequency can become comparable or larger than the cavity linewidth $\kappa$ and can hence drastically influence the detuning between pump laser and hot resonance frequency.

\begin{table}[b]
\caption{Thermo-refractive and thermal expansion coefficients (in K$^{-1}$) for key integrated photonics materials. Sources: Si: \cite{komma2012, okada1984}.  
Si$_3$N$_4$: \cite{arbabi2013MeasurementsRefractiveIndices, kaushik2005WaferlevelMechanicalCharacterization}.  
SiO$_2$: \cite{rego2023TemperatureDependenceThermoOptic,bachmann1988ThermalExpansionCoefficients}.  
LiNbO$_3$: \cite{browder1977ThermalExpansionData, moretti2005TemperatureDependenceThermooptic}. 
GaP: \cite{cheng2025CharacterizingThermoopticCoefficient, wang2024ReviewGalliumPhosphide}.  
Ta$_2$O$_5$: \cite{wu2019TantalumPentoxideTa2O5, yoon2005ComparisionResidualStress}.  
Diamond: \cite{sato2002ThermalExpansionHigh, yurov2017NearinfraredRefractiveIndex}. 
AlN: \cite{watanabe2008TemperatureDependenceRefractive}. \label{tab:ch6:thermal_coeff} }
\begin{ruledtabular}
\begin{tabular}{|c|c|c|}
\hline \text{Material} & $1/n \cdot \mathrm{d}n/\mathrm{d}T$ 10$^{-5}$ & $1/L \cdot \mathrm{d}L/\mathrm{d}T$ 10$^{-6}$ \\
\hline Ta$_2$O$_5$            & 0.3 & 4.7\\
\hline SiO$_2$  & 0.6 & 0.5\\
\hline Diamond                & 0.6 & 1.6\\
\hline AlN                    & 1.1 & 4.2\\
\hline Si$_3$N$_4$ (LPCVD)    & 1.2 & 2\\
\hline LiNbO$_3$ (e)     & 1.6 & 4\\
\hline GaP                    & 3.9 & 5.3\\
\hline Si        & 5.2 & 2.4\\
\hline
\end{tabular}
\end{ruledtabular}
\end{table}

Table~\ref{tab:ch6:thermal_coeff} lists the coefficients of thermo-refractivity and thermal expansion for common optical materials. In most listed materials $1/n\cdot\mathrm{d}n/\mathrm{d}T$ and $1/L\cdot\mathrm{d}L/\mathrm{d}T$ are of the same positive sign, however, in some materials, or for some temperatures, $1/n\cdot\mathrm{d}n/\mathrm{d}T$ can be negative. This can be exploited to reduce the thermal shifts~\cite{guha2013AthermalSiliconMicroring, djordjevic2013CMOScompatibleAthermalSilicon, lopez-rodriguez2025MagicSiliconDioxide}. 

In contrast to the (almost) instantaneous Kerr-effect (self- and cross-phase modulation), thermally induced resonance shifts occur on much longer time scales that are defined by the geometry-dependent heating (through absorption) and heat dissipation. If competing resonance shifting effects with different time constants are present (e.g. Kerr-nonlinear resonance shifts), thermal instability and oscillation can occur~\cite{he2009OscillatoryThermalDynamics,wang2013MidinfraredOpticalFrequency}.

A change in the resonance frequencies as described by Eq.~\ref{eq:thermal_res_shift} also implies a similar relative change in the free-spectral range.
The small relative change in $D_1$ (and potentially higher order dispersion coefficients) can be leveraged to control the repetition rate of a microcomb (Section~\ref{sec:stabilization}), however, it will not usually have a qualitative impact on the dynamics of the microresonator. Therefore, to describe the resonator dynamics within the framework derived in Sections~\ref{sec:basics_lin} to \ref{sec:aspects_DKS} we introduce a normalized temperature fluctuation
\begin{equation}
\label{eq:ch6:norm_thermshift}
\Theta = \frac{2 \omega_0}{\kappa}\left(\frac{1}{n}\frac{\mathrm{d}n}{\mathrm{d}T}  + \frac{1}{L}\frac{\mathrm{d}L}{\mathrm{d}T}\right) \Delta T
\end{equation}
evolving according to 
\begin{equation}
\label{eq:ch6:temp_dynamics}
        \frac{\partial \mathrm{\Theta}}{\partial \tau} =\frac{2}{\kappa \,\tau_T}\left(g_T\sum_\mu\left|a_\mu\right|^2-\Theta\right)
\end{equation}
and with the steady state
\begin{equation}
    \label{eq:ch6:steady_state_temp}
    \Theta = g_T \sum_\mu \left|a_\mu\right|^2
\end{equation}
where $\tau_T$ is the thermal relaxation time (heat dissipation), which is proportional to the heat capacity and inversely proportional to the heat conductivity~\cite{carmon2004DynamicalThermalBehavior}, and $g_{T}$ is the coefficient of \textit{thermal nonlinearity} that lumps together the absorptive heating dynamics of the resonator\footnote{This definition is in analogy to $g_\mathrm{K}$ in Eq.~\ref{eq:eff_resonance} causing resonance shifts proportional to $|a|^2$ via the Kerr-effect. }. Accurate values for $\tau_T$ and $g_T$ may be obtained through finite element modeling, or experimentally~\cite{gao2022ProbingMaterialAbsorptiona}.

Thermal effects can be included in the formalism developed in Section~\ref{sec:basics_kcomb:models_of_OFC} by replacing the detuning $\zeta_0$ by the \textit{hot} detuning $\zeta_0^T$:
\begin{equation}
\label{eq:ch6:thermal_detuning}
    \zeta_0^T = \zeta_0 - \Theta
\end{equation}
and solving the modified CMEs or LLE equations along with Eq.~\ref{eq:ch6:temp_dynamics}.

\subsubsection{Thermal triangle, thermal bistability and thermal locking}
\label{sec:thermal_bistab_lock}
Thermal resonance shifts are commonly observed when scanning a pump laser across a resonance. Assuming a positive $dn/dT$, the resonance takes on a triangular shape (\textit{thermal triangle}) when the laser is scanned from blue- to red-detuning. In the opposite scan direction, from red- to blue-detuned, the same thermal effect causes a compressed resonance shape, which can only reach a stable equilibrium for low power levels.
This is an immediate consequence of the thermal resonance shift and analogous to the resonance shape observed due to the Kerr-effect as shown in Fig.~\ref{fig:ch4:Bistability}. In analogy to the Kerr-induced bistability discussed in Section~\ref{sec:basics_nln:bistab}, there is also a thermal bistability where for certain detuning $\zeta_0$ two different power levels are possible. Which solution is realized depends on the scanning direction of the laser. The thermally induced resonance shift can by far exceed the Kerr-induced resonance shift and, due to its slower dynamics, is usually dependent on the rate with which the laser is scanned.

The interplay between temperature and effective detuning (Eq.~\ref{eq:ch6:thermal_detuning}) of the pump laser can cause a stable feedback mechanism \cite{carmon2004DynamicalThermalBehavior} effectively stabilizing the detuning $\zeta_0^T$ against fluctuations in $\zeta_0$ (e.g. from fluctuations in $\omega_\mathrm{p}$).
This \textit{thermal locking} occurs as long as the pump laser is effectively blue-detuned and is strongest where the resonance slope is steepest. It is an essential mechanism for the practical operation of nonlinear microresonators. Similar locking would be achievable with a red-detuned pump laser if $n_{2T}$ is negative, however, in this case thermal instability and oscillation can occur due to competing resonance shifts (cf.~\ref{sec:thermal_res_shift}). Importantly, the concept of thermal locking applies generally to any position in a complex resonance shape (such as those characteristic of DKS), where $\partial \sum_\mu |a_\mu|^2/\partial \zeta^T_0>0$, for the same reasons as described above. In particular, this implies that thermal locking is also possible on the \textit{soliton step feature}.

\subsubsection{Access to soliton states in the presence of thermal effects}
\label{sec:ch6:exp_DKS:thermal_shadowing}
In the absence of thermal effects, DKS can be accessed by setting $\zeta_0$ within the DKS existence range (Eq.~\ref{eq:ch5:sol_ex_range}), after transiting through a state that breaks the stability of the CW waveform (usually via MI when tuning the laser from blue to red), as shown in Fig.~\ref{fig:ch5:Tuning_LLE}. 

In contrast, in an experimental implementation, thermal effects can hinder access to the DKS regime when one attempts to tune the laser into the resonance (from blue- to red-detuned), by continuously increasing $\zeta_0$. 
As the intracavity power in a DKS precursor state is usually significantly higher than in a DKS state, the power drop upon transitioning into a DKS state will cause a cooling and an increase of the $\zeta^T_0$. Once the system reaches thermal equilibrium $\zeta^T_0$ can exceed the DKS existence range~\cite{herr2014TemporalSolitonsOptical}.

\begin{figure}[h]
\includegraphics[width=\columnwidth]{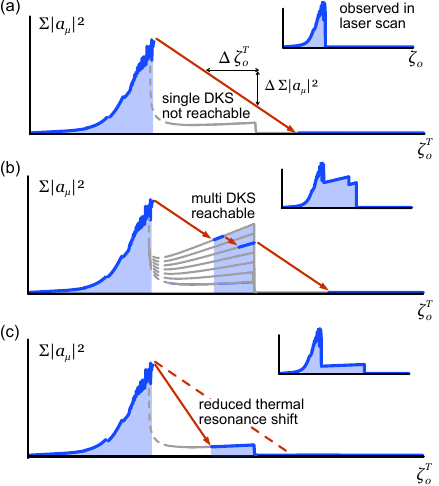}
\caption{\textbf{Thermal shadowing.} (a) Intracavity power as function of the effective detuning $\zeta_\mathrm{0,eff}$ showing CW/MI state (black) and DKS states (blue) (b) When thermal effects are present, DKS states may be shadowed by CW/MI states that can exist at the same experimentally set laser frequency with detuning $\zeta_0$} \label{fig:thermal_shadowing}
\end{figure}

This can be visualized~\cite{stone2018ThermalNonlinearDissipativeSoliton} by considering a resonance shape as a function of $\zeta^T_0$ as shown in Fig.~\ref{fig:thermal_shadowing}, where the increase in $\zeta^T_0$ due to the power drop is indicated by a red line. Only if this line intersects with the line describing the DKS state, the soliton regime be accessed stably (i.e. in thermal equilibrium); otherwise there is a ``thermal shadowing'' effect. Multi-DKS states are thermally easier to access (Figure~\ref{fig:thermal_shadowing}b), which explains, in part, thir prevalence in experiments. 
Reducing thermal shadowing is thus key to accessing DKS when working with an external laser that is to be tuned into a DKS state (Figure~\ref{fig:thermal_shadowing}c). This highlights the importance of low thermo-refractivity of resonator materials and low-absorption high-purity materials~\cite{pfeiffer2018UltrasmoothSiliconNitride}. 

\begin{figure}[h]
\includegraphics[width=0.9\columnwidth]{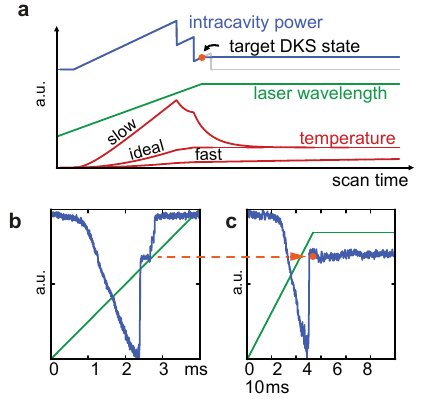}
\caption{\textbf{Rapid laser scan.} (a) Time evolution of intracavity power (blue) with target DKS state, laser wavelength (green), and temperature (red) for a laser scans that ideal, as well as fast/slow in comparison to the thermal time constant of the system. (b) Left: transmission signal when scanning over resonance, revealing ``soliton step''. Right: Experimental implementation of DKS initiation via rapid laser scan~\cite{herr2014TemporalSolitonsOptical}.}
\label{fig:ch6:rapid_scan}
\end{figure}

In addition, to mitigate thermal effects, several methods have been developed:

\textit{(i) Rapid laser actuation.} If the pump laser is scanned rapidly across the resonance into the DKS state, then the build-up of $\Theta$ can be minimized. Ideally the scan speed is chosen such that the DKS state is reached with the resonator being at the equilibrium temperature for that state, as illustrated in Figure~\ref{fig:ch6:rapid_scan}).
Instead of rapidly scanning the laser, the microresonator resonance frequency may be scanned e.g., through electric heaters~\cite{joshi2016ThermallyControlledComb} or pump power modulation~\cite{brasch2016PhotonicChipBased}. Moreover, feedback schemes, including the \textit{Pound-Drever-Hall} (PDH) technique, ~\cite{black2001IntroductionPoundDrever} may be employed in conjunction with fast laser actuation to stabilize the DKS state~\cite{weng2020HeteronuclearSolitonMolecules, yi2016ActiveCaptureStabilization}.

\textit{(ii) Auxiliary laser and auxiliary resonance.} The destabilizing drop of total intracavity power can be avoided if an additional \textit{auxiliary} laser (not participating in the parametric comb generation) is operated with sufficient power and an effective blue detuning to an auxiliary resonance. Auxiliary resonances can be resonances that are spectrally separated from the DKS, or resonances that are in an orthogonal mode family~\cite{zhang2019SubmilliwattlevelMicroresonatorSolitonsa,li2017StablyAccessingOctavespanninga, weng2022DualmodeMicroresonatorsStraightforwarda}.
In some implementations a sideband is derived from the pump laser through e.g., Brillouin-lasing or electro-optic modulation creating a blue-detuned laser for stability and a red-detuned laser driving the DKS\cite{li2017StablyAccessingOctavespanninga, wildi2019ThermallyStableAccess}.

\textit{(iii) Self-injection locking.} Self-injection locking relies on an intrinsic fast feedback mechanism between microresonator and driving laser that can stabilize the detuning $\zeta^T_0$. A detailed treatment of SIL and references are provided in Sec.~\ref{sec:exp_DKS:low_noise_ucombs:SIL}.
 
\textit{(iv) Synchronous pulsed driving.} The optically induced thermal resonance shifts are related to the average circulating power. This average power can be reduced, when the DKS is driven by a synchronized external pulse train whose repetition rate is synchronized with the DKS \cite{obrzud2017TemporalSolitonsMicroresonators}. Integer fraction driving repetition rates (\textit{sub-harmonic} driving) may also be used to drive the DKS with a lower repetition rate source~\cite{obrzud2019MicrophotonicAstrocomb, xu2020HarmonicRationalHarmonic}. If the driving pulse is temporally matched the driving efficiency can, in principle, approach unity.

\textit{(v) Photo-refractive effect.} It has been shown that the strong photo-refractive effect in lithium niobate acting in the opposite direction than Kerr- and thermo-optic resonance shift can create a stability for a red-detuned pump laser, enabling access to DKS states~\cite{he2019SelfstartingBichromaticLiNbO3}.

\subsubsection{Reducing the number of solitons via backward tuning}
\label{sec:exp_DKS:n_soliton}
\begin{figure}[h]
\includegraphics[width=0.95\columnwidth]{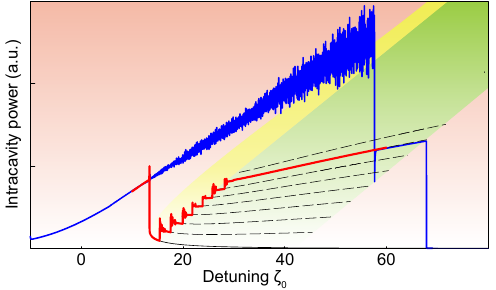}
\caption{\textbf{Backward tuning.} Simulation of forward (blue) and backward tuning (red) including thermal effects. The green area indicates the stable soliton states, yellow indicates breathing soliton states. Solitons cannot exist in the red area. The dashed lines show an analytical description of the soliton steps, with the analytical solution of soliton states in the system. Adapted from ~\cite{guo2017UniversalDynamicsDeterministica}. \label{fig:ThermalSwitchingSolitons}} 
\end{figure}

For many applications and studies of DKS the generation of exactly one DKS pulse is desired. In cases where multiple DKS are generated, \textit{backward tuning} of the driving laser, i.e., tuning the driving laser towards smaller detuning $\zeta^T_0$ may be utilized to gradually reduce the soliton number of DKS: When backward tuning, the driving laser approaches the DKS stability boundary, resulting in the decay of a DKS pulse. The concomitant reduction of the intracavity power and resonator temperature will lead to an increase of $\zeta^T_0$ stabilizing the remaining DKS. The procedure can be repeated until the desired number of DKS is reached. A numerical simulation of a forward scan into a multi-DKS state and subsequent backward tuning including thermal effects according to Eq.~\ref{eq:ch6:thermal_detuning} is shown in Fig.~\ref{fig:ThermalSwitchingSolitons}.

\subsection{Noise in microcombs}\label{sec:exp_DKS:noise}
Noise in soliton microcombs emerges from both fundamental and additional technical origins.
Fundamental sources of noise include  thermorefractive noise of the resonator\footnote{In the case of ferroelectric materials such as LiNbO$_3$, LiTaO$_3$, photorefractive effect~\cite{xu2021MitigatingPhotorefractiveEffect,taya1996PhotorefractiveEffectsPeriodically} and Johnson–Nyquist noise~\cite{zhang2025FundamentalChargeNoise} play an important role as well.}, coupling to the quantum mechanical vacuum and quantum noise in the pump laser. In addition, technical noise, including mechanical and electrical noise, as well as, excess phase and intensity noise in the pump laser may be present. These noise sources can impact the microcomb through various transfer mechanisms, including processes such as DW radiation and SRS.

\subsubsection{Tape model and microcomb linewidth}

\begin{figure}[h]
\includegraphics[width=0.99\columnwidth]{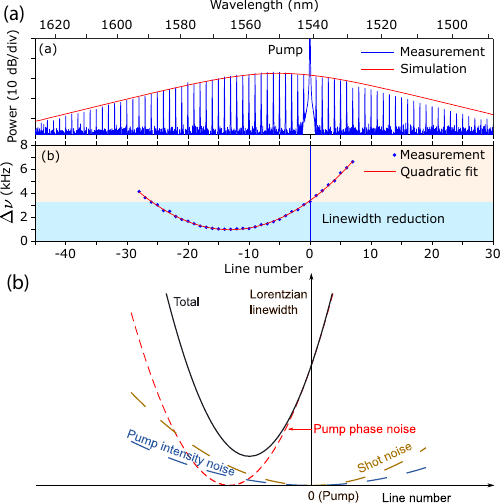}
\caption{\textbf{Spectral dependence of linewidth in soliton microcombs.} (a) Optical spectrum of a single-soliton microcomb (top) and the measured Lorentzian linewidth of comb lines  (bottom) according to the mean value of frequency noise power spectral density at high offset frequencies ranging from 3 to 5 MHz. (b) 
The shot noise and pump intensity noise affect directly the timing jitter of the soliton pulse train, which results in a linewidth distribution symmetrically located around the pump. These three effects together set the lowest achievable Lorentzian linewidth of soliton microcombs.
From~\cite{lei2021FundamentalOpticalLinewidth}. \label{fig:ch6:comb_noise}} 
\end{figure}

Noise in frequency comb spectra may be described via the \textit{elastic tape model}, which describes the symmetric breathing of a frequency comb about a spectral fix point~\cite{liehl2019DeterministicNonlinearTransformations, lei2022OpticalLinewidthSoliton}, as well as translation of this fix point. In this model, the phase noise of the comb lines $S_\mu(f)$ is described by
\begin{equation}
    S_\mu(f) = S_\mathrm{fix}(f) + (\mu-\mu_\mathrm{fix})^2 S_\mathrm{rep}(f)
\end{equation}
where $S_\mathrm{fix}$ and $S_\mathrm{rep}$ represent the phase noise power spectral densities (PSD) at the fix point $\mu_\mathrm{fix}$ (not necessarily an integer), and the repetition rate phase noise, respectively. In microcombs the fix point is usually close to the pump laser frequency; due to correlations between pump laser frequency noise and repetition rate noise it may however be shifted to $\mu_\mathrm{fix}\neq0$. The linewidth of the $\mu^\mathrm{th}$ comb line may be obtained via integration of $S_\mu(f)$~\cite{didomenico2010SimpleApproachRelation}. An example of the spectral dependence of soliton microcomb's linewidth and an illustration of the contribution of various noise processes is shown in  Fig.~\ref{fig:ch6:comb_noise}; the key noise processes are further discussed below.

\subsubsection{Fundamental quantum noise}
\label{sec:ch6:noise:quantum}
Quantum noise in microcombs originates from fundamental vacuum and photon-number fluctuations entering the cavity through all coupling channels (coupling waveguide and intrinsic loss), as those introduced in Eq.~\ref{eq:sec3:one_mode_CME} and that can be added to the full CMEs~\ref{eq:ch5:CME_rf}, and correspondingly to the LLE~\ref{eq:ch5:LLE}~\cite{chembo2016QuantumDynamicsKerra,matsko2013TimingJitterMode}.
As long as all baths and the pump remain in coherent states (i.e., no squeezing), the contribution from waveguide and intrinsic loss channels (as in Eq.~\ref{eq:sec3:one_mode_CME}) can be combined into a single $\delta$-correlated white-noise\footnote{More general Gaussian white noises (e.g.\ squeezed vacuum) can be defined by choosing different quadrature variances and correlations.} term $\sqrt{\kappa}\hat\xi_\mu(t)$ on the RHS of Eq.~\ref{eq:ch5:CME_rf} with
\begin{equation}
\bigl\langle \hat \xi_{\mu} (t) \hat \xi_{\mu'}^\dagger (t')\bigr\rangle
= \delta(t-t')\,\delta_{\mu\mu'},\qquad
\bigl\langle \hat \xi_{\mu } (t)\bigr\rangle = 0.
\end{equation}
For the LLE-based time-domain description, the same quantum noise appears as an additive stochastic field along the resonator coordinate, i.e. an extra term $\sqrt{\kappa} \hat{\Xi}(\phi,t)$ on the RHS of Eq.~\eqref{eq:ch5:LLE} with 
\begin{equation}
\bigl\langle \hat \Xi (\phi,t) \hat \Xi^\dagger (\phi',t')\bigr\rangle
= \delta(t-t')\,\delta_{2\pi}(\phi-\phi'),\qquad
\bigl\langle \hat \Xi (\phi,t)\bigr\rangle = 0.
\end{equation}
where $\delta_{2\pi}$ denotes a $2\pi$-periodic Dirac-$\delta$. 

The fundamental quantum noise defines a linewidth and timing jitter limit of soliton microcombs.
Stochastic methods and numeric simulation can be applied to compute the impact of quantum noise (and more generally white Gaussian noise) on microcomb~\cite{chembo2020FluctuationsCorrelationsKerr, liu2023StochasticApproachPhase}.
Specifically for DKS microcombs, a set of stochastic ordinary differential equations describing frequency and temporal shift of a DKS can be found based on the Lagrangian formalism (see Sec.~\ref{sec:basics_kcomb:DKS}) and using statistical properties of the added noise the fundamental soliton jitter power spectral density can be estimated as~\cite{matsko2013TimingJitterMode}:
\begin{equation}
S_t(\omega) \simeq \frac{\hbar \omega_\mathrm{p} \kappa}{W \omega^2}\left[\frac{\pi^2 \tau_{\mathrm{s}}^2}{12}+\frac{\left|\beta_2\right|^2 v_{\mathrm{g}}^2}{3 \tau_{\mathrm{s}}^2\left(\omega^2+\kappa^2\right)}+\frac{\omega\left|\beta_2\right| v_{\mathrm{g}}}{\omega^2+\kappa^2}\right]
\label{eq:ch6:timing_jitter}
\end{equation}
where $\omega_\mathrm{p}$ is the pump frequency, $W$ is the intracavity pulse energy, $\tau_{\mathrm{s}}$ represents the soliton pulse duration ($0.57$ of the FWHM pulse duration, c.f. Eq.~\ref{eq:ch5:sol_duration_and_width}), and $v_{\mathrm{g}}$ is the group velocity of the DKS. 
Eq.~\ref{eq:ch6:timing_jitter} has been verified experimentally in~\cite{bao2021QuantumDiffusionMicrocavity}.

\subsubsection{Fundamental thermodynamic noise}
\label{sec:ch6:noise:trn}
Small mode volumes enable efficient microcomb generation (see Eq.~\ref{eq:ch4:g0}), but also bring fundamental thermal fluctuation of the resonator (coupled to a thermal bath) to a relevant level. 
The thermal fluctuations cause thermorefractive noise (TRN) and dimensional fluctuation, setting the fundamental limitation of the resonance frequency stability in a microresonator~\cite{matsko2007WhisperinggallerymodeResonatorsFrequency}. 
The variance of thermodynamic fluctuations of temperature $\delta T$ is given by:
\begin{equation}
\left\langle\delta T^{2}\right\rangle=\frac{k_{B} T^{2}}{\rho C V},
\label{eq:Tfluc}
\end{equation}
where $k_\mathrm{B}$ is the Boltzmann's constant, $C$ is the specific heat capacity, $\rho$ is the density, and $V$ is the volume.

An expression for the power spectral density of the associated microresonator temperature fluctuations can be derived based on the fluctuation-dissipation theorem~\cite{kubo1966FluctuationdissipationTheorem}. Under the assumption of the infinite heat bath an analytic expression for whispering-gallery mode resonators was found~\cite{kondratiev2018ThermorefractiveNoiseWhispering}:

\begin{equation} 
S_{\delta T}(\omega)\approx\frac{k_{B} T^{2}}{\sqrt{\pi^{3} \kappa \rho C \omega}} \frac{1}{R \sqrt{d_{a}^{2}-d_{b}^{2}}} \frac{1}{\left[1+\left(\omega \tau_{d}\right)^{3 / 4}\right]^{2}},
\label{eq:TRN_analytic}
\end{equation}
where $R$ is the radius of the microresonator and, $d_a$ and $d_b$ are half widths of the modes transverse intensity profiles in horizontal and vertical direction assigned to indices $a$ and $b$ such that $d_{a}>d_{b}$, $\tau_d=\frac{\pi^{1 / 3}}{4^{1 / 3}} \frac{\rho C}{\kappa} d_b^2$. Eq.~\ref{eq:TRN_analytic} loses validity for $d_a\approx d_b$; in this case an estimate can be found by artificially multiplying $d_a$ with a factor and rescaling according to eq.~\ref{eq:Tfluc}.
Generally, accurate computation of $S_{\delta T}(\omega)$ for waveguide based resonators and their complex material composition necessitates FEM modeling
~\cite{huang2019ThermorefractiveNoiseSiliconnitride}.

\begin{figure}[h]
\includegraphics[width=0.95\columnwidth]{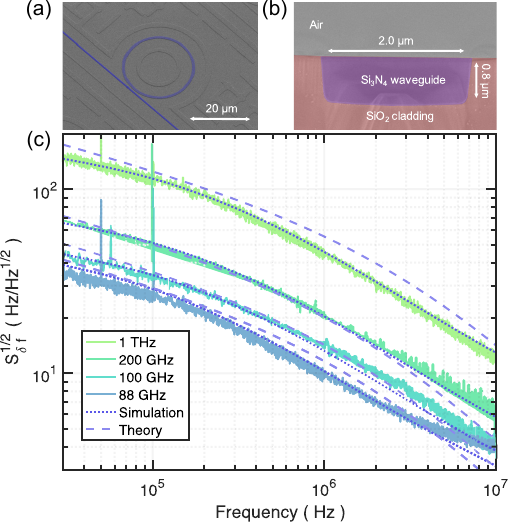}
\caption{\textbf{Thermorefractive noise.} (a) An image of a chip-integrated microresonator. (b) Corresponding cross-sections. (c) Comparison of thermorefractive noise in integrated Si$_{3}$N$_{4}$ microresonators of different sizes compared with numerical FEM simulations and theory. From~\cite{huang2019ThermorefractiveNoiseSiliconnitride}. \label{fig:ch6:TRN}} 
\end{figure}

The thermal fluctuations influence the microresonator resonance frequency and FSR mainly through the temperature dependence of the refractive index (thermo-optic/thermo-refractive effect) and thermal expansion (thermo-elastic effect)~\cite{chijioke2012ThermalNoiseWhisperinggallery}, as expressed by Eq.~\ref{eq:thermal_res_shift}, implying for the power spectral densities of resonance frequency $\omega_0$ and free-spectral range $D_1$ noise
\begin{equation}
\frac{S_{\delta \omega_0}}{\omega_0^2} = \frac{S_{\delta D_1}}{D_1^2}=\left(\frac{1}{n_\mathrm{eff}(\omega_0)} \frac{d n}{d T}\right)^{2} S_{\delta T}.
\end{equation}
Thus TRN directly impacts the effective detuning and the free-spectral range of the resonator. Figure~\ref{fig:ch6:TRN} compares analytical predictions with measurements, highlighting the trade-off between noise performance and mode volume.

TRN also impacts the repetition rate stability in microcombs~\cite{liu2020PhotonicMicrowaveGeneration}. Different approaches have been explored to reduce its impact, including counteracting thermal fluctuation through an effective blue-detuned pumping situation (cf. Sec.~\ref{sec:thermal_bistab_lock}), implemented through an auxiliary `cooling' laser, or a auxiliary `cooling mode' ~\cite{drake2020ThermalDecoherenceLaser,lei2022ThermalNoiseReduction}, dispersion-based TRN supression~\cite{stone2020HarnessingDispersionSoliton}, as well as bi-chromatic pumping, or other generalized pumping schemes that impose an externally defined repetition rate (Sec.~\ref{sec:generalized_pump}).

\subsubsection{Pump laser noise}
\label{sec:pump_noise}

In addition to fundamental quantum noise, there can be classical technical sources of noise. These can emerge for instance from technical pump laser noise, mechanical vibrations or fluctuating temperature. Indeed, microcomb performance is limited by the quantum noise only at high frequency offsets. Here, we consider technical noise of a CW pump laser noise as an example; other technical noise sources can be treated similarly. The laser noise can be represented by a Langevin noise term that can be added to the coherent driving field amplitude leading to the following replacement in Eq.~\ref{eq:sec3:one_mode_CME} and Eq.~\ref{eq:ch5:CME_rf}:
\begin{align}
 \sqrt{\kappa_{\mathrm{ex}}}\,\delta_{0\mu} s_\mathrm{in}\rightarrow \sqrt{\kappa_{\mathrm{ex}}}\,\delta_{0\mu}\,(s_\mathrm{in} + \xi_{\mathrm{las}}(t)),
\end{align}
where $\xi_{\mathrm{las}}(t)=\tfrac{1}{\sqrt{2}}\left[\delta x(t)+i\,\delta y(t)\right]$. The quadratures $\delta x(t)$ and $\delta y(t)$ are real-valued and  describe amplitude (intensity) and phase noise of the pump-laser, respectively. Their (two-sided) power spectral densities (PSDs) $S_x(f)$ and $S_y(f)$ are related to the Fourier-transforms of their autocorrelation functions, i.e.
\begin{align}
  S_x(f)& =\int_{-\infty}^{\infty}\langle \delta x(t)\,\delta x(t+\tau)\rangle \, e^{-i2\pi f\tau}d\tau
\end{align}
and analogously for $S_y(f)$ and $\delta y(t)$. In contrast to white quantum noise, $S_x(f)$ and $S_y(f)$ are generally non-white spectra.
Intensity noise is often characterized by relative intensity noise (RIN), which can be related to the amplitude quadrature noise PSD $S_x(f)$ via
\begin{align}
    \operatorname{RIN}(f) =  \frac{S_\mathrm{P}(f)}{P_{\mathrm{in}}^{2}} = \frac{4S_{x}(f)}{|s_\mathrm{in}|^2} 
\end{align}
, where $S_{P}(f)$ is the one-sided power spectral density of laser intensity noise in units of W$^2$/Hz. Frequency noise of the pump laser is often characterized by the power spectral density of the instantaneous frequency fluctuations and is usually expressed in units [Hz$^2$/Hz]. Equivalently, one may use the (one-sided) \textit{phase noise} PSD~\cite{riehle2004FrequencyStandardsBasics,rubiola2010PhaseNoiseFrequencyb}
\begin{equation}
    S_{\phi}(f) = \frac{S_{\nu}(f)}{f^2}
\end{equation}
which is conventionally expressed in units [dBc/Hz] or [rad$^2$/Hz], see~\cite{rubiola2023CompanionEnricoChart} for a discussion. For small phase excursions, it can be related to the phase quadrature noise PSD $S_{\delta y}(f)$ via
\begin{align}
    S_{\phi}(f) = \frac{S_{\delta y}(f)}{|s_\mathrm{in}|^2}.
\end{align}

When coupled to the resonator the pump laser noise experiences filtering and at the same time pump phase noise is converted into amplitude noise and vice versa (quadrature rotation). To provide an approximate description we consider only a single mode $a_0$ and neglect nonlinear effects. Considering the steady state of Eq.~\ref{eq:sec3:one_mode_CME} we find for the intracavity noise quadratures $\delta x^{(c)}$ and $\delta y^{(c)}$:
\begin{align}
    \frac{\partial}{\partial t}
    \begin{bmatrix}
    \delta x^{(c)} \\
    \delta y^{(c)}  
    \end{bmatrix} = 
    \begin{bmatrix}
    -\kappa/2 & \Delta  \\
    -\Delta & -\kappa/2  
    \end{bmatrix} 
    \begin{bmatrix}
    \delta x^{(c)} \\
    \delta y^{(c)}  
    \end{bmatrix} + \sqrt{\kappa_\mathrm{ex}}
    \begin{bmatrix}
    \delta x \\
    \delta y  
    \end{bmatrix}
\end{align}
where $\Delta=\omega_0-\omega_\mathrm{p}$ denotes the detuning between pump laser and resonance frequency, which could be modified through nonlinear effects or linear mode coupling. Fourier transformation and transitioning to noise PSDs under the assumption of uncorrelated input noise quadratures enables relating the intracavity and input noise PSDs:
\begin{align}
    \begin{bmatrix}
    \delta S_x^{(c)}(\omega) \\
    \delta S_y^{(c)}(\omega)  
    \end{bmatrix} = 
    \begin{bmatrix}
        \kappa_\mathrm{ex}\frac{(\kappa/2)^2 + \omega^2}{|D(\omega)|^2} & \kappa_\mathrm{ex} \frac{\Delta^2}{|D(\omega)|^2} \\
        \kappa_\mathrm{ex} \frac{\Delta^2}{|D(\omega)|^2} & \kappa_\mathrm{ex}\frac{(\kappa/2)^2 + \omega^2}{|D(\omega)|^2}
    \end{bmatrix}
    \begin{bmatrix}
    \delta S_x(\omega) \\
    \delta S_y(\omega)  
    \end{bmatrix}
    \label{eq:noise_PSD_intravacity}
\end{align}
where $D(\omega) = \Delta^2 + (\tfrac{\kappa}{2} -i\omega)^2$.
The general case, of a nonlinear cavity with microcomb formation, the systems response to noise is non-trivial and can be understood via the NDR (see~\ref{sec:ndr}).

\subsubsection{Pump noise transduction and quiet point}
\label{sec:ch6:noise:soliton}
In most of the integrated microresonator platforms, pump noise and TRN transferred to the microcombs define the microcomb noise levels and linewidth achievable in the experiment~\cite{lei2022OpticalLinewidthSoliton}. The amplitude and phase noises of the pump laser is transferred to the intracavity field, where noise filtering and quadrature rotation occurs (see Eq.~\ref{eq:noise_PSD_intravacity}). Generally, the noise properties of a soliton microcomb depend on both the effective detuning and the pump power~\cite{lucas2017DetuningdependentPropertiesDispersioninduced}.

Pump noise as well as other intrinsic noise sources can be transferred to the microcomb in complex pathways including Kerr-nonlinear effects (SPM, XPM), thermal effects, spectral reshaping and modified coupling to other mode-families. Two important mechanism for DKS are the Raman effect and the dispersive wave formation, both leading to a spectral shift of the comb's center frequency and thus to a change in the repetition rate~\cite{yi2017SinglemodeDispersiveWaves}:
\begin{equation}
    \omega_{\text {rep }}=D_1+\frac{D_2}{D_1}\left(\Omega_{\text {Raman }}+\Omega_{\text {DW}}\right),
    \label{eq:ch6:rep_rate}
\end{equation}
where $\Omega_{\text{Raman }}$ and $\Omega_{\text{DW}}$ are defined by Eq.~\ref{eq:ch6:raman_recoil} and Eq.~\ref{eq:ch6:dw_recoil}, respectively.

\begin{figure}[h]
\includegraphics[width=1.0\columnwidth]{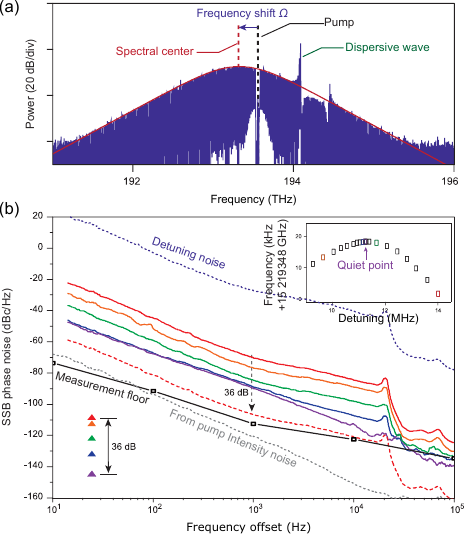}
\caption{\textbf{Quiet point in soliton microcombs.} (a) Soliton optical spectrum showing spectral envelope (red solid line), the attenuated pump
(black dashed line) and a strong dispersive wave. The spectral center of the soliton (red dashed line) is shifted in frequency relative to the pump frequency.
(b) Single-sideband (SSB) soliton microwave phase noise (solid curves) and calibration tone power (triangles) at different detuning
frequencies showing 36 dB of reduction at the quiet point. Inset shows the measured soliton repetition rate versus laser-cavity detuning, where the existence of a quiet point is revealed. Colors are preserved.
From~\cite{yang2021DispersivewaveInducedNoise}. \label{fig:qp1}}
\end{figure}

If the Raman and dispersive wave formation induced frequency shifts compensate each other, a particularly low-noise regime can be attained (\textit{quiet point})~\cite{yi2017SinglemodeDispersiveWaves,lucas2020UltralownoisePhotonicMicrowave, yang2021DispersivewaveInducedNoise}.
If pump laser frequency noise is the dominant source of noise, then the quiet point corresponds to the condition where the derivative $d\omega_\mathrm{rep}/d\delta\omega$ vanishes, with $\delta\omega = \omega_0 - \omega_\mathrm{p}$ representing the detuning between the cavity resonance frequency $\omega_0$ and the pump frequency $\omega_\mathrm{p}$. A reduction in the phase noise of the soliton repetition rate near the quiet point is illustrated in Fig.~\ref{fig:qp1}. Numerical simulations have shown that quiet point can be engineered and optimized, for instance in coupled resonators~\cite{triscari2023QuietPointEngineering}.

Additional noise transduction mechanism can arise from frequency dependence of intrinsic and extrinsic loss rates~\cite{matsko2015NoiseConversionKerr}, higher-order dispersion terms~\cite{stone2020HarnessingDispersionSoliton}.

\subsubsection{Stabilization of microcombs}
\label{sec:stabilization}
For many metrological applications of frequency combs such as precision spectroscopy, optical clocks or optical frequency division, the ability of phase stabilization of the comb to external frequency references is essential. Practically, for a microcomb this means stabilization of both $f_\mathrm{rep}$ and $f_p$, or equivalently, any pair of modes $(f_m, f_n)$ with $m\neq n$. To stabilize both degrees of freedom two non-collinear actuators are needed 
\begin{equation} 
    \binom{\delta f_\mathrm{rep}}{\delta f_p}=M\binom{\delta u_1}{\delta u_2}
    \label{eq:stabilization_and_control}
\end{equation}
where $M$ is a non-singular matrix describing the effect of the two actuation inputs $(\delta u_1,\delta u_2)$ on the $f_\mathrm{rep}$ and $f_p$~\cite{delhaye2008FullStabilizationMicroresonatorBased}.
Due to cross-coupling between both actuators the off-diagonal elements of $M$ are not generally zero. For microresonators pumped by an external laser actuation of the pump frequency and pump power (implying absorptive heating and thermo-refractive index change) are natural actuators~\cite{delhaye2008FullStabilizationMicroresonatorBased}. Moreover, electrical heaters~\cite{joshi2016ThermallyControlledComb,wildi2024PhasestabilisedSelfinjectionlockedMicrocomb}, auxiliary laser heating~\cite{jost2015AllopticalStabilizationSoliton}, piezo electric~\cite{papp2013MechanicalControlMicrorodResonator, liu2020MonolithicPiezoelectricControl} and electro-optic actuators~\cite{he2023HighspeedTunableMicrowaverate} have been implemented and in a self-injection locked system the pump diode current may be utilized~\cite{wildi2024PhasestabilisedSelfinjectionlockedMicrocomb}.
Moreover, sideband injection or pulse driving are also possible~\cite{obrzud2017TemporalSolitonsMicroresonators,weng2019SpectralPurificationMicrowave, brasch2019NonlinearFilteringOptical,stern2020DirectKerrFrequency, obrzud2019MicrophotonicAstrocomb}.   
To derive the error needed to stabilize $f_\mathrm{rep}$ and $f_p$ different techniques have been developed. Direct detection of the repetition rate is only possible in low-FSR resonators~\cite{yi2015SolitonFrequencyComb}, however, optical division of larger $f_\mathrm{rep}$ through electro-optic combs~\cite{delhaye2012HybridElectroOpticallyModulated} or lower repetition rate microcombs~\cite{spencer2018OpticalfrequencySynthesizerUsinga} have been implemented. Two combs with slightly different repetition rates in Vernier-configuration permit relating their repetition rates to their much smaller and detectable repetition rate difference. 
Measuring $f_p$ or the offset frequency has been demonstrated based on $f-2f$ and $2f-3f$ interferometry, which has so far usually involved complex setups~\cite{jost2015CountingCyclesLight,jost2015AllopticalStabilizationSoliton,delhaye2016PhasecoherentMicrowavetoopticalLink,brasch2017SelfreferencedPhotonicChip,spencer2018OpticalfrequencySynthesizerUsinga,drake2019TerahertzRateKerrMicroresonatorOptical,newman2019ArchitecturePhotonicIntegration,liu2021AluminumNitrideNanophotonics,wu2023VernierMicrocombsHighfrequency,wu2025VernierMicrocombsIntegrated}. Alternatively, the microcomb may be stabilized to an atomic frequency comb or to an atomic frequency standard~\cite{niu2023AtomreferencedStabilizedSoliton,qu2024CompactSolitonMicrocomb}.

\section{Conclusions and Outlook}
\label{sec:FuturePerspectives}

The field of microcombs combines the fascinating science of nonlinear systems, ultrafast photonics, and nonlinear photonics. The integration of high-Q nonlinear microresonators on a chip provides access to new nonlinear regimes and offers unprecedented control over the underlying nonlinear processes. This positions microcombs as a key enabling photonic technology. As this technology matures and progresses toward high-impact applications and commercial exploitation, new research frontiers emerge. These include, for example, extending to higher dimensionality in coupled systems, integrating with laser gain media, and exploring broadband, many-mode quantum correlations. This review, is intended to introduce new researchers to the field and provide a useful reference to support future research.

\bibliographystyle{apsrmp}
\bibliography{references}

\end{document}